\documentclass[11pt, showpacs]{article}
\pdfoutput=1
\usepackage{amsmath,amssymb,amsfonts,bm,marginnote,cite,graphicx}
\usepackage[colorlinks=true, pdfstartview=FitV, linkcolor=black, citecolor=black, urlcolor=black]{hyperref}
\usepackage[usenames,dvipsnames]{xcolor}
\newcommand{\hlt}[1]{{\color{WildStrawberry}{\em #1}}\index{#1}}

\edef\marginnotetextwidth{\the\textwidth}

\newcommand{\thistitle}{
	Metastring Theory and Modular Space-time
	}
\newcommand{\addresspi}{
	Perimeter Institute for Theoretical Physics, 
	31 Caroline St. N.,  Waterloo ON, N2L 2Y5, Canada
	}
\newcommand{\addressuiuc}{
	Department of Physics, University of Illinois,
 	1110 West Green St., Urbana IL 61801, U.S.A.
	}
\newcommand{\addressvt}{
	Department of Physics, Virginia Tech,  
	Blacksburg VA 24061, U.S.A.
	}

\newcommand{\be}{\begin{equation}}
\newcommand{\ee}{\end{equation}}
\newcommand{\beq}{\begin{eqnarray}}
\newcommand{\eeq}{\end{eqnarray}}
\newcommand{\bea}{\begin{eqnarray}}
\newcommand{\eea}{\end{eqnarray}}
\newcommand{\beqn}{\begin{eqnarray}}
\newcommand{\eeqn}{\end{eqnarray}}
\renewcommand{\Im}{{\rm Im }}
\renewcommand{\Re}{{\rm Re }}

\newcommand{\X}{\mathbb{X}}
\newcommand{\Pm}{\mathbb{P}}

\def\pa{\partial}
\def\om{\omega}
\newcommand{\rd}{\mathrm{d}}
\def\dbr{]\!]}
\def\dbl{[\![}
\def\II{\mathrm{I\!I}}

\def\dd{\!\cdot \!}
\def\S{\mathbb{S}}
\def\s{\sigma}

\def\P{{\cal{P}}}
\def\Q{{\mathbb{Q}}}
\def\Pol{{polarization }}
\def\Pm{{P-metric} } 
\def\Qu{{quantum }}
\def\Qm{Q-metric }
\def\bra{\langle}
\def\ket{\rangle}

\def\nn{\nonumber}

\newcommand{\myfig}[3]{
	\begin{figure}[ht]
	\centering
	\includegraphics[width=#2cm]{#1}\caption{#3}\label{fig:#1}
	\end{figure}
	}

\begin{document}

\title{\thistitle}
\author{
	{Laurent Freidel$^{a}$, Robert G. Leigh$^{b}$ and Djordje Minic$^{c}$ }\\
	\\
	{\small ${}^a$\emph{\addresspi}}\\ 
	{\small ${}^b$\emph{\addressuiuc}}\\ 
	{\small ${}^c$\emph{\addressvt}}\\
\\}
\date{\today}
\maketitle\thispagestyle{empty}

\begin{abstract}
String theory is canonically accompanied with a space-time interpretation which determines S-matrix-like observables, and connects to the standard physics at low energies in the guise of local effective field theory. Recently, we have introduced a reformulation of string theory which does not rely on an {\it a priori} space-time interpretation or a pre-assumption of locality. This \hlt{metastring theory} is formulated in such a way that stringy symmetries (such as T-duality) are realized linearly. In this paper, we study metastring theory on a flat background and develop a variety of technical and interpretational ideas. These include a formulation of the moduli space of Lorentzian worldsheets, a careful study of the symplectic structure and consequently consistent closed and open boundary conditions, and the string spectrum and operator algebra. What emerges from these studies is a new quantum notion of space-time that we refer to as a quantum Lagrangian or equivalently a \hlt{modular space-time}. This concept embodies the standard tenets of quantum theory and implements in a precise way a notion of {relative locality}. The usual string backgrounds (non-compact space-time along with some toroidally compactified spatial directions) are obtained from modular space-time by a limiting procedure that can be thought of as a correspondence limit. 
\end{abstract}

\newpage
\section{Introduction}

After more than 40 years \cite{Cappelli:2012cto} the deep nature of string theory \cite{Polchinski:1998rq} remains largely hidden. In its conventional formulation, space-time is taken to be the target space of a worldsheet sigma model. 
It is widely taken for granted that the raison d'\^etre for string theory is to provide local effective field theories on a (non-compact) space-time in a setting that incorporates quantum gravity. These theories are complete from this field theory point of view in the sense that they are apparently ultraviolet finite. 

Whenever one pushes the theory to its limits, by looking for example at high energies or short distances, there are indications that the structure of local quantum field theory in a fixed space-time cannot be correct. Certainly the UV finiteness fits with this. More generally, presumably in any theory of quantum gravity, one expects cross-talk between short and long distances and thus some form of non-locality. This is manifested in a variety of ways. It is well-known that there are no local observables in gravity, a fact that was so crucial in the development of holographic space-times. But perhaps even more fundamentally, if one probes quantum gravity theory at very short distances, of the order of the Schwarzchild radius of some probe, then it has been suggested that some sort of `classicalization' may emerge, involving large scale physics. Conceptually, this feels consistent with one of the avatars of string theory, T-duality, in which under certain conditions, short and long distance physics are swapped --- a new notion of space-time emerges at short distances (at least along compactified dimensions).  Presumably all of these exotic properties of string theories are tied to the fact that what we conceive of as classical geometries are  fully discoverable only by particle-like probes. So if we ask any question of string theories that gets at some non-particle aspect, we are likely to lose contact with an understanding within local effective field theory. There are many examples of this sort of effect, involving either perturbative or non-perturbative string physics. A central issue going hand in hand with the emergence of space-time, is the emergence and nature of locality.

In two recent letters \cite{Freidel:2013zga,Freidel:2014qna} we introduced a new formulation of string theory as a quantum theory living outside of the usual space-time framework.
Our motivation for developing such a theory, which we now call \hlt{metastring theory}, is manyfold. It is based on the same fundamental concepts as is the usual string theory, departing from it in its initial assumptions about physical space-time. In the present paper, we will explore some aspects of this theory, establishing  a number of foundational principles and interpretations. 
Some of the structure of the theory that we construct is shared by double field theory \cite{Siegel:1993th,Siegel:1993xq,Siegel:1993bj,Hohm:2010pp,Hull:2009zb,Hohm:2010jy,Hull:2009mi,Zwiebach:2011rg,Aldazabal:2013sca,Berman:2013eva, Blumenhagen:2014gva} and the so-called generalized geometries \cite{Hitchin,Gualtieri:2003dx, Weinstein}. As we move through the paper, we will  be specific about the differences between our formulation and those treatments. 

Classically, our starting point will be the Tseytlin action. The form of this action, at least for flat backgrounds (which we mostly confine ourselves to in this paper), can be derived directly from the Polyakov path integral. One of the main features of this formulation is that it is chiral; a second feature is that the target space of this formulation is a phase space and not space-time. 
The utility of this formulation is that T-duality acts linearly on the target space coordinates, which also explains its role in double field theory. As we described in \cite{Freidel:2014qna}, our interpretation is more general than just implementing T-duality, but touches on  
the foundations of quantum theory as it relates to string theory. In quantum gravity, there are a number of distinct ways to formulate theories, differing in what is taken as the set of fundamental objects. Are the fundamental objects the smallest (particles, strings) or the largest (space-time itself)? Making either choice means that that choice must define the other.  In the worldsheet path integral formulation of string theory, the fundamental {\it probes} are strings;\footnote{Of course, string theory contain other objects that become visible at finite coupling. These are expected to play a vital role in a complete theory.} in the usual formulation we regard them as probes of a given space-time theory. But another point of view is that they {\it define} what we mean by space-time, that the geometry is determined by how probes interact with one another. 

In the usual formulation of string theory, all the probes agree on a notion of space-time, as space-time is the target space. This of course is ambiguous when (spatial) dimensions are compactified, but becomes unambiguous in a given limit (such as large  or small radius).  
In the chiral phase-space formulation, T-duality gives an action on the phase-space coordinates. At least classically, a choice of a space-time can be thought of as a choice of polarization, in that we identify space-time with a (Lagrangian) submanifold\footnote{We emphasize that when we talk about phase space, we always mean the phase space of probes of space-time and momentum space, such as strings, and {\it not} of the phase space
of gravitational fields, which are emergent in string theory.}. In double field theory, one imposes a constraint that is equivalent to identifying a particular submanifold of this phase space as space-time. 

In the absence of interactions amongst strings, it is perhaps not obvious that different strings should view the same Lagrangian submanifold as space-time. We think of this as an implementation of Born reciprocity ($X\to P, P\to -X$). This interpretation is particularly clear if we think in terms of string wave-functionals whose natural basis specifies the position in space-time of string loops. In this context, passage to other Lagrangian submanifolds is obtained by Fourier transformation. In fact, this Fourier transform implements generalized T-dualities in the compact case.
In ordinary quantum mechanics we may, depending on convenience, choose a position or momentum basis of states; it is a fundamental property of quantum theory that this choice of polarization is immaterial. In quantum gravity, if all probes agree on what we mean by space-time, then we have broken Born duality --- there is a preferred choice of polarization, the space-time one. Thus, we emphasize that a suitable notion of quantum gravity is not as a quantization of a space-time theory, but rather should be viewed in a broader context in which space-time is a choice of polarization. This is the structure that metastrings provides. From this point of view the fact that there is a preferred interpretation of space-time in the usual string theory implies some degree of classicality. 
 We refer to this as \hlt{absolute locality}: the same space-time is shared by all probes, independently of their energy state or their history. It is worth pointing out that absolute locality is an assumption that underlies the interpretation of all cosmological observations, as well as all high energy experiments.

We distinguish absolute from \hlt{relative locality} \cite{AmelinoCamelia:2011bm,Freidel:2011mt}, the idea that each probe has, in a sense, its own notion of space-time. Colloquially, it is only when probes talk to one another, through interactions, that they compare their choices.  One manifestation of this idea is that the dual momentum-energy space becomes curved (indeed, in quantum field theories, absolute locality is implemented by the linearity of momentum space), an idea that goes back to Max Born \cite{born1938suggestion,born1949reciprocity}\footnote{As well, a few attempts have been made  
to incorporate momentum space curvature as a regulator in quantum field theory, without any definite success. The efforts of Snyder \cite{Snyder:1946qz} and Golfand \cite{golfand10} are particularly noteworthy. Curved momentum space plays a central role also in 3d quantum gravity\cite{Freidel:2005me,Freidel:2005bb}.}. Another motivation for introducing metastring theory is to implement the idea of relative locality in a theory that has a chance to be a complete theory of quantum gravity. We will see that indeed there is a notion of relative locality that emerges in the metastring.

Fixing a specific submanifold as space-time can be thought of within the process of quantizing the string as a choice of specific boundary conditions, constraining the form of string zero modes, in particular, the monodromies. This is the first primary difference between the usual string and the metastring: in the metastring, we do not impose such constraints from the outset, but merely ask the metastring to be consistent with its gauge symmetries and with worldsheet locality. Thus our first task in this paper will be to formulate the Tseytlin theory allowing for generic monodromies. 

Such a formulation requires us to consider carefully the general problem of summing over worldsheets. Because the Tseytlin theory does not possess manifest worldsheet Lorentz invariance at the level of the action, we consider the formulation of {\it Lorentzian} worldsheets, extending old work of Giddings and Wolpert \cite{Giddings:1986rf}, Krichever and Novikov \cite{Krichever:1987,Krichever:1987qk}, and Nakamura \cite{nakamura2000calculation}. The Lorentzian formalism allows us to consider a more generalized notion of closed string boundary conditions, based not on the vanishing of monodromies, but on the continuity of symplectic flux. 

The relaxation of the zero mode sector to allow for general monodromies cannot be implemented without restrictions. Consistent with the diffeomorphism constraint, we will in general have `dyonic states' in  the spectrum. Thus, the imposition of worldsheet locality on the algebra of vertex operators is a non-trivial condition. Remarkably, we find that this constraint implies that there is a unique\footnote{As we will clarify later, the uniqueness applies to a certain class of boundary conditions which do not include, for example, orbifolds.} Lorentzian lattice dual to the target space. 

The usual interpretation in ordinary string theory would be that this lattice is the Narain lattice of a string theory on a fully compactified Lorentzian space-time. It seems unlikely\footnote{We note that Moore \cite{Moore:1993zc} has previously tried to make sense of such a compactification.} that such an interpretation gives rise to a sensible theory (causality for example, would seem hard to implement). Note that in such an interpretation, the space-time is a Lagrangian submanifold of the target space. By studying the quantum algebra of vertex operators, we find that in fact another interpretation comes to the forefront involving a quantum notion of Lagrangian submanifold, which we refer to as \hlt{modular space-time}. In fact, this interpretation fits well with ideas in ordinary quantum mechanics formulated by Aharonov and Rohrlich \cite{aharonov2008quantum}.  These authors have shown that modular observables are the ones that allow to observe quantum interferences. They have no classical analog and obey non-local equations of motions. They also argue that, remarkably, thanks to the uncertainty principle, this dynamical non-locality does not lead to a violation of causality.
This dynamical non-locality is the source of some of the most striking quantum mechanical effects, such as the Aharonov-Bohm or Aharonov-Casher effects \cite{Aharonov:1959fk,Aharonov:1984xb}. We establish here that the modular space-time experienced by the metastring is colloquially obtained  by replacing classical coordinates by modular coordinates
which form a commutative sub-algebra of the quantum phase space algebra. 
The appearance of modular space-time is fundamentally non-perturbative, and even if it contains in some sense a doubling of the target it cannot be understood in terms of   $\alpha'$ corrections as considered in the context of double field theory. 
Some of the key features of modular space-time have been already discussed on the other hand, although not in our terms, in the context of the 
`non-geometrical backgrounds' such as monodrofolds or T-folds  \cite{Hellerman:2002ax, Hellerman:2006tx, Dabholkar:2002sy,Hull:2004in, Hull:2006va, Hull:2007jy}.

It is of interest to consider the notions of `quantum' and `classical' in what we have described here. Even in the usual string theory, there are many layers to these notions; certainly, the worldsheet theory is quantum in the usual sense (being a (path-integral) quantization of  a well-defined classical theory). From the space-time point of view (even if we confine attention to string perturbation theory), it is also quantum in the sense of the S-matrix interpretation in (asymptotically-) flat backgrounds, and perturbative in the corresponding expansion in powers of $g_{str}$. Clearly, given the progress over the last 20 years, it is not enough to describe string theory as an S-matrix theory, and this is even more clear if, as in the metastring, there is no a-priori notion of space-time.
The metastring is formulated as a worldsheet theory, and so there is a definite notion of quantum from the worldsheet point of view. However, it has long been known \cite{Veneziano:1986zf} that in the Polyakov string there is no direct notion of $\hbar$; instead there is a length parameter $\lambda$ that sets the scale of length on the target space. In the Tseytlin form of the action, there are actually two scales $\lambda$ and $\varepsilon$ whose product and quotient correspond to $\hbar$ and $\alpha'$ respectively. 

In fact, given our notion of modular space-time we should ask in what sense the usual string backgrounds can be recovered. In fact, as we will now summarize, they can be recovered from the metastring via `classical' (for lack of a more precise term) limits. Modular space-time corresponds to a cell in phase space whose size is set by $\lambda$ and $\varepsilon$. It reduces to the classical notion of Lagrangian submanifold in a limit, such as $\lambda\to 0$, in which the cell is squashed (preserving volume) in half the directions. Depending on how this squashing is done, one may obtain a theory identical to any compactification of the usual bosonic string (and presumably any superstring as well) with any number of non-compact directions. The low energy physics of such a compactification is local and causal. 

 Another consequence of these ideas is that they inevitably lead to a certain \hlt{gravitization of quantum theory}. This notion has been suggested before \cite{Penrose:2014nha}, but such discussions have always been hampered by the necessity of discussing it within (semi-)classical GR. It seems natural in unifying the geometrical nature of general relativity and the rigid algebraic structure of quantum theory that both must learn from each other. In the context of the metastring, the rigidity of the quantum theory is encoded 
into the flatness of the polarization metric $\eta$, a metric in phase space that tells us how to define the notion of Lagrangian submanifolds. In order to make the metastring consistent on general backgrounds this metric needs to be curved and hence the rigidity of quantum mechanics will be relaxed once  we  show that the metastring theory is a consistent quantum theory.
Trying to quantize the metastring and keep the flatness of the polarization metric leads to inconsistent truncations and presumably explains some of the tensions and difficulty inherent to double field theory, for example.  Indeed we will later see that the metastring admits in its spectrum vertex operators which are the seeds of deformation of the \Pol  metric $\eta$.

In the future we intend to develop the theory of metastrings on arbitrary backgrounds. To begin, in this paper we will consider the semi-classical structure of the simplest example, involving only a flat background. Although this is far from our ultimate goals, it is important to establish a firm foundation, based on free worldsheet field theory techniques. 

The organization of this paper is thus as follows. In Section 2, we recall the derivation of the Tseytlin $\sigma$-model, which we interpret as a chiral theory on a $2d$-dimensional target that we call phase space. In this section, we also discuss some geometrical aspects of this target and the symmetries and constraints of the $\sigma$-model. In particular we show how the chiral $\sigma$-model  necessitates the introduction of a quantum metric $H$ (also called generalized metric) and \Pol metric $\eta$ and a phase space 2-form $\omega$. The absence of worldsheet Lorentz invariance of the $\sigma$-model action leads us in Section 3 to consider the formulation of Lorentzian worldsheets.  
In Section 4, we consider the canonical analysis of the metastring. In particular we construct the symplectic structure on a strip geometry and show that there is a consistent notion of closed string boundary conditions. A more thorough analysis of the gluing of arbitrary genus Lorentzian worldsheets is reserved for a future publication. 
In Section 5, we briefly summarize some features  of quantum amplitudes of the metastring, culminating in the derivation of the unique Lorentzian lattice $\Lambda_d = \II_{1,d+1}\times \II_{1,d+1}$ as a label of the zero modes of the metastring  states. In Section 6, we discuss metastring observables and their canonical bracket. We also show how the classical metastring observables are the canonical generators of phase space diffeomorphism symmetry. The imposition of mutual locality at the quantum level leads us to the realization that the classical notion of projecting to a Lagrangian submanifold must be replaced by the notion of $\Lambda$-periodicity. In Section 7, we elaborate on this idea and argue that $\Lambda$-periodicity can be interpreted in terms of modular variables. We finish this section with a brief discussion of `classical' limits of modular space-time and how the effective description of strings can be done in terms of fields defined on a modular space-time. In an Appendix, we briefly extend our previous discussion of symplectic structure to worldsheets with time-like boundaries and thus establish a few notions of the open metastring. In Section 8, we conclude with comments on the present status of the metastring theory and future investigations.

\section{Sigma Model in Phase Space}

We are now ready to formulate the metastring theory. As we mentioned above, our aim is to establish a theory that is capable of describing curved space-times and momentum space simultaneously. We review here the passage to such a theory, which we obtain by deforming the usual Polyakov path integral formulation. 
We begin our discussion \cite{Freidel:2013zga} by examining the Polyakov action coupled to a flat metric $h$, 
\beq
S_{P}(X) =\frac1{4\pi} \int_{\Sigma} h_{\mu\nu} (*\rd X^{\mu} \wedge  \rd X^{\nu}),
\eeq
where $*,\rd$ denote the Hodge dual and exterior derivative on the worldsheet, respectively.
We generally will refer to local coordinates on $\Sigma$ as $\sigma,\tau$, while it is traditional to interpret $X^\mu$ as local coordinates on a target space $M$, here with Minkowski metric $h_{\mu\nu}$. Since we are in Lorentzian signature,  
$*\rd \tau= \rd \sigma$ and $*^{2}=1$.
Note that $S_P$ has dimensions of length-squared if we take $X^\mu$ to have dimension of length, so appears in the path integral as $e^{iS_P/\lambda^{2}}$. $\lambda$ is the string length which is related to the slope parameter by 
 $\lambda^2 \equiv \alpha'\hbar$, where $\hbar$ is the Planck constant of the worldsheet quantum theory. With this definition $S_P/\lambda^2$ has the usual coefficient $\frac{1}{4\pi\alpha'}$ in units of $\hbar$.
In order for the Polyakov action to be well-defined,
one must demand that the integrand be single-valued on $\Sigma$. 
For example, on the cylinder $(\sigma,\tau)\in [0,2\pi]\times \mathbb{R}$ it would be sufficient
that $\rd X^{\mu}(\sigma,\tau)$ is periodic\footnote{The most general condition would be to ask that $\rd X^{\mu}(\sigma+2\pi) = \Lambda^\mu_\nu \rd X^{\nu}(\sigma)$ where $\Lambda$ is a Lorentz transformation. In this work we only consider the case where $\Lambda=1$. This restriction is a fundamental limitation of our analysis that excludes, in particular,  orbifolds.}  with respect to $\sigma$ with period $2 \pi$.
However, and this is a crucial point, this does not mean that $X^{\mu}(\sigma,\tau)$ has to be a periodic function, even if $M$ is non-compact.
Instead,  it  means that $X^{\mu}$ must be a {\it quasi-periodic} function which satisfies 
\be\label{defdel}
X^{\mu}(\sigma + 2\pi, \tau) = X^{\mu}(\sigma,\tau) + \delta^{\mu}.
\ee
Here $\delta^{\mu}$ is the \hlt{quasi-period}, or monodromy,  of $X^{\mu}$.
If $\delta^{\mu}$ is not zero, there is no {\it a priori} geometrical interpretation of a closed string propagating in a flat space-time -- periodicity goes hand-in-hand with a space-time interpretation.
{Of course, if $M$ were compact and spacelike then $\delta^\mu$ would be interpreted as winding, and it is not in general zero \cite{Freidel:2013zga}.} However since we want ultimately to generalize the $T$-duality to curved backgrounds, we do not want to impose the restriction that there is a space-time interpretation of the monodromies. Instead we want to find what 
conditions these monodromies have to satisfy.
As we stressed in \cite{Freidel:2013zga} the string can be understood more generally to propagate inside a portion of a space that we will refer to as phase space $\cal P$. 
What matters here is not that string theory possesses or not a geometrical interpretation but whether it can be defined consistently. This is no different than the usual CFT perspective, in which there are only a few conditions coming from quantization that must be imposed; a realization of a target space-time is another independent concept.  It has always been clear that the concept of T-duality must change our perspective on space-time, including the cherished concept of locality, and so it is natural to seek a relaxation of the space-time assumption.

In order to present our perspective on T-duality, let us
consider the dimensionless first order action\footnote{The passage from the usual Polyakov formulation to this can be performed straightforwardly in the full worldsheet path integral.}
\be
\hat{S} = \frac1{2\pi} \int_{\Sigma} \left( \frac{1}{\lambda \varepsilon} \bm{P}_\mu \wedge\rd X^\mu+ \frac{1}{2\varepsilon^2} h^{\mu\nu} (*\bm{P}_\mu \wedge \bm{P}_\nu) \right),
\ee 
where $\varepsilon$ is a momentum scale, $\lambda$ is a length scale and $\bm{P}_\mu$ is a one form with dimension of mass.
If we integrate the one form  $\bm{P}$ we get back the space-time Polyakov action, and if we integrate $X$ we get the momentum space Polyakov action. 
Indeed, if we integrate out $\bm{P}_\mu$, we find
$*\bm{P}_\mu=\tfrac{\varepsilon}{\lambda} h_{\mu\nu} \rd X^\nu
$
and we obtain the Polyakov action
$
{\hat{S}} \to\frac{1}{\lambda^2} S_{P}(X).
$

Now, the reader may come to the conclusion that  $\lambda,\varepsilon$ are not independent scales, and this would be true within the confines of this flat non-interacting theory. However, the introduction of $\varepsilon$ here is an important step conceptually \cite{Veneziano:1986zf}. In any theory of quantum gravity, we expect to find {\it three} dimensionful constants, $c$, $\hbar$ and $G_N$. Putting $c=1$ aside, this implies that 
quantum gravity depends both on a length scale $\lambda_{P}\propto \sqrt{\hbar G_{N}}$ and an energy scale $\varepsilon_{P}\propto \sqrt{\hbar}/\sqrt{G_{N}}$ (here we are using the language of  dimension $4$ for simplicity). As was emphasized by Veneziano long ago, the usual formulation of string theory as a theory of quantum gravity contains a puzzle:  there is apparently only one dimensionful scale, $\lambda$ (or equivalently, $\alpha'$) that appears directly in the quantum phase factor. In the presence of both a length scale $\lambda$ and an energy scale $\varepsilon$, we can reconstruct 
\be
\hbar = \lambda \varepsilon,\qquad  \alpha'= \lambda/ \varepsilon.
\ee
The Newton constant is proportional to the latter scale, $G_{N}=\rho \alpha'$, depending on the dimension and the details of compactification. 

Of course, in the present context, these constant scales can be reabsorbed into a redefinition of the fields $(X, \bm{P})$.
The significance of the parameters are only seen when we ask questions about specific probes in the phase space target theory (e.g., we compare a probe momentum to $\varepsilon$), or if we consider backgrounds that have their own inherent length scales (such as a curvature scale). 

Now, on the other hand, if we integrate out $X$ instead, we get
$\label{intXflat}
\rd \bm{P}_\mu=0,$ and so we can locally write 
 $
 \bm{P}_\mu= \rd Y_\mu,
 $
 where $Y$ can be thought of as a momentum coordinate. 
 It is in this sense that there is ``one degree of freedom" in  $\bm{P}$ even though it is a worldsheet 1-form-- on-shell, $\bm{P}$ is locally equivalent to the scalar $Y$.
 Notice though that this is true only locally, and in order to interpret it globally we must allow
 $Y_{\mu}$ to be multi-valued on the worldsheet. That is,  even if we assume that $X^\mu$ is single-valued to begin with, $Y_{\mu}$ should carry additional monodromies associated with each non-trivial cycle of $\Sigma$.
 This means that the function $Y$ is only quasi-periodic with periods given by the momenta
 \be\label{defp}
 \oint_{C} \bm{P}_{\mu} = \oint_{C}\rd Y_{\mu}= 2\pi p_{\mu}.
 \ee
 The action for $Y$ (obtained by integrating out $X$) becomes  essentially the Polyakov action, with the addition of a boundary term
 \be\label{dualPoly}
{\hat{S}}= \frac1{2\pi \lambda \varepsilon} \int_{\pa\Sigma}    Y_\mu \rd X^\mu   +\frac1{\varepsilon^{2}} S_{P}(Y).
 \ee
Note that $X$  has the dimension of length while $Y$ has the dimension of momentum.
 We recover the Polyakov action for the momentum variable, with $\varepsilon$ playing the role of $\lambda$ for this dual theory. 
The presence of the boundary term in (\ref{dualPoly}) is related to the fact that the transformation $X^\mu \to Y_\mu$ corresponds to the string Fourier transformation \cite{flmq}.
Indeed, as we will see, for a boundary located at fixed  $\tau$, $\pa_\s X^\mu$ and $Y_\mu$ are conjugate variables satisfying
\be
\{ \pa_\s X^\mu(\s), Y_\mu(\s')\} = 2\pi \delta_P(\s,\s'),
\ee
where $\delta_P$ is the periodic delta function.

In order to obtain a formulation where we are left with a phase space action, a natural idea is to partially integrate out $\bm{P}$. 
In a local coordinate system on the worldsheet, we write
the decomposition of the momentum one-form
\be
\bm{P}_{\mu} = P_{\mu} \rd \sigma +  Q_{\mu}\rd \tau.
\ee
 In conformal co-ordinates the first order action then reads\footnote{Our conventions are such that in the conformal frame the 2d metric is $-d\tau^{2} + \rd \sigma^{2}$ and  
$*\rd\sigma =\rd \tau$, $*\rd \tau=\rd \sigma$ and $\rd\sigma \wedge \rd \tau = \rd^{2}\sigma$. Here $\int [\cdot] $ means $\int \rd^2\s[\cdot] $.}
\be
\hat{S} = \frac{1}{2\pi\lambda \varepsilon}\int \left( P_{\mu} \pa_{\tau}{X}^{\mu} -  Q_{\mu} \pa_{\sigma}X^{\mu} + \frac{\lambda}{2\varepsilon} (Q_{\mu}Q^{\mu}- P_{\mu}P^{\mu} )\right).
\ee
The equations of motion for $P, Q$ are simply 
\be
P = \frac{\varepsilon}{\lambda}\pa_{\tau}{X},\qquad  Q= \frac{\varepsilon}{\lambda}\pa_{\sigma}X.
\ee
By integrating out $Q$, we insert the $Q$ equation of motion and get the action in Hamiltonian form:
\be
\hat{S}
=\frac{1}{2\pi \lambda \epsilon}\int P_{\mu}\cdot\pa_\tau X^{\mu} -\frac{1}{4\pi}\int\left(\frac{\lambda}{\varepsilon}h^{\mu\nu} P_{\mu} P_{\nu} +
\frac{\varepsilon}{\lambda} h_{\mu\nu}
\partial_{\sigma}X^{\mu} \partial_{\sigma}X^{\nu} \right).
\ee
Now we  locally introduce a momentum space coordinate $Y$ such that 
$\pa_\sigma Y = P$. Like $X$, this coordinate is not periodic, its quasi-period $2\pi p\equiv Y(2\pi) -Y(0) $ is proportional to the string momentum.
Using this coordinate the action becomes simply
\be
\hat{S}\to \frac1{2\pi}
\int \left[\frac1{\lambda \varepsilon} \partial_{\tau}{X}^{\mu} \partial_{\sigma}Y_{\mu} 
- \frac{1}{2\varepsilon^{2}} h^{\mu\nu}\partial_{\sigma}Y_{\mu} \partial_{\sigma}Y_{\nu} 
-\frac{1}{2\lambda^{2}} h_{\mu\nu}\partial_{\sigma}X^{\mu} \partial_{\sigma}X^{\nu}
\right].
\ee
The main point is that in this action both $X$ and $Y$ are taken to be quasi-periodic.
The usual Polyakov formulation is recovered if one insists that $X$ is single-valued, and the usual T-dual formulation is recovered if one insists that quasi-periods of $X$ appear only along space-like directions and have only discrete values.

It is convenient, as is often used in the double field theory formalism \cite{Siegel:1993th,Siegel:1993xq,Siegel:1993bj,Hohm:2010pp,Hull:2009zb,Hohm:2010jy,Hull:2009mi,Zwiebach:2011rg,Aldazabal:2013sca,Berman:2013eva, Blumenhagen:2014gva}, to unify both 
$X^\mu$ and $Y_\mu$ in one space $\P$ (that we often refer to as phase space) and introduce a dimensionless coordinate $\X$ on $\P$
\be\label{XY}
	\X^{A} \equiv \left( \begin{array}{c} {X^{\mu}}/\lambda \\  Y_{\mu}/\varepsilon\end{array} \right).
	\ee
To write the action, we introduce 
 a  constant neutral\footnote{Here neutral means that $\eta$ is of signature $(d,d)$, while $H$ is of signature $(2,2(d-1) )$.} metric $\eta^0$,  a constant metric $H^0$
 and a constant symplectic form $\omega_{AB}^0$
\be\label{etaH0}
	\eta_{AB}^0 \equiv \left( \begin{array}{cc} 0 & \delta \\ \delta^{T}& 0  \end{array} \right),\quad
H_{AB}^0 \equiv  \left( \begin{array}{cc} h & 0 \\ 0 &  h^{-1}  \end{array} \right),
\quad \om_{AB}^0 = \left( \begin{array}{cc} 0 & \delta \\ -\delta^{T}& 0  \end{array} \right) ,
\ee
where $\delta^{\mu}_{\nu}$  
 is the $d$-dimensional identity matrix and $h_{\mu\nu}$ is the $d$-dimensional Lorentzian metric, $T$ denoting transpose.
 The presence of a symplectic structure $\omega^0$ expresses the fact that $\cal P$ is a symplectic manifold. 
 The space-time vectors of the form $ (X^\mu,0)$ defines a subspace $ L$ of $\P$.
 Similarly  momentum space vectors of the form $(0,Y_\mu)$ form defines another {\it transversal}  subspace $\tilde{L}$ of $\P$.
 Moreover, we see that  both the  space-time subspace   $L$ or momentum-space subspace $\tilde{L}$ are Lagrangian subspaces of $\P$ of maximal dimension.
  That is the 
 symplectic structure $\omega^0$ vanishes on both of them and $\P =L\oplus \tilde{L}$.
 We can also see that both $L$ and $\tilde{L}$ are  null subsets of $\P$ with respect to $\eta^0$. That is $ \eta^0_{AB} \X^A\X^B=0$ if $\X \in L$ and similarly for $\tilde{L}$.
 The $\eta^0$ metric has therefore the property that its 
 null subspaces are Lagrangian manifolds of maximal dimension.
 A choice of Lagrangian subspace of phase space is called a choice of polarization. We therefore refer to the metric $\eta^0$ as a \hlt{\Pol metric} or \Pm\!\!.
 This metric is of signature $(d,d)$ and it is therefore neutral. The subscript $0$ refer to the fact that the metric is constant in the present discussion.
 
 The metric $H^0$ already appears in the context of double field theory  and generalized geometry \cite{Gualtieri:2003dx} and is often referred to as the generalized metric.
 We feel however that this denomination misses the point that $\P$ is a phase space and that this metric can be understood as descending from the quantum probability 
 metric applied to coherent states \cite{Freidel:2014qna,flmq}.
 Therefore we refer to this metric as the \hlt{\Qu metric} or \Qm\!\!.
 This metric is of signature $(2,2(d-1))$, the two negative eigenvalues corresponding to the time direction in space-time and the energy direction in energy-momentum space.
 When restricted onto the space-time Lagrangian subspace $L$ it provides the space-time metric $g = H|_L$.

 The \Pm and the \Qm  are not independent in the present context: if we define 
\be
J_0\equiv (\eta^{0})^{-1}H^0,
\ee
we see that $J_0$ is an involutive  transformation preserving $\eta^{0}$,
that is, 
\be
J^{2}_0=1,\quad \mathrm{ and} \quad  J^{T}_0\eta^0 J_0=\eta^0.\label{Jprops}
\ee
  $(\eta^0,J^0)$ defines a \hlt{chiral structure}\footnote{Also called a para-complex structure in the mathematical literature  \cite{cruceanu1996survey}.} on phase space $\cal P$.
  We also introduce the constant symplectic form:
\be\label{PolyakovFlatomega}
\om_{AB} = \left( \begin{array}{cc} 0 & \delta \\ -\delta^{T}& 0  \end{array} \right) ,
\ee
which expresses the fact that $\cal P$ is a symplectic manifold.

Using these definitions, the action is written as a $\sigma$-model on ${\cal P}$:
\be\label{flatTseytlin}
S= \frac1{4\pi}\int  \Big( \partial_{\tau}{\X}^{A} (\eta_{AB}^0 + \omega_{AB}^0)\partial_{\sigma}\X^{B} -   \partial_{\sigma}\X^{A} H_{AB}^0\partial_{\sigma}\X^{B} \Big).
\ee
The term proportional to $\omega^0$
is a total derivative. However
since there are monodromies, it will be relevant, as we will see, to keep track of it.
One sees that the Hamiltonian $H_{AB}^0 \partial_{\sigma}\X^{A} \partial_{\sigma}\X^{B}$ is ultra-local -- it depends only on the space derivatives. In view of the pioneering work \cite{Tseytlin:1990nb,Tseytlin:1990va,Tseytlin:1990hn}, we call this expression the Tseytlin action\footnote{See also \cite{Floreanini:1987as}. }.
The Tseytlin action is such that its target is ${\cal P}$.

This space is equipped as usual with a symplectic structure, and in order to carry a string we emphasize that it contains {\it two} metrics, $(\eta^0, H^0)$. The \Qm can be thought of as being an extension to $\P$ of  the space-time metric, while as we will see more precisely later, the \Pm is related to the decomposition of phase space into space-time $L=\{(X,0)\}$ and energy-momentum $\tilde{L}=\{(0,Y)\}$.
A point that will become important later is the fact that space-time $L$ can be characterized as the kernel of $(\eta^0 +\omega^0)$ while energy-momentum $\tilde{L}$ is the kernel of $(\eta^0 -\omega^0)$.  In the case at hand we also have that the momentum Lagrangian is the image of the space-time one by the chirality map $\tilde{L}=J_0(L)$. As we will see, this last property is specific to a geometry with vanishing $B$-field.

At first one might wonder how one can double the target space dimension without doubling the degrees of freedom. This is related to the fact that the metastring is \hlt{chiral}: i.e., there are no terms quadratic in time derivatives. This is achieved thanks to the presence of the chiral structure $J_0$ and, in particular, the fact that it squares to unity.
While the Polyakov string contains both left- {\it and} right-movers, the metastring contains only left- and right-movers that are chiral in the target.
As we will see, the left-movers have negative chirality while the right-movers have positive chirality. 

\subsection{More General Backgrounds and Born Geometries}

Although in this paper we will work exclusively with the flat theory described by (\ref{flatTseytlin}), it is instructive to consider the generalizations to which we will turn our attention in future publications. One might  expect that $\eta^0, \omega^0, H^0$ can be replaced by more general structures. 

In fact, it is a simple extension of the above construction to include a curved background in the Polyakov action
\be
S_{P}(X) =\frac1{4\pi } \int \left( {G}_{\mu \nu}(X) {*\rd} X^{\mu}\wedge \rd X^{\mu} + {B}_{\mu\nu}(X) \rd X^{\mu}\wedge \rd X^{\mu}\right).
\ee
We can recast this action in the first order form by introducing dual $\hat{G}$ and $\hat{B}$ fields by $[\hat{G} + \hat{B}] \equiv [({G} + {B})^{-1}]$
or equivalently
\be\label{mets}
\hat{G}^{-1}= G-BG^{-1}B,\qquad \hat{G}^{-1}\hat{B}=- B G^{-1}. 
\ee
The first order dimensionless action  reads
\be
\hat{S} = \frac{1}{2\pi}\int \left(\frac{1}{\lambda\varepsilon}\bm{P}_{\mu}\wedge \rd X^{\mu} + \frac{1}{2 \varepsilon^{2}} (\hat{G}^{\mu\nu}*\bm{P}_{\mu}\wedge \bm{P}_{\nu}  + \hat{B}^{\mu\nu}\bm{P}_{\mu}\wedge \bm{P}_{\nu}) \right).
\ee
Following the same procedure as above, we obtain
\be
\hat{S}=\frac{1}{2\pi}\int d^2\sigma  \left(\frac{1}{\lambda\varepsilon}\pa_{\sigma}Y \pa_{\tau}X 
-\frac{1}{2\varepsilon^{2}} \pa_{\sigma}Y [G^{-1}]\pa_{\sigma}Y 
+\frac{1}{\lambda\varepsilon} \pa_{\sigma}Y [{G}^{-1} B ] \pa_{\sigma }X 
-\frac1{2\lambda^{2}} \pa_{\sigma }X [G-BG^{-1}B] \pa_{\sigma }X \right).
\ee
As before, by introducing the  dimensionless coordinates $\X^{A}\equiv (X^{\mu}/\lambda,Y_{\mu}/\varepsilon)$, we write the action as
\be\label{curvedTseytlin}
\hat{S}=  \frac{1}{4\pi}\int_{\Sigma}d^2\sigma \Big(  \pa_{\tau}{\X}^{A} (\eta_{AB}^0+\omega_{AB}^0)\pa_{\sigma}\X^{B} -  
  \pa_{\sigma}\X^{A}  H_{AB}\pa_{\sigma}\X^{B}\Big) ,
\ee
where now
\be\label{etaH}
\eta_{AB}^0 = \left( \begin{array}{cc} 0 & \delta \\ \delta^{-1} & 0  \end{array} \right),\quad
H_{AB} \equiv  \left( \begin{array}{cc} [G-BG^{-1}B] & [BG^{-1}] \\ -[{G}^{-1}B] &  [G^{-1}]  \end{array} \right),
\quad
\om_{AB}^0 = \left( \begin{array}{cc} 0 & \delta \\ -\delta & 0  \end{array} \right)
.
\ee
Let us finally remark that the general metric $H$ can be obtained from the trivial one $H^0$
by an O$(d,d)$ transformation: $H=O^{T}H^0 O$, where 
\bea\label{OH}
O^{T}=\left( \begin{array}{cc} 1 & B \\ 0 & 1  \end{array} \right)\left( \begin{array}{cc} e^{T} & 0 \\ 0 & e^{-1}  \end{array} \right)
\eea
is an $O(d,d)$ matrix and $e$ is the frame field corresponding to $G=e^{T}h e$.

Thus, the usual string theory in curved backgrounds corresponds to making the \Qm $H$ dynamical (but not the \Pm $\eta$ or the symplectic structure $\omega$). Let us discuss further generalizations. Suppose that we first generalize $\eta^0, \omega^0, H^0$ to general structures $\eta, \omega, H$. 
Furthermore, given the existence of $\omega$ and $H$, there is a natural way to understand this geometrical structure $(\omega,H)$  from the point of view of quantum mechanics. If one takes the point of view of geometric quantization \cite{geoq, blau}, the construction of the Hilbert space associated with a phase space $(P,\omega)$ requires the introduction of  a \hlt{complex structure} $I$ compatible with $\omega$. 
Such a complex structure defines the notion of coherent states as holomorphic functionals and equips the phase space with a  quantum-metric 
 via the relation $H I= \omega$ \cite{Freidel:2013zga}. 
 This structure is, in effect, what Born suggested to be part of quantum gravity in the 1930's \cite{born1938suggestion}.
In the string case if the $B$ field vanishes $H$ is related to $\omega$ via a complex structure. This is no longer true if $B$ does not vanish. We can still define the map $I \equiv H^{-1}\omega $ in this case, but the Q-metric $H$ and the symplectic structure are no longer compatible.

However, the Born proposal is not enough. As we have pointed out in \cite{Freidel:2013zga}, in the metastring theory we must take $\eta$ to be  dynamical as well. As we have seen above, it is $\eta$ that governs
the splitting of phase space into space and momentum space.
In particular, one can think of space-time as a Lagrangian submanifold, that is a manifold of maximal dimension on which the symplectic structure vanishes.
Analogously, momentum space is just another Lagrangian submanifold $\tilde{L}$ in this description, which is transverse to the space-time Lagrangian submanifold.
Thus we end up with a bilagrangian structure on ${\cal P}$. That is a decomposition of $\P$ into two transverse Lagrangian manifolds:
 $T\P=TL \oplus T\bar{L}$ and $TL \cap T\tilde{L} =\{0\}$.
 What is remarkable is the fact that a bilagrangian structure is uniquely characterized by a \Pol metric $\eta$. This metric is characterised by the fact that $L=\ker(\eta+\omega)$ and $\tilde{L}=\ker(\eta-\omega)$. 
 In other words, the geometrical notion of $\eta$ is to provide a 
 bilagrangian decomposition of phase space. The neutral metric $\eta$ that seems like a purely stringy metric is in fact a very natural object from the point of view of phase space, in that it labels its decomposition into space and momentum. 
 In order to prove this,  let us introduce a structure $K$ which is $+1$ on the vectors tangent to the space-time Lagrangian $L$ and $-1$ on the momentum space Lagrangian $\tilde{L}$.
This is a real structure which satisfies $K^2=1$. 
Since $L$ and $\tilde{L}$ are Lagrangians $K$ also satisfies an anti-compatibility condition with $\omega$: $ K^T\omega K=-\omega$.
These two properties in turn show that $\eta= \omega K$ is a neutral metric .

We have already emphasized the importance of the endomorphism $J=\eta^{-1}H$, which relates the two metrics. Its  properties enforce the chirality of the $\sigma$-model. We thus suppose that the geometry of ${\cal P}$ should be constrained by the property $J^2=1$. 

It is relative locality that suggests that both $\eta$ and $H$ be dynamical. 
In particular, in canonical quantum theory $H$ is a purely kinematical structure and $\eta$, which describes the choice of polarization, can be modified by
unitary dynamics. Conversely, in the context of gravitational dynamics, $\eta$ is a purely kinematical structure (because space-time provides the preferred basis or 
polarization), while $H$, through its space-time part, can be made dynamical. 
According to Born, when we introduce gravity into quantum theory we have to make $H$ into a truly dynamical quantity. When we introduce quantum theory into gravity, we have to make the neutral metric $\eta$ dynamical, and thus in the context of quantum gravity, both $H$ and $\eta$ have to be dynamical.

The neutral metric $\eta$ is, together with the generalized phase space metric $H$, indispensable for the definition of space-time
as a maximally null subspace of $\eta$ with the space-time metric given by the restriction of the $H$ metric to this
$\eta$-null subspace \cite{Freidel:2013zga}. 

The structure $(\omega,\eta,K)$ can also be described in terms of the two real structures $J,K$ and the map $I$. We can check that the relation between these maps is given by 
\be
JK=I.
\ee
If, in addition, we assume that $B$ vanishes we have that $I$ is a complex structure and that $JK=-KJ$.
Phase space geometries that have  $I,J,K$  satisfying these conditions were referred to as Born geometries in \cite{Freidel:2013zga}. These possess para-quaternionic structure (because $I^2=-1$, but $J^2=1=K^2$ and they anticommute). Born geometry represents a natural unification of quantum and space-time and phase space geometries, and it implies a new view on the kinematical and dynamical structure of quantum gravity.
This structure is natural in a quantum particle theory. In string theory it is also natural to allow for a  non zero $B$-field, in which case $I$ is no longer a complex structure\footnote{As a side comment, note also that in the mathematical literature the Born reciprocity idea has been at the root of the invention of quantum groups.
Indeed, quantum groups, originally designed by Drinfeld \cite{drinfeld1987,Majid:1990gp} as doubles, are self-dual algebraic structures and the famous R-matrix is the kernel of the Fourier transformation. Another independent invention of a subclass of quantum groups \cite{majid1990physics,Majid:1988we}, the bi-crossproduct ones,  directly stems from the algebraic  implementation of the Born self-dualization idea, a principle at play in 3d gravity \cite{Freidel:2005ec}.
Finally let us note that the canonical quantization of curved momentum space has also been discussed in  other contexts as well \cite{Batalin:1989dm,Batalin:1989xu,Bars:2010xi,Chang:2010ir,Chang:2011jj}.
}.

\subsection{T-duality}

The expression of T-duality in the Polyakov formulation of constant backgrounds appears as the worldsheet symmetry 
\be
\rd X^{\mu} \to *\rd X^{\mu} ,
\ee
which exchanges $\sigma$ and $\tau$ in the conformal gauge.
The phase space formulation on the other hand breaks the symmetry between $\sigma$ and $\tau$.
The T-duality symmetry does not appear as a worldsheet symmetry, but instead  appears as a target space symmetry.
This manifest transfer of the symmetry property from worldsheet to target is one of the main advantages of this formulation. In order to see this let's consider,
given $\eta^0$ and $H$,  the chiral operator $J\equiv (\eta^0)^{-1}H$.
It can be written explicitly as 
\be
J^{A}{}_{B}= \left( \begin{array}{cc} -G^{-1}{B} &    {G}^{-1}\\ (G- BG^{-1}B) & BG^{-1}  \end{array} \right).
\ee
What is remarkable about this operator is the fact that it is an $O(d,d)$ transformation leaving the \Pm invariant and that, as we have mentioned above,  
it is a chiral structure which squares to the identity:
\be
J^{T}\eta J = \eta,\qquad J^{2}=1.
\ee
From its definition it can be seen that $J^{T}H = H J= H \eta^{-1}H$, so it also preserves $H$:
\be
J^{T}H J = H.
\ee
These properties imply that the map
\be
\X \mapsto J(\X),
\ee
is a symmetry of the bulk action, and 
it  expresses the T-duality symmetry.

Note however that  $J$ does not preserve $\omega^{0}$. When the $B$ field vanishes it maps $\om^{0}$ into $-\om^{0}$,
while if the $B$-field is non-zero it rotates non-trivially the Lagrangian subspaces. An explicit computation gives $ J^{T}\om^{0} J  = \tilde{\om}^{0}$ with
\bea
\tilde{\om}^{0} 
 &=&- \om^{0} +   2
\left( \begin{array}{cc}   B(1- (BG^{-1})^{2} &   (BG^{-1})^{2}   \\  
- ({G}^{-1}B)^{2} &  G^{-1} B {G}^{-1}\end{array} \right).
\eea
In the constant background case this breaking of T-duality appears only as a change of the boundary conditions via the boundary term.
Another way to express this is to notice that when the $B$-field is non-zero, the momentum Lagrangian $\tilde{L}$ is no longer aligned with the  subspace $L^\perp$ orthogonal to $L$ with respect to $H$. Indeed, this space is simply given by $L^\perp = J(L)$ since $H(L, J(L))=\eta(L,L)=0$.

\subsection{Usual string viewed from phase space}

It is clear from the previous analysis that the formulation (\ref{curvedTseytlin}) begs for a natural generalization where 
$\X$ possesses arbitrary monodromy and where not only the constant \Qm $H^0$ is promoted to an arbitrarily curved metric $H$, but also the \Pm and symplectic structure are allowed to be dynamical. In the general case we promote 
$(\eta^0,H^0,\om^0) \to (\eta,H,\om)$ to be  functions of $\X$.
Such a generalization aims to provide a string theory formulation where T-duality is manifest even in the curved context \cite{flmcurved}. The action is given by the consequent generalization of (\ref{flatTseytlin}) and we call such a generalization the metastring.
Double field theory, on the other hand, usually considers the effective field theory based on the 
restricted structure $(\eta^0,H,\om^0)$, where both the symplectic structure and the \Pm are treated as background structures, while $H$ is allowed to have a specific type of  $\X$ dependence\footnote{The fields are demanded to be projectable. A recent exception \cite{Blumenhagen:2014gva}  considers a non-trivial $\omega$ while still keeping a flat \Pm $\eta^0$. 
Another notable exceptions are in the context of beta function calculations \cite{Berman:2007xn, Berman:2007yf, Avramis:2009xi}. }.

Before doing so, it is necessary to pause for a moment and understand what specific conditions characterize the Polyakov string within  the metastring.
Let us start by listing the necessary and sufficient 
conditions that the Tseytlin string has to
satisfy in order to be a Polyakov string in disguise. There are 5 conditions:
\begin{itemize}
\item $J\equiv \eta^{-1}H$ is an involution  preserving $\eta$.
\item $\eta \pm \omega$  are maximally degenerate, i.e., of rank $d$.
\item $\omega$ is a closed form.
\item The fields $\Phi=(\eta, H,\om)$ only depend on the degenerate directions of
$\eta - \omega$; that is  
\be\label{phirestr}
(\eta - \om)_{AB} \eta^{BC} \pa_{C}\Phi =0.
\ee
\item The fields possess  monodromy only in the degenerate direction of $\eta -\omega$;
that is \be\label{mono}
(\eta - \om)_{AB}\Delta^B=0,
\ee
where $\Delta^A(\tau) \equiv \left[\X^{B}(\sigma + 2\pi,\tau) - \X^{B}(\s,\tau)\right]$ is the monodromy.
\end{itemize}
In the case where $(\eta, \omega)=(\eta^0, \omega^0)$ are constant and given by (\ref{etaH}), the matrix $(\eta -\omega)_{AB} $ projects to zero the  energy-momentum vectors $\X^A=(0,Y_\mu)$, that is, the vectors belonging to the Lagrangian $\tilde{L}$.
On the other hand $(\eta^0 -\omega^0) (\eta^0)^{-1}_{A}{}^B $ projects out the space-time derivative $\pa_A=(\pa_X,0)$. 
This means that the conditions
(\ref{mono}) and (\ref{phirestr}) read respectively $X^\mu(\sigma + 2\pi,\tau) =X^{\mu}(\s,\tau)$ and $ \pa_{Y_\mu} \Phi =0$.
They imply that the fields depend on $L$ while monodromy is only in the momentum Lagrangian $\tilde{L}$.
We will analyze what happens when we relax these conditions.

The mildest condition to relax is the last, in which  we allow monodromy in all directions.
In the case where all the fields are constant and extra monodromies are allowed only in space-like directions, this corresponds to the torus compactification of the Polyakov string. If monodromy is allowed in the time-like direction, the usual interpretation is in terms of thermal solitons and gives rise (under Euclidean continuation) to the string free energy, etc. \cite{Polchinski:1985zf, Moore:1986rh}. 
In a later section, we will carefully consider this generalization and show that there is a consistent but non-trivial notion of closed string boundary conditions.
 
Next we can relax the condition (\ref{phirestr}) by allowing the fields themselves to depend on all coordinates in ${\cal P}$. This generalization is one of the most interesting and 
will need to be dynamically constrained in order to give admissible backgrounds.
In particular, it implies considering the new possibility where $\eta$ is no longer a flat metric. This entails relaxing the condition that the splitting between space-time and energy-momentum is universal. That is, it relaxes the hypothesis of absolute locality  and allows us to have a framework in which locality is relative, or,  in 
colloquial terms, a framework where each string can carry a different space-time. 

Another level of relaxation is to allow $\omega$ to not be closed. This would impede its interpretation as a symplectic form in Born geometry. Although this generalization deserves study, it is beyond the scope of our present discussion.
As we will see \cite{ flmq, flmbeta, flmcurved} these three levels of relaxation are admissible both at the  classical and the quantum level.

The next level of generalization would be to consider a string where 
$\eta \pm \om $ is not maximally degenerate.
For instance, as we will see later, if $\eta-\omega$ is invertible, there is no propagating open string. For simplicity, we will keep the condition of maximal degeneracy for now.
We have seen that in the Polyakov case the kernel of $(\eta + \om)$ plays the role of the space-time Lagrangian $L$. By keeping the property of maximal degeneracy, we keep the concept of a preferred Lagrangian defined by the metastring fields. 
Moreover we will see that the open metastring boundary naturally propagates inside $L=\mathrm{ker}(\eta + \om)$. If we want to keep the compatibility condition between open and closed string in the sense that the open string possesses half the closed string degree of freedom, we have to keep the condition of maximal degeneracy.

Finally, we are also going to see in this work that it is inconsistent to relax the first condition: we always need $J$ to be a chiral structure if we want to keep 
 the conformal symmetry of the theory.
 In summary, the metastring action is given by 
 \be\label{metastringAction}
\hat{S}=  \frac{1}{4\pi}\int_{\Sigma}d^2\sigma \Big(  \pa_{\tau}{\X}^{A} (\eta_{AB}+\omega_{AB})\pa_{\sigma}\X^{B} -  
  \pa_{\sigma}\X^{A}  H_{AB}\pa_{\sigma}\X^{B}\Big) ,
 %+\frac1{4\pi} \int_{\pa\Sigma} \omega_{AB}\rd\X^{A} \X^{B}.
\ee where  the fields $\Phi=(\eta, H,\om)$ which correspond respectively to a neutral \Pm\!\!, a \Qm and a 2-form,  are all dynamical and depend on $\cal P$.  We demand however that $\eta -\omega$ is maximally degenerate and that $J\equiv \eta^{-1}H$ is a chiral structure.

\subsection{Global Symmetries}

We now comment on the global symmetries of the flat Tseytlin action (\ref{flatTseytlin}).
We still assume in this section that $\eta$, $H$ and $\omega$ are constant matrices. 
Let us first use the fact that since $\eta$ is a neutral metric, we can always choose a frame where it assumes the form given in (\ref{etaH}), that is $\eta = e^T \eta^0 e$. 
As we have seen in (\ref{OH}) we can, in this frame, trivialize $H$ by an O$(d,d)$ transformation. Without loss of generality we can therefore take for illustration $(\eta,H,\omega)$ in the form (\ref{etaH0}).

\subsubsection{Double Lorentz symmetry}

The first global symmetry of the action is the double Lorentz group $O(\eta,H)$, preserving {\it both} $\eta$ and $H$.
That is, we define 
\be
O(\eta,H) \equiv \left\{ \bm{\Lambda} \in GL(2d) \Big|\, \bm{\Lambda}^{T}\eta \bm{\Lambda} = \eta, \, \bm{\Lambda}^{T}H \bm{\Lambda}=H\right\}.
\ee
This group is isomorphic to 
$ O(1,d-1) \ltimes \mathfrak{so}(1,d-1)\times \mathbb{Z}_{2}$.
The $ O(1,d-1) \ltimes \mathfrak{so}(1,d-1)$ component is generated  by matrices of the form
\be
\bm{\Lambda}= \left( \begin{array}{cc} \Lambda \sqrt{\delta+\beta^{2}}  &\Lambda \beta h^{-1} \\ h \Lambda \beta & h \Lambda\sqrt{\delta+\beta^{2}}h^{-1}   \end{array} \right) ,
\ee
where $\Lambda \in O(1,d-1) $, $\Lambda^{T} h \Lambda= h$ and $\beta\in \mathfrak{so}(1,d-1)$, i.e., $h \beta+ \beta^{T}h=0$. 
There are two types of `boosts' here. First, we have the usual ones $\Lambda$, that act in the usual  way $(X,Y)\to (\Lambda X , (\Lambda^T)^{-1} Y) $ on space-time  and energy-momentum space defined as Lagrangian subspaces of $\omega$. Secondly, the $\beta$ boosts $(X,Y)\to (\sqrt{\delta+\beta^2} X + \beta h^{-1} Y, 
h\beta X + \sqrt{\delta+\beta^2}Y) $ mix space-time and energy-momentum space in a 
non-trivial manner.
This is the group of symmetries of the metastring theory, the action being invariant under $\X \to \bm{\Lambda}\X$. Thus the group of Lorentz transformations is generalized to its double since its Lie algebra 
  is locally isomorphic to $\mathfrak{so}(1,d-1)\times \mathfrak{so}(1,d-1)$. 
This fact can be clearly seen if we look at the action of this group on the chiral components $\frac12(1\pm J)\X$ of $\X$.
We find that it acts diagonally:
 \be
\bm{\Lambda}(1\pm J) 
= (1\pm J) \left( \begin{array}{cc} \Lambda U^{\pm1}&0 \\  0 &h \Lambda U^{\pm1}h^{-1}  \end{array} \right) ,
\ee
where $U^{\pm1}=(\sqrt{1+\beta^{2}}\pm \beta h )$ is a Lorentz transformation.

The $\mathbb{Z}_{2}$ component of the symmetry group is generated by $ J$. This corresponds to the exchange of two Lagrangian subspaces. 

\subsubsection{Discrete symmetries}

The metastring possesses three distinct discrete symmetries.\footnote{There are also discrete elements of the double Lorentz group acting locally on $\Sigma$, such as the inversion $\X\to -\X$, corresponding to $\Lambda=-1, \beta=0$.}
The first one that we have already seen is the duality symmetry
\be
D: \X(\s,\tau) \mapsto J\X(\s,\tau).
\ee
We also have  the PT symmetry
\be
\mathrm{PT}: \X(\s,\tau) \mapsto \X(2\pi-\s,-\tau) ,
\ee
and  the time reversal symmetry
 \be
\mathrm{T}: \X(\s,\tau) \mapsto K \X(\s,-\tau) ,
\ee
where $K$ is a matrix such that $K^{2}=1$ and it also satisfies 
$K^{T}HK = H$ and $K^{T}(\eta+\omega)K 
= - (\eta+\omega) $.
This $K$  is given by 
\be
K=K^0= \left( \begin{array}{cc} \delta & 0 \\ 0& -\delta^T   \end{array} \right).
\ee
It is interesting to note that this matrix anti-commutes with $J$
\be
K^{2}=1,\quad J^{2}=1,\quad KJ+JK =0.
\ee
This means that the combination of time reversal and duality symmetry 
is implemented by the map \beq \mathrm{DT}:\X(\s,\tau) \mapsto I\X(\s,-\tau), \eeq where 
  $I\equiv JK$ is a complex structure which preserves $H$:
  \be
  I^{2}=-1,\quad 
  I^{T}HI = H.
  \ee
 The $I,J,K$ found here are those of the corresponding (trivial) Born geometry. Here we have seen that they are involved in symmetries of the flat Tseytlin model that act on both worldsheet and target space.
 
\subsection{Time translation symmetry}\label{sec:timetrans}

Another important symmetry of the Tseytlin action is the time translation symmetry.
We consider the transformation, described in the local conformal frame  
\be\label{timesym}
\delta_f \X^A(\tau,\sigma)\equiv f^A(\tau).
\ee
This corresponds to a translation along a $\sigma$-independent vector field. In the case where $f^A(\tau)$ are constant, we are just describing a global translation of the flat target space. We emphasize that there is a larger symmetry here under certain conditions on $f^A(\tau)$. Indeed,
under such a $\tau$-dependent transformation the action transforms by a boundary term
\be
\delta_f S
= \frac1{4\pi} \int_{\tau_i }^{\tau_o} \rd\tau \,  \Delta^A(\tau)(\eta -\omega)_{AB}\dot{f}^B(\tau) ,
\ee
where $\Delta(\tau) = \X(2\pi,\tau) - X(0,\tau)$ is the monodromy.
We see that this variation vanishes if $\dot{f}$ belongs to the kernel of $\eta - \omega$.
It is important to note that this necessarily implies that $\dot{f}^A$ is null with respect to the 
\Pm $\eta$, $ \dot{f}^A\eta_{AB}\dot{f}^B=0$. In other words, $\dot{f}$ belongs to the momentum space Lagrangian $\tilde{L}$. 
We will analyze later the consequences of this extra symmetry.

\subsection{Constraints}
Let us now understand the nature of the Virasoro constraints in this formulation.
In string theory we integrate over all worldsheet metrics, that is we integrate over all  conformal structures and quotient by the action of 2d diffeomorphisms. This imposes Hamiltonian and diffeomorphism constraints on the 
data. For now, we focus on a given cylinder, in which the worldsheet metric is conformally equivalent to $\rd s^{2}=-\rd \tau^{2} +\rd \sigma^{2}$, coming back to general worldsheets later. 
If we change the conformal frame infinitesimally, we have to introduce a new time and space coordinate frame. A variation of the conformal structure can be encoded in two functions $\alpha, \beta$ via
\be
\delta \rd s^{2} = 2\alpha (\rd \tau^{2} + \rd \sigma^{2})  + 4 \beta \rd \tau \rd \sigma.
\ee
A new conformal frame is obtained by a redefinition of the local frame coordinates $ \rd\sigma^{a}\to \rd\sigma^{a} + \delta \rd\sigma^{a}$
with
\be
\delta \rd \tau = -\alpha \rd \tau -\beta \rd \sigma,\qquad \delta \rd \sigma = \alpha \rd\sigma +\beta \rd \tau.
\ee
The variation of the space and time derivatives due to this local change of frame is given by
\be
\delta \partial_{\tau} = \alpha \partial_{\tau} + \beta \partial_{\sigma},\qquad 
\delta \partial_{\sigma } = -\alpha \partial_{\sigma } - \beta \partial_{\tau}.
\ee
We can now determine the Hamiltonian and diffeomorphism constraints  from the variations
$H=  \delta_{\alpha}S $, $\hat{D} = \delta_{\beta} S$, which in local coordinates read
\bea
\hat{H} &\equiv&      \pa_{\sigma}\X^{A}  \pa_{\sigma}\X^{B} H_{AB} \\
\hat{D}&\equiv&   \frac12 (   \pa_{\sigma}\X^{A}  \pa_{\sigma}\X^{B} -  \pa_{\tau}{\X}^{A}  \pa_{\tau}{\X}^{B})\eta_{AB} 
+  \pa_{\tau}{\X}^{A}  \pa_{\sigma}\X^{B} H_{AB}.
\eea
Finally, it is also important to consider variations of the coordinate frames that do not change the conformal structure.
These are given by the Weyl and Lorentz transformations: $\delta_{W} \rd \tau = \lambda\rd \tau$, $ \delta_{W}\rd \sigma =\lambda\rd \sigma $ and $\delta_{L} \rd \tau = \omega\rd \sigma$, $ \delta_{L}\rd \sigma =\omega\rd \tau $ respectively. Demanding invariance under these variations  leads to the (classical) constraints
\bea
\hat{W} &\equiv& 0 \nonumber \\
\hat{L}&\equiv& \frac12 (  \partial_{\sigma}\X^{A} \partial_{\sigma}\X^{B} + \partial_{\tau}{\X}^{A} \partial_{\tau}{\X}^{B})\eta_{AB} - \partial_{\tau}{\X}^{A}\partial_{\sigma}\X^{B} H_{AB}.
\eea
In order to see that these reduce on-shell to the usual Hamiltonian and diffeomorphism constraints of string theory, and that the Lorentz and Weyl constraints are trivially satisfied,
let us first write  these constraints in a slightly different form.
Consider the vectors
 \be
 \S^{A} \equiv \partial_{\tau}{\X}^{A}-(J\partial_{\sigma}\X)^{A},
 \ee
and  rewrite all the constraints 
 in terms of $\S$ and $\partial_{\sigma}\X$.
 (In the following we denote by $\cdot$ the contraction with the metric $\eta$.)
The constraints are then
\bea 
\hat{W}&=& 0 \nn \\
\hat{L}&=& \frac12 \S\dd \S + \frac12 \pa_{\sigma}\X\dd (1-J^{2})\pa_{\sigma}\X. 
 \eea
 Note that in the flat case the constraint $J^{2}=1$ is identically satisfied. 
 In this case, the Lorentz condition simply becomes $\hat{L}= \frac12 \S\dd \S=0$.
  In the flat case the equation of motion implies that $\pa_\sigma\S=0$.
  This means that $\S$ depends only on $\tau$. 
  The Lorentz condition means that $\S(\tau)$ belongs to a Lagrangian subspace $\tilde{L}$, that is a null subspace of the \Pm $\eta$. Choosing $\omega$ such that $\tilde{L}$ is  the kernel of $\eta - \omega$, we can use the time symmetry described earlier to fix the gauge where  $\S=0$. This is the gauge in which we will now work.   Notice that this gauge choice, given $J$, fixes a relationship between chirality on the worldsheet and $J$-chirality in the target space.
  
Also, in this language, the Hamiltonian and diffeomorphism constraints are given by:
  \bea
 \hat{H} =   \partial_{\sigma}\X\dd J \partial_{\sigma}\X, \qquad D =\partial_{\sigma}\X\dd \partial_{\sigma}\X,
  \eea
where we have denoted $D = \hat{D} +\hat{L}$. In terms of the phase space coordinates
$\X=(X,Y)$, the constraints read $\hat{H}= (X'^{2} + Y'^{2})$, $D=2X'\cdot Y'$. These reduce to the usual form 
\bea
H^{\mathrm{red}} =   (X'^{2} + \dot{X}^{2}),\qquad 
D^{\mathrm{red}} = 2 (\dot{X}\cdot X' ),
\eea
once we impose the duality equations $\pa_{\tau}{Y} = \pa_{\sigma}X$, $\pa_{\sigma}Y=\pa_{\tau}{X}$.

\subsubsection{Energy momentum Tensor}

We would like to write the phase space action in a more covariant manner in order to clarify the constraints.
Indeed, so far we have heavily relied on the space-time splitting which assumes a conformal frame on the worldsheet.
We now introduce a fully covariant formulation of metastring theory that does not assume a particular choice of coordinates on the worldsheet.

In order to find a covariant formulation, we introduce the co-frame field 
\be
e^{a}\equiv e^{a}_{\tau}\rd \tau + e^{a}_{\sigma}\rd \sigma,
\ee
with $a=0,1$ and the 
corresponding frame fields which we denote as 
\be
\pa_{a}\equiv e_{a}^{\tau}\pa_{\tau}+e_{a}^{\sigma}\pa_{\sigma}.
\ee
They are such that $\pa_{a}(e^{b})=\delta_{a}{}^{b}$.
Given these definitions the metric can be written as $\rd s^{2} = -e^{0}\otimes e^{0}+ e^{1}\otimes e^{1}$.
It is convenient to write everything in terms of a chiral frame: $ e^{\pm} \equiv  e^{0}\pm e^{1}$
and $ \pa_{\pm}= \frac12(\pa_{0}\pm \pa_{1})$ in which the metric reads  $\rd s^{2} = - \frac12 e^{+}\otimes e^{-}-\frac12 e^{-}\otimes e^{+}$.
The Tseytlin action can be now written 
\beqn\label{Tseytlin}
S &=& \frac1{4\pi}\int \det(e) \left[\pa_{0}\X(\eta +\omega)\pa_{1}\X - \pa_{1}\X H \pa_{1}\X  \right] .
\eeqn
This action is manifestly diffeomorphism and Weyl invariant, but not manifestly locally Lorentz invariant.
 
We define the energy momentum tensor as
$
T^{a}{}_{b} \equiv \frac{2  \pa^{\alpha}_{b}}{\det(e)} \frac{\delta S}{\delta \pa_{a}^{\alpha}}. 
$
We make this definition rather than the usual one involving the variation with respect to the metric, because in the absence of Lorentz symmetry, the stress current is not automatically symmetric. 
We then find
\bea
T^{0}{}_{0} &=& \pa_{1}\X \dd J\pa_{1}\X,\qquad 
T^{0}{}_{1} =  \pa_{1}\X \dd\pa_{1}\X ,
 \\
 T^{1}{}_{1} &=& - \pa_{1}\X \dd J \pa_{1}\X
,\qquad 
T^{1}{}_{0} =\pa_{0}\X\dd \pa_{0}\X - 2\pa_{0}\X\dd J\pa_{1}\X \quad 
.
\eea
The generators of Weyl and Lorentz transformation act on the frame fields as :
\bea
W: (e^+,e^{-})&\mapsto& (e^{\rho} e^{+}, e^{\rho } e^{-}), \\
L: (e^+,e^{-})&\mapsto& (e^{-\theta} e^{+}, e^{ \theta } e^{-}),
\eea
and the Tseytlin action transforms as 
$
\delta S = \int \det(e) ( \delta \rho\ W + \delta \theta\ L ) ,
$
where the Weyl and Lorentz generators are given by 
$W= \tfrac12(T^{0}{}_{0}+T^{1}{}_{1})$, $L = \frac12(T_{10}-T_{01})$. This gives explicitly:
\be
W= 0, \qquad L= \frac12 \S \dd \S+\frac12 \pa_1\X\dd (1-J^2)\pa_1\X ,
\ee
where $\S= \pa_{0}\X - J(\pa_{1}\X)$.
The generators of conformal transformations are then given by
$H= -(T_{00}+T_{11})/2$ and $D= -(T_{01}+T_{10})/2$ and read as follows
\be
H = \pa_{1}\X \dd J\pa_{1}\X,\qquad D = \pa_{1}\X \dd \pa_{1}\X-L,
\ee
in agreement with our previous derivation.

The new feature of this formulation is the fact that worldsheet Lorentz invariance is {\it not} manifest;
 under an infinitesimal Lorentz transformation the action transforms as (assuming $J^2=1$)
\be
\delta_{\theta} S = \theta \S\dd \S ,
\ee
and the constraint $\S\dd \S=0$ has to be imposed, 
in other words, $\S$ has to be null with respect to the neutral metric.
It is only after the imposition of this constraint, which implies $\S=0$ on-shell after use of the time symmetry, that we recover the usual Polyakov formulation where this  
symmetry  is satisfied on-shell for the flat background.
As we will see the non manifest Lorentz symmetry is akin to the non manifest Weyl invariance of the massive  deformations of Polyakov string.
It is one of the most challenging but also one of the most interesting and fruitful aspects of this new formulation.
The deep quantum implications of this fact will be explored in \cite{flmq, flmbeta}. See also \cite{Berman:2007xn,Sfetsos:2009vt,Avramis:2009xi}.

\subsubsection{Euclidean form and Level Matching}

The  description given here may look unfamiliar since it is intrinsically Lorentzian and refer to a particular time slicing $\tau$. As we will see in the next section the Lorentzian nature of the metastring is one of its key features. That is, once the Lorentzian structure and the proper time $\tau$  is given, it is possible to do a Wick rotation and write the previous expressions in terms of Euclidean coordinates. We do this here for the reader's convenience in order to connect to the more usual notation. To do so,  we switch to Euclidean worldsheet  coordinates $\sigma \to x_{1}$, $\tau\to ix_{2}$, and  $z\equiv x_{1}+ix_{2}$.
With this convention we can replace $\partial_{\tau} \to (\partial_{z} -\partial_{\bar{z}})$ and $\partial_{\sigma}\to (\partial_{z}+ \partial_{\bar{z}})$.
In general, the frame field can be decomposed in terms of a conformal factor $\phi$ and imaginary internal rotation parameter $\theta$ and a Beltrami differential $\mu$:
$ e = e^{\phi+i \theta}( \rd z +\bar{\mu} \rd \bar{z})$.
We denote by $(\partial, \bar{\partial})$  the components of the inverse frame field  and its conjugate,
which is explicitly given, in this parameterization, by
 $\partial \equiv e^{a} \partial_{a} = e^{-\phi- i \theta}( \partial_{z}  - {\mu} \partial_{\bar{z}})/(1-|\mu|^{2})$.
It is illuminating to write down the constraints in terms of the Euclidean variables. The first quantity to consider is the equation of motion $\S =0$. Since $\S=(1-J)\pa\X-(1+J)\bar\pa\X$ this equation imposes a  {\it soldering} between the worldsheet chirality determined by the choice of holomorphic coordinates and the target space chirality determined by $J$ and it implies that 
\be\label{solder}
\pa \X = \frac12 (1+J) \pa_\s \X,\qquad \bar{\pa}\X = \frac12 (1-J) {\pa}_\s\X.
\ee
These equations relate the worldsheet notion of chirality (LHS) with the target space notion as eigenspaces of value $\pm1$ of $J$. Note that the RHS does not contain reference to the worldsheet chiral structure.
This is the essential soldering phenomenon happening in the metastring that allows us to promote the worldsheet notion of T-duality $ \pa\X \to \pa\X$ and  $ \bar{\pa}\X \to -\bar{\pa}\X$ to a linear target space operation $ \pa_\s\X \to J(\pa_\s\X)$ and this will eventually allow us to promote T-duality to a symmetry valid in general backgrounds.
 
Once we assume the chiral soldering to be in place, we can easily write the constraints in the usual form 
\be
L_+\equiv\frac12(H+D)\hat{=} 
\pa\X\cdot \pa\X,\qquad L_-\equiv\frac12(H-D) \hat{=}
-\bar\pa\X\cdot \bar\pa\X , 
\ee
where $\hat{=}$ is the equality once we impose the chiral soldering.
It is also interesting to express the action in chiral coordinates
\bea
S 
&\equiv& - \frac1{2\pi}\int \rd^2 z \left[  
  \partial \X^{A} (H -\omega)_{AB}\bar{\partial} \X^{B} 
+\tfrac12  \partial \X^{A}(H-\eta)_{AB}\partial \X^{B}+ \tfrac12  \bar\partial \X^{A} (H+\eta)_{AB}\bar\partial \X^{B}  \right] . \label{TseytlinEuc}
\eea

\section{Lorentzian Worldsheets}

In the rest of the paper, we will focus on the Tseytlin action (\ref{flatTseytlin}) for flat backgrounds, that is backgrounds for which, $\eta, \omega$ and $H$ are all constant. We have seen that the Tseytlin action is not Lorentz invariant; one expects that the full quantum theory is nevertheless Lorentz invariant, certainly at least for the flat background. Nevertheless, one should be concerned in this context with the veracity of the usual Euclidean continuation, and thus we are motivated to revisit the construction of the moduli space of Lorentzian worldsheets. This, of course, is an old problem even in the context of the usual string \cite{Witten:2013pra}; it was initially studied within light-cone string theory \cite{Mandelstam:1974fb, Mandelstam:1974hk, Mandelstam:1973jk }, but the program was never satisfactorily completed 
\cite{Hata:1985zu, Hata:1985tt, Hata:1986vq, Hata:1986jd, Hata:1986zq, Hata:1986ke, Hata:1986kj, Hata:1986mz, Hata:1987qx, Kugo:1992md,Ishibashi:2008xu, Nielsen:2012kg, Nielsen:2014wna }. Although naively the formalism seems non-covariant, we will establish that with some minor modifications it is in fact covariant and modular invariant, and that it can be applied to arbitrary conformal field theories. 
A feature of the  Lorentzian formulations is the fact that  the string splitting-joining interaction  which is associated with spatial topology change, is singular. One may worry that this may lead to the loss of finiteness.
  On the contrary this singular point acts as an anchor for the insertion of the dilation which provides the natural weight for these singularities.
 Such singular points are an integral part of the Lorentzian worldsheet construction and they act as a string interaction vertices. 
 Finally, we will just touch on the fact that this Lorentzian formulation suggests a new and simpler version of string field theory in which there is only one type of vertex to all orders.
These subjects are however beyond the scope of the present paper.

In this section we will investigate an explicit construction for decomposing general Lorentzian worldsheets (of genus $g$ and $n$ boundaries) into a collection of strips, each of which can be coordinatized by locally flat coordinates. This construction is possible due to a simple but powerful result of Giddings and Wolpert \cite{Giddings:1986rf} (also derived by Krichever and Novikov \cite{Krichever:1987,Krichever:1987qk}). Recently, some combinatoric aspects have been investigated in \cite{Freidel:2014aqa}. The decomposition of the worldsheet $\Sigma_{g,n}$ gives rise, as we will describe below, to a \hlt{Nakamura graph} $N$, such that $\Sigma_{g,n}\backslash N$ is connected and simply connected. Such graphs 
correspond to a cell decomposition of the moduli space of Riemann surfaces, ${\cal M}_{g,n}$; points in a given cell parameterize distinct Riemann surfaces with the same Nakamura graph, and these parameters are encoded in an Abelian differential that we refer to as the Giddings-Wolpert one-form. This one-form possesses simple poles, one for each boundary (interpreted as incoming or outgoing states), and zeroes, one for each singular interaction point. In a later section, we will begin a study of how to sew strips back together to form closed worldsheets; the principal tool that we employ in this sewing procedure is the continuity of symplectic flux across any cut in a surface.

\subsection{Giddings-Wolpert-Krichever-Novikov (GWKN) Theorem}

In order to formulate the Tseytlin action we introduced local coordinates on the worldsheet and, in particular, distinguished between $\sigma$ and $\tau$.
The worldsheet $\Sigma$ is equipped with a \hlt{causal structure}:
 that is, we assume that there exists a time function $\tau: \Sigma \to \mathbb{R}$ 
such that  $\tau$ is a Morse function and such that\footnote{This assumption focuses the discussion on closed strings. We will comment on open string observables in the appendix to this paper.}  $\pa\Sigma = \pa\Sigma_{-}\cup \pa\Sigma_{+}$
where $\pa\Sigma_{\pm}= \tau^{-1}(\pm \infty)$.
We denote by $x_{i}\in \Sigma$, $ i=1,\cdots, \ell$ the critical points of $\Sigma$ and by
$C=\{x_{1},\cdots,x_{\ell}\} $ the critical set.
$\Sigma\backslash C$ is equipped locally with a flat Lorentzian metric 
$\rd s^{2}=-\rd \tau^{2} +\rd \sigma^{2}$. 
Two flat  metrics related by a global conformal transformation 
are considered equivalent.
This constitutes a causal structure.

Note that $\tau$ has only a finite number of non-degenerate\footnote{ Points at which $\rd\tau=0$ and for which the Hessian is non-degenerate.} critical points $x_{i}\in \Sigma$,
$i=1,\cdots, \ell$  at which $\rd \tau=0$. The corresponding critical values are 
$\tau_{i}=\tau(x_{i}) $. For all $t \in \mathbb{R}\backslash \{ \tau_{1},\cdots, \tau_{\ell}\}$,
we have $\tau^{-1}(t) = S^{1}\times \cdots \times S^{1}$, the product of $k$ 
copies of $S^{1}$; $k$ is constant within each interval $t \in (\tau_{i},\tau_{i+1})$.
Here
$\tau_{i}$ are the interaction times at which the circles join or split and $x_{i}$ 
are the interaction points.
A version of the Riemann-Hurwitz theorem shows that the
number of critical points $\ell$ is bounded by  $2g + n-2$, where $g$ denotes the surface genus and $n$ the total number of external circles. This simply states that in the generic case, adding a handle adds two interaction points and adding an external circle adds one.
The moduli space of Riemann surface $\Sigma_{(g,n)}$ is of complex dimension $3g-3+n$; this space admits a cellular decomposition which respects the interaction data, and the cells of maximal dimensions all have the maximal number of interaction points  $\ell = 2g + n-2$, \cite{nakamura2000calculation, Freidel:2014aqa}.

\myfig{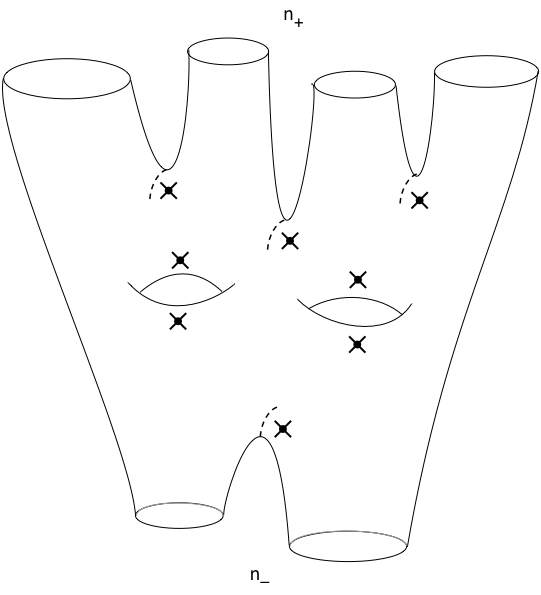}{7}{A typical Lorentzian worldsheet, this with $n_-=2, n_+=4$ and $g=2$. $2g+n-2=8$ critical points are present, which are marked by a cross.} 

In the following we denote by $n_{\pm}$ the number of components of $\pa\Sigma_{\pm}$. Obviously we have that $n_{+}+ n_{-}=n$. To each of these, we associate a real number $r_\alpha$, $\alpha=1,...,n$ such that $\sum_\alpha r_\alpha=0$; the $r_\alpha$ may be thought of as corresponding to the oriented lengths of each circle. These parameters are associated with the specification of a local coordinate system around each external states, but decouple on-shell. (Note that in light-cone gauge, these becomes related to $P_{+,\alpha}$, but an association with specifics of a CFT is not required in general -- we can make any choice. For example, a simple democratic choice is to take $r_\alpha=2\pi n_{+}$ for all in-circles and $r_\alpha=- 2\pi n_{-}$ for each out-circle.)
A remarkable theorem developed independently by Giddings-Wolpert \cite{Giddings:1986rf} and by Krichever-Novikov \cite{Krichever:1987,Krichever:1987qk}  states that  given a 
complex structure on $\Sigma$ and given a splitting of the boundary into $n_{-}$
in-boundaries and $n_{+}$ out-boundaries we can assign a 
{\it unique} causal structure $(\tau,\sigma)$. Vice-versa, there exists a unique Riemann surface with a given causal structure.
What is remarkable about this theorem is the fact that it means  
once we have fixed an in-out splitting of the boundaries, 
the map from Riemann surface to causal structure is modular invariant.
This means that a given locally flat splitting of space and time 
 amounts to a choice of a complex structure on a corresponding 
 Riemann surface and this complex structure is uniquely determined.

 More precisely, the Giddings-Wolpert-Krichever-Novikov (GWKN) result 
 is stated as follows:
 First, given a causal structure we construct on $\Sigma $ an Abelian differential 
 given by $e =\rd \tau + i\rd \sigma$ outside the critical points. 
 The imaginary periods of $e$ around the in- or out- circles are identified with $r_\alpha$; they are thus the residues of the poles corresponding to each in- or out- state.
 What is less obvious but nevertheless true, is that the Abelian differential necessarily has imaginary periods along any
 closed curve on $\Sigma$.
 The GWKN theorem is the expression that the reverse statement is true:
 given a complex structure on $\Sigma$ there exists a {\it unique} Abelian differential $e$ which possesses only imaginary periods and which is such that the periods around the in-circles (resp. out-circles) are all equal to $r_\alpha$. 
 Given such an Abelian differential we can construct a time function by $\rd \tau = \Re(e)$. This equation is integrable since $e$ possesses only imaginary periods. We also construct a locally Lorentzian structure by
 $\rd s^{2}= -\Re(e)^{2} +\Im(e)^{2}$. In other words, the GWKN differential defines a locally flat complex frame $e = e^0 + i e^1$.
In summary, these results imply that we can consider the chiral phase space action (\ref{flatTseytlin}) and preserve modular invariance. In particular, the slicing of the worldsheet that we have described is actually invariant under large diffeomorphisms. This is in distinction to the usual slicing along Torelli cuts (cycles $(a_i,b_i)$) of Riemann surfaces, in which modular transformations act non-trivially on the slicing and thus invariance under the modular group becomes non-trivial. This then is a significant advantage of the Nakamura formulation.

Using the GWKN differential $ e $ associated with a complex structure on $\Sigma$ we can construct locally flat coordinates $\rd z=  e $.
These coordinates are related to the  Lorentzian flat coordinates by replacing  $\sigma \to x_{1}$, $\tau\to ix_{2}$, and  $z\equiv x_{1}+ix_{2}$.
With this convention we can replace $\partial_{\tau} \to (\partial_{z} -\partial_{\bar{z}})$ $\partial_{\sigma}\to (\partial_{z}+ \partial_{\bar{z}})$. The Tseytlin action can be written, as shown previously, in a Wick rotated form as in eq. (\ref{TseytlinEuc}). 

\subsection{The scattering differentials}

Our next and  central point  is  that in order to demand that the action 
 is well-defined on $\Sigma$,  we have to impose that $\pa_{\tau} \X$ and $\pa_{\sigma}\X$ are single-valued on $\Sigma$, i.e., periodic.
But as we have already emphasized, this {\it does not mean} that $\X$ is a periodic function.
The proper mathematical implementation of this idea is that instead of parametrizing the action by 
a set of coordinates $\X^{A}$ on $\cal P$, we need to parametrize it by a {\it closed one form}
$\delta^{A}=\delta^{A}_{\s}\rd\s +\delta^{A}_{\tau}\rd \tau$ valued in $T\cal P$, $\rd \delta^{A}=0$.
Such a form possesses {\it monodromies}; for each cycle $\gamma$, we have 
\be\label{gendefDelta}
P^{A}_{\gamma} = \frac{(J\Delta_{\gamma})^A}{2\pi}  = \oint_{\gamma} \delta^{A},
\ee
where we define $ \oint = \tfrac1{2\pi} \int_{0}^{2\pi}$.
Since $\delta^{A}$ is closed, the monodromy depends only on the homology of $\gamma$.
In particular this means that if the string goes through an interaction point and splits, then
the monodromy before the splitting is the sum of the monodromies after, 
$\Delta_{\gamma_{12}}=\Delta_{\gamma_{1}} +\Delta_{\gamma_{2}}.$
This means that the set of monodromies should form a lattice.
We denote by  $\Lambda$ the lattice formed by rescaled monodromies $\Delta/2\pi$.
The normalization appears for future convenience.
In other words,  if $\Delta/2\pi ,\Delta'/2\pi \in \Lambda$, then $m\Delta/2\pi +n\Delta'/2\pi \in \Lambda$ for $n,m \in \mathbb{Z}$.

From this point of view the Polyakov theory in a large space-time corresponds to a lattice that is continuous in half the directions and of infinite lattice spacing in the others. This can be related to a particular limit\footnote{See eqs. (\ref{defdel},\ref{defp},\ref{XY}). $\Delta^A=(\delta^\mu/\lambda , 2\pi p_\mu/\varepsilon)=(2\pi n^\mu,2\pi m_\mu)$, where $n^\mu,m_\mu\in\mathbb{Z}$  label monodromy lattice points. In the limit $\lambda\to \infty$, $\varepsilon\to 0$, we have $\delta \to \infty$ and $p\to 0$. This should be interpreted as corresponding to a continuous space-time with coordinates $X^\mu$.} $\lambda\to \infty$, $\varepsilon\to 0$ (holding $\hbar$ fixed). Later we will refer to this as a sort of classical limit.

In the following we denote the space of closed one-forms with monodromies in the lattice $\Lambda$ by 
$C^1_\Lambda(\Sigma)$. It will be convenient to additionally fix the value of the external monodromies. If $\Sigma$ possess external points $i$,  the external monodromies are $\Delta_i^A/2\pi = \oint_i \delta^A  \in  \Lambda$. By construction we have that $\sum_i \Delta_i=0$. The space of such closed differentials is denoted $C^1_\Lambda(\Sigma,\Delta_i)$.
This means that the  Tseytlin action on a generic surface depends on the  monodromies $\Delta_i/2\pi \in \Lambda$ and that  the flat Tseytlin action on a generic surface should be written as 
\be\label{Tactionform}
S_{\Delta_i}\equiv \frac1{4\pi} \int_{\Sigma}d^2\sigma \Big( (\eta_{AB}+ \omega_{AB}) \delta^{A}_{\tau} \delta_{\s}^{B} -  
 H_{AB}  \delta_{\s}^{A} \delta_{\s}^{B}\Big),
\ee
where $ \delta^{A}$ is a closed form with fixed monodromy $\delta^A \in C^1_\Lambda(\Sigma,\Delta_i)$.

\subsection{Nakamura strips}

It will be convenient in the following to write the action in a more familiar manner in 
 terms of coordinates $\X^{A}$.
Locally the one-form can be written as $\delta^{A}=\rd \X^{A}$ and the coordinate $\X^{A}$ is recovered as 
\be\label{intX}
\X^{A}(p) =\int_{p_{0}}^{p} \delta^{A},
\ee
where $p_{0}$ is a reference point in $\Sigma$.
Since $\delta^{A}$ have monodromies, $\X^{A}$ is multivalued on $\Sigma$.
In order to define $\X$ we therefore need to refer to a simply connected domain $D_{\Sigma}$ whose closure covers $\Sigma$. Such a domain is obtained by cutting open $\Sigma$ along a graph
$N$ where $D_{\Sigma}=\Sigma\backslash N$. One usually chooses the Torelli graph consisting of a homology basis $(a_{i},b_{i})$ with $i\in\{1,\cdots,g\}$. 
Such a choice is simple but inconvenient since it breaks the explicit modular invariance of the theory. The question is therefore whether or not there exists a cutting graph which preserves modular invariance.  Remarkably the answer is yes!
 
Such graphs were first proposed by Nakamura \cite{nakamura2000calculation} and we will call them Nakamura graphs. In fact, they provide a covering of the moduli space, in which each Nakamura graph corresponds  to an open cell in moduli space. This is more economical than the usual Penner decomposition \cite{penner2003}. 
 The idea for these graphs is very natural: we have seen that the GWKN theorem establishes an isomorphism between the moduli space of complex structures of a Riemann surface with in-out splitting and the moduli space of causal structures.
 The causal structure possesses interaction points $x_{i}$ which are the critical points of the time function.
 We take these interaction points as  vertices of the Nakamura graph $N$. From these vertices we draw the curves that are purely real trajectories of the GWKN differential. 
 That is, we draw trajectories along which $\Im(e)=0$, where $e$ is the GWKN differential.
 
\begin{figure}[h]
\begin{center}
$\begin{array}{ccc}
\includegraphics[width=2in]{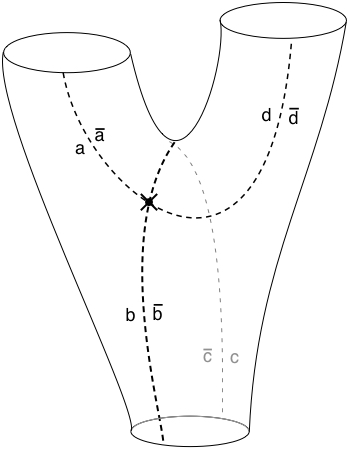}&\qquad
\includegraphics[width=1in]{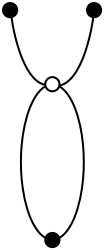}&\qquad
\includegraphics[width=2in]{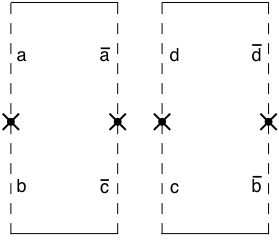}\\ 
~~~~~(a)&~~~~~~~~~~~~(b)&~~~~~~~~~~~~(c)
\end{array}$
\end{center}
\caption[FIG. \arabic{figure}.]{The Nakamura graph for the pants diagram is drawn on the surface in (a), and displayed in (b). The corresponding domain $\Sigma\backslash N$, consisting of two strips, is shown in (c). The interaction point is marked by a cross in (a) and (c) and by an open circle in (b).}
   \label{Naka1}
\end{figure}

 These real trajectories can end only at other interaction points or at the boundary of $\Sigma$. Therefore they provide the edges of the Nakamura graph of $\Sigma$.
 These edges are time oriented and it can be checked that at the interaction points an incoming edge (coming from the past) always alternates with an outgoing (or future directed) one.
 In summary, the Nakamura graph of $\Sigma$ consists of internal vertices which are the interaction points, external vertices which are associated with the external circles and edges 
 which are the real trajectories of the GWKN differential. A detailed study of this structure is given in \cite{Freidel:2014aqa}.
 By construction this graph is uniquely determined from the complex structure of the Riemann surface and the edges of this graph are purely timelike.
 Away from the interaction points, the causal structure is the usual one where each point has one  past and one future light cone.
 At the interaction points the causal structure is modified; we can have now several future light cones (equal to the number of past light cones). Fig. \ref{fig:fig4.jpg} displays the causal neighborhood of a typical interaction point. Such an interaction point is obtained (see Fig. \ref{Naka1}(a)) by considering future directed time-like curves in the neighborhood of a pants fixture; thus the interaction points are associated with the topology change of spatial sections involved in the string splitting-joining interaction.
 The interaction points coincide with the critical points of the Morse function $\tau$, the vertices of the Nakamura graph and the zeroes of the GWKN differential. In the gauge for the worldsheet metric that we are using, the worldsheet curvature is singular there. Thus in general $\sigma$-model backgrounds, dilaton degrees of freedom couple to the worldsheet at these interaction points, and thus it is through the interaction points that the string coupling will enter the theory.
 
 \myfig{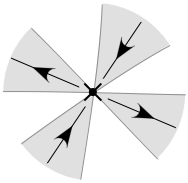}{5}{Typical causal structure at an interaction point with two past and two future light cones. The shaded regions are timelike and the future and past light cones alternate. At a regular point, of course, there is a single light cone with forward-directed time-like curves.   }
Using this graph, we can define the  domain $\Sigma\backslash N$.
This domain is non-connected but each connected component is simply connected.
Let's denote each connected component by $S_{i}$ so that $\Sigma\backslash N= \cup_i \mathrm{Int}({S}_{i})$.
In each domain $S_i$ we can choose a base point $z_i$ and use the GWKN differential $e$ to construct flat coordinates 
$\rd\tau +i \rd \s \equiv \int_{z_{i}}^{z} e$. 
The $\tau$ function is well-defined inside each $S_i$ since the critical points of $\tau$ are 
by definition the vertices of $N$ and are therefore in the boundary of a domain.
The causal structure is also well-defined inside the strip and near the boundary.
One boundary of the strip includes only future oriented edges while the other only past oriented edges.

Since the boundary of each domain $S_i$ is a real trajectory the value of $\s$ is fixed on the boundary. This shows that $S_i$ is isometric to a strip $S=[\sigma^{-}_{i},\sigma^{+}_{i}]\times \mathbb{R}$ where $\mathbb{R}$ is the time interval and the strip size is $\Delta\sigma_{i}=\sigma^{+}_{i}-\sigma^{-}_{i} =\int_{\gamma_{S_i}} \Im(e)$, where the integral is along any curve that goes from one boundary of the strip to the other.
There are constraints on the sets of admissible strip widths $\Delta\sigma_{i}$, appropriate to a given value of $r_\alpha$, once the strips are sewn together. 
We can also assign the interaction time differences  $\tau_{j}\equiv \int_{x_{0}}^{x_{j}}\Re(e)$, where $x_{j}$ are the interaction points and $x_{1}$ is the first  interaction point.
The collection of strip widths and interaction times $(\Delta\sigma_{i},\tau_{j})$ modulo the normalization conditions for each external leg represents the set of moduli parameters.
The number of parameters is therefore the number of strips plus the number of interaction points minus the number of boundary circles.
It can be checked that for the top-dimensional cell,  this is exactly 
$ 6g-6 + 2n$, the dimension of moduli space. 
This  leads to a very simple representation of the integral over the moduli space
as first a sum over all Nakamura graphs and then an integral over the strip parameters (see \cite{Freidel:2014aqa}). We note in passing that the top dimensional cells are special in that they have no edges linking internal vertices. It seems likely that this structure would have an important impact on a string field theory formulated in this way. 

We now assume that a point in the moduli space and 
 a corresponding Nakamura graph associated with $\Sigma$ has been chosen.
 As we just have seen, this amounts to a flat strip decomposition of $\Sigma$.  The boundary of each strip can be decomposed as 
 $ \pa S = \pa S_{+}\cup \pa S_{-}$, where $\pa S_{+}$ consists of edges ${\bm e}_{+}$ oriented from 
 the past to the future and $\pa S_{-}$ consists of edges ${\bm e}_{-}$ oriented from  the future  to the past. The simplest example of this construction is shown in Fig. \ref{Naka1}.
 
Given the strip decomposition we can now construct a set of coordinates 
$\X_{i}^{A}$ for each strip $S_i$ by first choosing a set of base points $x_{i}\in S_i$ and then defining
\be\label{Xdef}
\X^{A}_{i}(x)\equiv \int_{x_{i}}^{x} \eta^{A},\quad \mathrm{for} \quad x\in S ,
\ee
so that $\eta^{A}(x)=\rd \X^{A}_{i}(x)$ for $x\in S_i$.
An edge ${\bm e}$ of $N$ belongs to two strips ${\bm e} \subset \pa S_{i+}\cap \pa S_{i'-}$
and we denote by ${\bm e}_{+}$ (resp. ${\bm e}_{-}$) a point approaching ${\bm e}$ from $\pa S_i$ (resp. $\pa S_{i'}$). 
Accordingly, we can compute the ``discontinuities'' across ${\bm e}$ to be
\be\label{disc}
\Delta_{{\bm e}}^{A} = \X^{A}_{i}({\bm e}_{+})-\X^{A}_{i'}({\bm e}_{-})= \int_{x_{i'}}^{x_{i}}\eta^{A},
\ee
where the integral is along a path that crosses $\pa S_{i+}\cap \pa S_{i'-}$ only once along ${\bm e}$.
 The sets of discontinuities $\Delta_{{\bm e}}$ determine the monodromies via 
\be\label{gendefmon}
\Delta_{\gamma}=\sum_{{\bm e}\cap \gamma\neq \emptyset} (\pm \Delta_{{\bm e}}).
\ee
The sign depends on whether the frame $(\gamma,{\bm e})$ at each intersection point agrees with
the orientation of $\Sigma$ or not.

\section{Canonical analysis of the metastring}

The Nakamura decomposition of a Lorentzian worldsheet allows for a complete study of generic quantum amplitudes, broken up into a series of strip geometries. In this section, we will consider one important piece of this construction, in which we focus on a cylindrical worldsheet geometry, cut open along $\sigma=0,2\pi$. We thus are considering a single strip $S$. What we intend to show is that there is a suitable notion of `closed string boundary conditions,' even though monodromies are present.\footnote{Specifically, we mean here that the usual notion of the closed string boundary condition, $X(\sigma+2\pi)=X(\sigma)$, must be replaced by a more general notion in the presence of monodromy. It is perhaps surprising that such a more general notion exists. The use of careful symplectic methods here ensures that we have consistent canonical evolution and consistent boundary conditions.} We construct this in a general symplectic formulation; the basic notion of the closed string boundary condition will be that the symplectic flux across the cut should be continuous, and in fact independent of where we make the cut.

As described earlier, the worldsheet action takes the form
\beq\label{actionpatch}
S=\frac1{4\pi} \int_{\tau_o}^{\tau_i}\rd \tau \int_{0}^{2\pi} \rd \s\left[  \pa_{\tau}\X^{A} (\eta_{AB}+ \omega_{AB})\pa_{\s}\X^{B} -  
  \pa_{\s}\X^{A} H_{AB} \pa_{\s}\X^{B}\right] ,
\eeq
where we have restricted to $S$, coordinatized by $\tau\in [\tau_i,\tau_0], \s\in[0,2\pi)$.
The fields $\X$ are assumed to be quasi-periodic $ \X(\tau,\s+2\pi)= \X(\tau,\s)+\Delta(\tau) $, with $\pa_\s \Delta=0$. Note that this implies that $\pa_\s\X(\tau,\s)$ is itself periodic. 
When we derived this action above, we were careful to not make use of integration by parts, or in other words discard any terms associated with the cut. If we had done so, then we could have, for example, eliminated the $\omega$-term in the action, as it is a total derivative (for constant $\omega$). Indeed, our generalization is to include monodromies, and naively this would seem to imply that the mappings of the worldsheet into the target space correspond to `torn' worldsheet embeddings. We now demonstrate that this is in fact {\it not} the case, despite appearances. To see this, we will carefully study the symplectic structure of the theory, and we will see that when the appropriate notion of closed string boundary conditions are implemented, $\omega$ does not appear in the symplectic structure. We take this as an indication that even though monodromy is present, the string embeddings should be regarded as closed. In this construction, the monodromy will appear as the discontinuity across the cut surface; the closed string boundary condition is the requirement that this cut be invisible, or that the choice we make for its position is immaterial. Ultimately then, we will find that the modes of the string are characterized by periodic chiral oscillator modes, and by a set of zero modes. The monodromy appears in the zero mode sector, and will be canonically conjugate to the center of mass of the string (in phase space). 

\subsection{Solutions}\label{sold}

The bulk variation of the action is given by $ \int \rd\tau \rd \s\ \delta \X\dd \pa_\s\S$. It  imposes the equation of motion 
\beq\label{eomS}
\pa_\s\S=\pa_\s(\pa_\tau-J\pa_\s)\X=0.
\eeq
 In order to analyze this equation on the strip geometry, we introduce  the center of mass coordinate $x$ and  the monodromy $\Delta$:
 \beq
\oint\rd\s\ \X(\tau,\sigma)=x(\tau) 
,\qquad 
\oint\rd\s\ \pa_\s\X(\tau,\sigma)=\frac{\Delta(\tau)}{2\pi},
\eeq
where $\oint \equiv \frac1{2\pi} \int_{0}^{2\pi}$.
The equation of motion implies that 
\be\label{eomS2}
\S(\tau)=\oint \S(\tau) = \pa_\tau x - \frac{J\Delta}{2\pi}.
\ee
Moreover, by integrating (\ref{eomS}) and using the periodicity of $\pa_\s\X$ we get that 
\be
0= \oint \pa_\s\S= \frac{\pa_\tau\Delta}{2\pi}.
\ee
Therefore $\Delta$ is time-independent, a result that follows from the equations of motion. This is the manifestation, in these local coordinates, that the general definition of monodromy, eq. (\ref{gendefDelta}), depends only on the homology class of $\gamma$. 

It will be convenient for us to introduce the coordinates 
\be
\bar{x}(\tau) \equiv \int_0^\tau\rd\tau\ \S(\tau),
\ee
so that we  can write the time-dependence of the center of mass position as 
\be
x(\tau)= x_c + \frac{J\Delta}{2\pi} \tau + \bar{x}(\tau).
\ee
Here $ x_c$ is time-independent. This decomposition is useful since we will see that $\bar{x}(\tau)$ is a gauge parameter, it can be fixed to any value without affecting the symplectic structure. 

From (\ref{eomS2}), we see that the general solution to the equation of motion can be written
\beq\label{on-shell}
\X(\tau,\sigma)=x_c + \frac{J\Delta}{2\pi} \tau  + \frac{\Delta}{2\pi}\s +{\Q}(\tau,\sigma) + \bar{x}(\tau),
\eeq
where ${\Q}$ is chiral: $(\pa_\tau-J\pa_\sigma){\Q}(\tau,\sigma)=0$ and periodic under $\sigma\to\sigma+2\pi$. Consequently, we can Fourier expand
\beqn
{\Q}(\tau,\sigma)&=&\sum_{n\in\mathbb{Z}_*}\Q^{+}_n e^{-in(\tau+\sigma)}
-\sum_{n\in\mathbb{Z}_*}\Q^{-}_n e^{-in(\tau-\sigma)}\\
&\equiv & {\Q}^{+}(\tau+\sigma)- {\Q}^{-}(\tau-\sigma).
\eeqn
 where 
 ${\Q}^{\pm}$ are the analogues of the left- and right-movers in the usual string.
 They are distinguished in this chiral formulation according to their chirality:
 \be
 J{\Q}^{+}=+{\Q}^{+},\qquad J{\Q}^{-}=-\Q^{-}.
 \ee

\subsection{Symplectic structure}
In order to construct the symplectic potential and the symplectic form we focus on variations of the action that preserve the equation of motion $\pa_\s \S=0$. 
Here we follow the method of \cite{Crnkovic:1986be,Crnkovic:1987tz}; see also \cite{Gawedzki:1990jc} and \cite{Kijowski:1973gi,Kijowski:1976ze}.
One introduces the notion of a differential on field space denoted as $\delta $, this differential 
satisfies the Leibnitz rule and squares to zero, i.e. $\delta^{2}=0$. It also acts on the space of solutions of the Lagrangian system, infinitesimally mapping solutions to solutions and it is such that the product of field differentials anti-commute. 
 The on-shell variation of the action is  a pure boundary term given by:
\beqn\label{varaction}
\delta S &\hat{=}& \frac12 \left[\oint \rd \s\  \delta \X (\eta+\omega) \pa_{\s}\X\right]^{\tau_{o}}_{\tau_{i}} +
\frac1{4\pi}\int_{\tau_{i}}^{\tau_{o}}\rd \tau\   \Big[\delta\X \left(2\eta\S-(\eta+\omega) \pa_{\tau}\X\right)\Big]\Big|^{2\pi}_{0}.
\eeqn 
By $\hat{=}$, we mean that the equality is on-shell.

From this, we can extract the  symplectic 1-form by writing
\beq
\delta S
=\int_{\pa M} *\bm{\theta}[\delta\X].
\eeq
The symplectic potential current $\bm{\theta}$ is determined only up to the addition of  a total derivative $\frac{1}{2\pi} \rd \alpha$.
In general this choice is associated with a choice of boundary condition.
Care must be taken to select $\alpha$ appropriately.
As we will see, demanding that the metastring is a {\it closed} string will determine the appropriate choice 
of $\alpha$, such that the resulting symplectic structure is time-independent.  Such a choice corresponds to the specification of boundary conditions, in particular, across the cut.

 In the case at hand,  the boundary has four components, $\pa M=\Sigma_o\cup\Sigma_i\cup R_0\cup R_{2\pi}$ (with $R_{0,2\pi}\equiv \{\s=0,2\pi\}$), each of which contributes  terms to $\delta S$. What we need to identify is a form in the bulk whose pull-back to the boundary reproduces (\ref{varaction}). By inspection, we have
\beqn
*\bm{\theta}&=&\frac{1}{4\pi}\left( -\delta\X (\eta+\omega) \rd\X+2 \delta\X\dd \hat\S-\rd\alpha\right) ,
\eeqn
where we have defined the form $\hat\S\equiv \S\ \rd\tau$ and $\X\dd \mathbb{Y} =\X^A\eta_{AB}\mathbb{Y}^B$. In component form,
$(*\bm{\theta})_a = \epsilon_{ab} \bm{\theta}^b$, where $\bm{\theta}^b$ is the symplectic current and 
$\epsilon_{\sigma\tau} =-\epsilon_{\tau\sigma}=1$, we can write this as
\beq
\epsilon_{ab}\bm{\theta}^b[\delta\X]=\frac{1}{4\pi} \left(-\delta\X (\eta+\omega) \pa_a\X
+2  t_a \delta\X\dd \S -\pa_a\alpha \right) ,
\eeq
where   $t_a=(t_\s,t_\tau)\equiv(0,1)$ is the one form tangential to the  the timelike  boundary components, $R_{0,2\pi}$.

The symplectic 2-form current $\bm{\omega}^a=\delta\bm{\theta}^a$  is given by 
\beqn\label{symp2form}
\epsilon_{ab}\bm{\omega}^b &=&\frac{1}{4\pi}\left(\delta\X (\eta+\omega) \pa_a\delta\X
-2 t_a\delta\X\dd\delta\S-\pa_a\delta\alpha\right)\\
&=&\frac{1}{4\pi}\left(\delta\X \dd \pa_a\delta\X
+\frac{1}{2}\pa_a(\delta\X \omega \delta\X)
-2 t_a \delta\X\dd\delta\S-\pa_a\delta\alpha\right).
\eeqn

Note the central property that the symplectic 2-form (\ref{symp2form}) is conserved on-shell 
\beqn
\nabla_a \bm{\omega}^a &=&\frac{1}{4\pi}\epsilon^{ab}\pa_a\delta\X\dd\pa_b\delta\X+ \frac{1}{2\pi} \pa_\sigma( \delta\X\dd\delta\S)=  \frac{1}{2\pi}  \delta\X\dd \delta(\pa_\sigma\S)\;\hat{=}\; 0.
\eeqn
Now let us look at the conservation of symplectic flux.
Since $d*\bm{\omega}=0$, we have\beqn
0&=&\int_{\pa M} *\bm{\omega} 
= -\int^{\tau_o}_{\tau_i} d\tau\ \bm{\omega}^\s\Big|^{2\pi}_{0} - \int^{2\pi}_{0} d\sigma\ \bm{\omega}^\tau\Big|^{\tau_o}_{\tau_i}.
\eeqn
The second term represents the difference of the symplectic structures between final and initial times  at initial and final times, while the first term is the net symplectic flux through the spatial boundary. Indeed, we define $\Omega(\tau) \equiv  \int^{2\pi}_{0} \rd \s\ \bm{\omega}^\tau $ to be the symplectic structure. The symplectic flux  across the cut is defined by $ \Phi_e \equiv -\int_{\tau_i}^{\tau_o}\rd\tau \left.\bm{\omega}^\s\right|^{2\pi}_{0} $, and the conservation equation reads 
\be
\Omega(\tau_o) -\Omega(\tau_i)= \Phi_e.
\ee
It is this net flux that must vanish in order for the symplectic structure to be time independent.
Interestingly, we can accomplish this on-shell by an appropriate choice of the ambiguity $\alpha$. Denoting $\alpha_e \equiv \alpha({2\pi})-\alpha(0)$, we see that the ambiguity $\alpha$  in the symplectic structure corresponds to the subtraction of a term
$\int_0^{2\pi} \pa_\s \delta \alpha = \delta \alpha_e$ to $\Omega$. 
We can now evaluate the symplectic flux across the cut.
Remarkably, using that $\pa_\tau \delta \Delta=0$, we obtain that 
$\left.\bm{\omega}^\s\right|^{2\pi}_{0}$  is a total time derivative, therefore the symplectic flux is given by
\beqn
\Phi_e=
\frac{1}{4\pi}\Big[ \frac12\delta\Delta(\eta+\omega)(\delta\X_0+\delta\X_{2\pi})-2 \delta \Delta \dd \delta \bar{x} - \delta\alpha_e\Big]_{\tau_i}^{\tau_o} ,
\eeqn
where we have 
used evaluation (\ref{eomS2}): $\left.[\delta \X\dd \delta \S]\right|_0^{2\pi}= \delta \Delta\cdot \pa_\tau \delta \bar{x}$.
Thus, we are led to take 
\bea
\delta\alpha_e
&=&\frac12\delta\Delta(\eta+\omega)(\delta\X_0+\delta\X_{2\pi}) -2 \delta \Delta \dd \delta \bar{x}. 
\eea 
From this expression we can extract a proposal for the potential $\alpha$:
\be
\alpha \equiv  \frac12 \X (\eta+\omega)(\delta\X_0+\delta\X_{2\pi}) -2  \X \dd \delta \bar{x},
\ee
to which we can add any closed expression $\delta\phi$ and any periodic one-form.
These ambiguities in the choice of $\alpha$ do not affect the choice of symplectic structure which is only characterized by $\delta \alpha_e$.
The symplectic structure is then
\beqn
\Omega=\int d\sigma\; \omega^\tau
&=&\frac12\oint\rd\s\ \delta\X \dd \pa_\s\delta\X+\frac14\oint\rd\s\ \pa_\s \big(\delta\X\omega\delta\X\big)- \frac12\oint \pa_\s \alpha\\
&=&\frac12\oint\rd\s\ \delta\X \dd \pa_\s\delta\X+\frac1{8\pi} \delta\Delta\omega(\delta\X_0 + \delta\X_{2\pi})- \frac1{4\pi}\delta\alpha_e\\
&=&\frac12\oint\rd\s\ \delta\X \dd \pa_\s\delta\X-\frac{1}{4\pi}\delta\Delta\dd\delta\X_0 + \frac1{2\pi} \delta \Delta \dd \delta \bar{x}.
\eeqn
Making use of the general solution given above, this can be rewritten as
\beqn\label{Om1}\boxed{
\Omega
=
\frac{1}{2\pi}\delta x_c\dd\delta\Delta
+\frac12\oint\rd\s\ \delta\Q\dd \pa_\s\delta\Q
}.
\eeqn
Here we 
have used that
\be
\left(\oint \delta\X -\delta \bar{x}\right)\dd\delta\Delta =\delta x_c \dd \delta \Delta.
\ee
Thus the symplectic geometry is coordinatized by the periodic chiral oscillator modes $\Q$, and a zero mode sector $\{x_c,\Delta\}$.  
This expression is remarkably simple and natural.
First, let us notice that the reference to the end points of the strips has disappeared. The symplectic structure is independent of the choice of the cut; it depends only on the cylinder topology, as it should.
 We also clearly see that the 2-form  $\omega$ does not enter the definition of the string symplectic form. It only depends on  the neutral metric $\eta$. This is consistent with the interpretation that despite the cut  the string is closed since the term proportional to $\omega$ is a closed form.

Finally, we see  that $\bar{x}$ is a gauge parameter: any change of $\S =\pa_\tau\bar{x}$ leaves the symplectic structure unchanged since $\Omega(\delta\S,\cdot)=0$. 
This  follows  from the fact that the theory is invariant under  the partially-local time shift symmetry, $\X(\tau,\s)\mapsto\X(\tau,\s)+f(\tau)$, that we discussed in Section \ref{sec:timetrans}.
In order to have an invertible symplectic structure we need to fix this symmetry. It is natural to choose 
\be
\S=0.
\ee
In this gauge the symplectic structure can be written in terms of $\X$ as 
\be\label{Om2}\boxed{
\Omega= \frac12 \oint\rd\s\ \delta\X \dd \pa_\s\delta\X-\frac{1}{4\pi}\delta\Delta\dd\delta\X_0}.
\ee
In this form the symplectic structure is reminiscent of the one used by Bowick and Rajeev on the loop space associated with the open string \cite{Bowick:1986rc,Bowick:1987pw}.
The addition of the boundary term can be seen to be necessary in order to make the full expression independent of the position $\s_0=0$ of the cut.
In this gauge we also have that the general solution is given by
\be
\X(\tau,\s)=x_c + \frac{J\Delta}{2\pi} \tau  + \frac{\Delta}{2\pi}\s + {\Q}(\tau,\s).
\ee
Note that due to the equations of motion, $P\equiv (J\Delta)/2\pi $ is  
the  velocity  of the center of mass of the strip. Indeed 
\be
 \pa_{\tau}x^{A}_c = \oint \rd\sigma\ \pa_{\tau}\X^{A}(\sigma) = \oint \rd\sigma\ \pa_{\s}(J\X)^{A} = \frac{(J\Delta )^{A}}{2\pi}.
\ee
This is a striking property of the metastring: its velocity 
 is proportional to its extension. 
This is reminiscent of the dyonic condition appearing in non-commutative field theories \cite{Grosse:2004yu, Douglas:2001ba} where the Fourier transform of a Wilson line at momentum $k$ is gauge invariant, if the length of the line is related to its momentum via $\ell = \theta k$, where $\theta$ measures the non-commutativity of coordinates.

From the expression of the symplectic form, we deduce the Poisson bracket\footnote{ The correspondence between Poisson bracket and symplectic form  from the fact that a Hamiltonian $H$ determines a vector field $X_H$ via 
$
\Big\{H , \cdot \Big\} = X_H
$. The relation with the symplectic structure $\Omega$ is given by 
\be
I_{X_H}\Omega (X_F) = \bm{\omega}(X_H,X_F) =- \Big\{H,F\Big\},
\ee where $I_X$ denotes the interior product. Since $X_F(\delta H) = \Big\{F,H\Big\}$ this correspondence can also be written as
\be
I_{X_{H}}\Omega =\Omega(X_H,\cdot)= \delta H \quad\Leftrightarrow\quad \Big\{H,\cdot \Big\}= X_{H}.
\ee }
\beq
\Big\{x^{A}_c, \Delta^{B}\Big\}=  2\pi  \eta^{AB},\qquad 
 \Big\{{\Q}^{A}(\sigma), {\Q}^{B}(\sigma') \Big\} = 2\pi \eta^{AB} \epsilon(\sigma-\sigma'),
\eeq
where ${\epsilon}$ is an antisymmetric and periodic  function
\beq \label{eps}
{\epsilon}(\sigma) \equiv \frac1{2\pi i} \sum_{n\neq 0} \frac1{n}e^{in\sigma},
\eeq
which satisfies $\partial_{\sigma}{\epsilon}(\sigma) = \delta(\sigma)- \frac{1}{2\pi} $ where  $\delta(\sigma)$ is the periodic Dirac distribution. In other words, ${\epsilon}$ is the inverse of $\pa_{\s}$ on the space of $2\pi$-periodic functions that have a vanishing average. 
We can see that $\theta(\s)\equiv {\sigma} + 2\pi  \epsilon(\sigma)$ is the staircase distribution, which is characterized by 
its normalization: $\theta(\s)=\pi $ for $\s\in ]0,2\pi [$, quasi-periodicity: $\theta(\s + 2n \pi )=\theta(\s)+2\pi n$ for $n\in\mathbb{Z}$ and skew-symmetry, $\theta(-\s) =-\theta(\s)$.  
The staircase distribution enters the equal time commutators of the coordinates 
\be
\Big\{ \X^A(\s),\X^B(\s')\Big\} =  \eta^{AB}\theta({\s -\s'}).
\ee
The commutation relation can be equivalently written in terms of the mode expansion
\be
{\Q}(\s) =\sum_{n\in \mathbb{Z}_*} ( \Q_{n}^+ e^{-in\s} -
 \Q_{n}^- e^{in\s} ). \ee
 This commutator involves the projectors on the chiral components:
 \be
 \Big\{ \Q_{n}^{+ A} , \Q_{m}^{+B} \Big\}= i\frac{\delta_{n,-m}}{  2 n} (H+\eta)^{AB},\qquad 
 \Big\{ \Q_{n}^{-A} , \Q_{m}^{-B} \Big\}= i\frac{\delta_{n,-m} }{2n}(H-\eta)^{AB},
 \ee
while the commutators $\Big\{\Q_{n}^+,\Q_{m}^-\Big\}$ vanish.

\subsubsection{ Shift invariance }

In the previous section we have seen that in order to be conserved, the symplectic structure 
has to include boundary term variations which promote some of the boundary data associated with the cut to dynamical degrees of freedom. The net effect is to compensate all  symplectic flux going through the cut showing that the cut is in fact irrelevant and the string closed.
This suggests that instead of choosing $\alpha$ we could already include some boundary terms
directly  in the action. This is what we now present.
The way to see that a boundary term is needed in the action is to leave the position $\s_0$ of the cut arbitrary. That is we consider 

\beq\label{actionpatchs}
S_T=\frac1{4\pi} \int_{\tau_o}^{\tau_i}\rd \tau \int_{\s_0}^{\s_0+2\pi } \rd \s\Big[  \pa_{\tau}\X (\eta+ \omega)\pa_{\s}\X - 
  \pa_{\s}\X H \pa_{\s}\X\Big]- s(\s_0).
\eeq
The first term is the bulk action we have already considered while $s$ is a boundary action. We can choose this boundary term by demanding the total action $S_T$ to be  independent of $\s_0$ and  invariant under change of orientation $\s \to 2\pi-\s$.
These conditions are satisfied by choosing   
\be
s(\s_0)  = \frac1{8\pi} \int^{\tau_o}_{\tau_i}\rd \tau\, \pa_\tau\Delta (\eta +\omega)\left(\X_{\s_0} +\X_{\s_0 + 2\pi}\right).
\ee
Since the total action is independent of $\s_0$ we can fix it  to be $\s_0=0$.
Combining the variation of the bulk action (\ref{varaction}) with the variation of $s$, we obtain after simplification that 
\bea
\delta S_T 
&=& \frac12\left[\oint \rd \s\  \delta \X (\eta+\omega) \pa_{\s}\X\right]^{\tau_{o}}_{\tau_{i}} 
- \frac1{8\pi}
\left[\delta \Delta (\eta+\omega) (\X_0 +\X_{2\pi})\right]_{\tau_i}^{\tau_o}+ \frac1{2\pi}\int_{\tau_{i}}^{\tau_{o}}\rd \tau\  \delta\X \eta\S\Big|_0^{2\pi}\\
& & -
\frac1{4\pi} \int_{\tau_o}^{\tau_i} \rd\tau \,
\left( \delta \left(\X_{0} +\X_{ 2\pi}\right)\eta \pa_\tau\Delta
 + 2 \oint\rd\s\,  \delta \X \eta \pa_\s\S\right),
\eea
this being valid off-shell.
If we now impose the equation of motion $\pa_\s \S=0$, $\pa_\tau \Delta =0$ and 
introduce the coordinate $\bar{x}=\int^\tau_0\rd\tau\ \S$ as before,  the 
on-shell variation of the total action becomes simply  $\delta S_T \hat{=} \Theta(\tau_o)-\Theta(\tau_i)$
where the symplectic potential is given by
\be
\Theta = \frac12 \oint \rd\s\,(\delta \X (\eta+\omega) \pa_{\s}\X) -\frac{1}{8\pi} \delta \Delta (\eta+\omega) (\X_0 +\X_{2\pi}) + \frac{1}{2\pi} \delta \Delta \eta  \bar{x}.
\ee
By taking its  variation we recover the symplectic potential $\Omega=\delta \Theta$ constructed in the previous section.
The conclusions reached previously are unchanged and this should not come  as a surprise since the additional boundary action $s$ vanishes on-shell. 
The advantages of this formulation are  twofold. First, we have that both the action and the symplectic potential are independent of the position of the cut from the outset. Moreover, one sees that the symplectic flux along the cut vanishes without the need to add an extra boundary contribution.

\subsubsection{Time Symmetry}

Recall that the metastring action possesses a time-translation symmetry
\be\label{timesym}
\delta_f \X^A(\tau,\sigma)\equiv f^A(\tau).
\ee
This corresponds to a translation along a $\sigma$-independent vector field.
Under such a transformation, the action transforms by a boundary term
\be
\delta_f S
= \frac1{4\pi} \int_{\tau_i }^{\tau_o} \rd\tau \, \dot{f}(\tau) (\eta +\omega) \Delta.
\ee
This term vanishes when $(\eta-\omega) f=0$ and thus  (\ref{timesym})
 is a symmetry.
 
 In order to compute the Noether current associated with the time translation symmetry we subtract 
 this previous variation to the on-shell variation, that is 
 \be
 \delta_f S- \hat{\delta}_fS = \int_{\pa{\Sigma}} J_f ,
 \ee
 where  $\hat\delta_fS =\int *\theta[\delta\X=f]$, which can be evaluated using (\ref{varaction}).
Since the translation parameter is time-dependent, it is clear that the time component of the current vanishes.
Overall, we find that 
\be
J_f^\tau=0,\qquad J_f^\s= \frac1{2\pi} (f\dd \S).
\ee
The vector $\S$ is therefore  the space component of the Noether current.
The equation of motion $\pa_\s \S=0$ can therefore be understood as the conservation of the Noether current associated with time translation.
Under a time translation symmetry we have that 
\be
\delta_f \S = \pa_\tau f
,
\ee
which implies that we can always fix a gauge where $\S=0$. This gauge leaves open the possibility to have constant translations, with $\pa_\tau f=0$.

\subsection{Constraint algebra}

We can now demonstrate that the constraint algebra is consistent with the symplectic structure. What we will see is that the constraints that we discussed previously are given correctly by Hamiltonian vector fields.\footnote{ The correspondence between Hamiltonian vector fields $X_{H}$ and Hamiltonians is taken to be 
\be
I_{X_{H}}\Omega =\Omega(X_H,\cdot)=  \delta H \quad\Leftrightarrow\quad \{H,\cdot \}= X_{H}.
\ee} In the derivation of this result, the full form of the symplectic structure (\ref{Om1}) will be required.
We start with the diffeomorphisms
\be
\delta_{N}\X^{A} = N \pa_{\s}\X^{A},\qquad {\delta}_{\tilde {N}}\X^{A}= \tilde{N}\pa_\tau\X^A,
\ee
where $N,\tilde{N}$ are $2\pi$-periodic functions.
The first transformation corresponds to the space reparameterization while 
the second to time reparameterization.
These vector fields are Hamiltonian:
\bea
I_{\delta_{N}} \Omega = \delta \left( \frac12\oint N \pa_{\s}\X^{A}\eta_{AB}\pa_{\s}\X^{B}\right)\equiv\delta D_{N},\\ 
I_{\delta_{\tilde {N}}} \Omega = \delta \left( \frac12\oint \tilde{N} \pa_{\s}\X^{A} H_{AB}\pa_{\s}\X^{B}\right)\equiv \delta H_{\tilde{N}}
\eea
 and $I_{\delta}$ denotes the interior product,
$(I_{\delta}\Omega)(\delta') = \Omega(\delta,\delta')$. For example, starting from (\ref{Om2}) we find
\beqn
\Omega(\delta_N\X,\delta\X)&=&
\frac12\oint\rd\s \left(\delta_N\X \dd \pa_\s\delta\X-\delta\X \dd \pa_\s\delta_N\X\right)
-\frac{1}{4\pi}\delta_N\Delta\dd\delta\X_0+\frac{1}{4\pi}\delta\Delta\dd\delta_N\X_0\\
&=&
\frac12\oint\rd\s \left(N \pa_{\s}\X \dd \pa_\s\delta\X-\delta\X \dd \pa_\s (N \pa_{\s}\X)\right)
+\frac{1}{4\pi}\delta\Delta\dd N_0\pa_\s\X_0\\
&=&
\frac12\oint\rd\s \left(2N \pa_{\s}\X \dd \pa_\s\delta\X-\pa_\s (\delta\X \dd N \pa_{\s}\X)\right)
+\frac{1}{4\pi}\delta\Delta\dd N_0\pa_\s\X_0\\
&=&\frac12
\oint\rd\s\ \delta \left(N\pa_{\s}\X\dd\pa_\s\X\right) =\delta D_N.
\eeqn
We have used here the result that $\delta_N\Delta=0$ and the periodicity of $N$
and $\pa_\s\X$.
Similarly,
\beqn
\Omega(\delta_{\tilde{N}}\X,\delta\X)
&=&
\frac12\oint\rd\s \left(\tilde{N} \pa_{\tau}\X \dd \pa_\s\delta\X-\delta\X \dd \pa_\s (\tilde{N} \pa_{\tau}\X)\right)
+\frac{1}{4\pi}\delta\Delta\dd \tilde{N}_0\pa_\tau\X_0\\
&=&
\frac12\oint\rd\s \left(2\tilde{N} \pa_{\s}\X \dd \pa_\tau\delta\X-\pa_\s (\delta\X \dd \tilde{N} \pa_{\tau}\X)\right)
+\frac{1}{4\pi}\delta\Delta\dd \tilde{N}_0\pa_\tau\X_0\\
&=&\frac12
\oint\rd\s\ \delta \left(\tilde{N}\pa_{\s}\X\dd J\pa_\s\X\right) = \delta H_{\tilde{N}},
\eeqn
where in the last line we used the gauge condition $\S=0$.

This confirms the interpretation of $H$ and $D$ as the canonical generators of local time and space reparameterizations.
 The Poisson algebra of constraints is isomorphic to $\mathrm{Diff}(S^{1})\times \mathrm{Diff}(S^{1})$ where the Virasoro generators are $L_{N}^{\pm}\equiv  (H_{N}\pm D_{N})/2 $.
 The (classical) algebra is
 \be
 \Big\{L_{N}^{\pm},L^{\pm}_{N'}\Big\}= L_{(N\pa_{\s}N'-N'\pa_{\s}N)}^{\pm}, \qquad \Big\{L_{N}^{+},{L}_{N'}^{-}\Big\}= 0. 
 \ee

\subsection{General strips}
In the previous section we have constructed the symplectic structure for the cylinder possessing one cut. Given a more general surface and the corresponding Nakamura decomposition into a collection of flat strips, it is necessary to describe how these are to be glued back together. When two time-like edges are glued together, we require that the 
symplectic fluxes across each cut match. The closed string boundary condition that we discussed carefully above is then just a special case of this more general assertion. 
Here we will just sketch some elements of this description since the full analysis deserves a separate study.

The dynamics for a general metastring  is simply given by a sum of Tseytlin actions for each strip
\be\label{stripaction}
S=\frac1{4\pi} \sum_{i}\int_{S_i}\rd \tau \rd \s\left[  \pa_{\tau}\X^{A}_i (\eta_{AB}+ \omega_{AB})\pa_{\s}\X^{B}_i -  
  \pa_{\s}\X^{A}_i H_{AB} \pa_{\s}\X^{B}_i\right].
\ee
The sum is over the Nakamura flat strips and $\X^A_i=\int_{\sigma_i}^\s\eta^A$, is the coordinate defined in (\ref{Xdef}).
Let us recall that for the moment we restrict our analysis to the case where $(\eta,\omega, H)$ are all constant and that each cut $e$ belongs to two strips $S_{{\bm e}_-}$ and $S_{{\bm e}_+}$.
The continuity of the action demands that across a cut located at $\s_{\bm e}$ we have that 
$\pa_\s\X_{{\bm e}_-}(\sigma_{\bm e}) = \pa_\s\X_{{\bm e}_+}(\sigma_{\bm e})$.
The discontinuities in the choice of coordinates are encoded into the edge monodromies which encode the translation involved with the change of 
coordinates $\X_{{\bm e}_-} \to \X_{{\bm e}_+}$
\be\label{De}
\Delta_e \equiv \X_{{\bm e}_+}(\sigma) - \X_{{\bm e}_-}(\sigma) = \int_{\sigma_{{\bm e}_-}}^{\s_{{\bm e}_+}}\eta.
\ee
The bulk variation of the action inside each strip implies the bulk equations of motion
\be
\pa_\s \S_i =0,\qquad \S_{iA}\equiv \eta_{AB}  \pa_{\tau}{\X}_i^{B}-H_{AB}\pa_{\s}{\X}_i^{B}.
\ee
From these equations and the discontinuity condition we conclude  that 
\be\label{taudisc}
\pa_\tau \Delta_{\bm e} = \S_{{\bm e}_+}-\S_{{\bm e}_-} .
\ee
Let us first assume that there are no interaction points in the developments of the strips. And
in order to not clutter the derivation 
we look at variations that vanish on the initial and final time slices, that is $\delta \X(\tau_i,\s)=\delta \X(\tau_o,\s)=0$. These variations do not enter the derivation of the equations of motion.
We find that the variation consists of bulk and boundary contributions and that the boundary contributions can be written in terms of the discontinuities (\ref{De},\ref{taudisc})
\bea\label{var1}
\delta S &=&  \frac1{4\pi}\sum_{{\bm e}}\int \rd \tau\ 
\left(\delta \Delta_{{\bm e}}^A  \left[ 
2 \eta \S_{{\bm e}_+} 
 -(\eta+  \omega) \pa_{\tau}\X_{{\bm e}_+} (\sigma_{\bm e})
\right]_A   +  \delta X_{{\bm e}_-}^A(\sigma_{\bm e}) [(\eta -\omega) \pa_\tau \Delta_{\bm e}]_A\right). \nonumber
\eea
Demanding that this variation vanishes imposes an additional boundary equation of motion:
\be
(\eta -\omega) \pa_\tau \Delta_{\bm e}=0 ,\qquad 2 \eta \S_{{\bm e}_+} =
 (\eta+  \omega) \pa_{\tau}\X_{{\bm e}_+}.
\ee
The first equation is an integrability condition for the second.

This derivation is valid as long as $\tau$ belongs to a range where there are no interaction points.
Around an interaction point $\tau =\tau_i$,   the discontinuities $\Delta_i$ are not arbitrary. In order to write this condition 
let's introduce for every vertex the quantity
\be
\Delta_{v}^{A}=\sum_{e| t_{e}=v}\Delta_{e}^{A} - \sum_{e|s_{e}=v}\Delta_{e}^{A}.
\ee
The sum is over all the edges that meet at the vertex $v$, $s_{e}$ is the source vertex of the edge $e$ and $t_{e}$ its target.
From the definition of the discontinuities, it can be checked that $\Delta_v=0$, so the total discontinuity is preserved across an interaction vertex. This generalizes the conservation of momenta. This follows from integrating $\eta$ around a small loop that encloses the interaction vertex and demanding that there are no residues.
Note that the discussion of string interactions outlined here is  schematic and needs to be developed further, in particular with the inclusion of the dilaton interaction.

\section{Metastring Quantum Amplitudes}\label{metamp}

In the present paper we are discussing mostly classical aspects of this theory. However, it is instructive to inject here some structural comments about the quantum theory.
The quantum amplitude  associated with $\Sigma$ is a functional that depends on 
$n_{-}$ in-configurations that we will denote $\bm{x}_{i}$ and $n_{+}$ out-configurations 
$\bm{x}_{o}$, where $\bm{x}(\sigma)=\int_{0}^{\s}\rd \s\delta_{\s}^{A} $ for each boundary circle.
It is then defined as
\be
A_{\Sigma}(\bm{x}_{i},\bm{x}_{o}) \equiv \sum_{\Delta_{i,o}/2\pi \in \Lambda }^{\sum_{j=i,o} \Delta_j=0} \int_{C_\Sigma} Dm_{\tau}  \int_{C^1_\Lambda(\Sigma,\Delta_{i,o})} [D\delta]  e^{i S_{\Delta}(\delta)}  .
\ee
Here the sum is over all the external monodromies $\Delta_{i,o}$ in $\Lambda$.
The first integral is over the moduli space of causal structures which is the Lorentzian analog of the moduli space of complex structures, as follows from the GWKN theorem. The last integration is over all closed one-forms with prescribed external holonomies and all internal holonomies in $\Lambda$.
This prescription is our generalization of the  Polyakov prescription for string amplitudes; it defines
what we mean by the quantum metastring theory.
We will study in detail this formal prescription in future publications  \cite{flmq, flmbeta, flmcurved}. 

There are however two interesting things about this amplitude that we wish to emphasize here.
The first is that $A_{\Sigma}(\bm{x}_{i},\bm{x}_{o})$   is necessarily  
a function periodic with respect to translations along the lattice $\Lambda$
\be
A_{\Sigma}(\bm{x}_{i} +\bm{\Delta}_i,\bm{x}_{o}+\bm{\Delta}_o) = A_{\Sigma}(\bm{x}_{i},\bm{x}_{o}), \qquad \forall \bm{\Delta}_i/2\pi ,\bm{\Delta}_o/2\pi  \in \Lambda.
\ee
Indeed, from  its definition we know that  $ \bm{x}_{i}(\sigma + 2\pi) = \bm{x}_{i}(\sigma ) +\Delta_i$;
since the amplitude involves  a sum over all external monodromies, it is invariant under this shift.
The second one is that the form of the lattice $\Lambda$ is restricted by the 
demand of worldsheet diffeomorphism invariance.
As we have already mentioned in the introduction, and as we will review in more detail later, the demand that the coupling of the 2-dimensional  causal structure to the metastring is invariant under worldsheet diffeomorphisms leads to {\it two} constraints, the Hamiltonian and diffeomorphism constraints. As we have seen these can be expressed simply at the classical level as 
\be
H= \delta_{\s}^{A}H_{AB}\delta_{\s}^{B} =0, \qquad D= \delta_{\s}^{A}\eta_{AB}\delta_{\s}^{B} =0.
\ee
It is convenient to use the convention where indices are raised and lowered with the \Pm and the pairing is denoted  $\X\dd \mathbb{Y} = \mathbb{X}^{A}\eta_{AB}\mathbb{Y}^{B}$.
At the quantum level these constraints in the flat background imply that for a state labelled by $(P,N_+,N_-)$
\be\label{HD}
\frac12 P\dd P= N_{-} - {N}_{+},\qquad 
\frac12 P\dd JP =2- N_{+} - {N}_{-},
\ee
Where $N_{\pm}\in \mathbb{N}$ are positive integers  which correspond respectively to the number of left- and right-moving oscillator excitations. We  have introduced the momenta
\be\label{Ppi}
P \equiv \frac{1}{2\pi}J\Delta.
\ee
 We also introduce the projected momenta $ P_{\pm}^{A} \equiv \tfrac12(H\pm \eta )^{AB}P_{B}$, so we can write the sum and difference of the constraints as 
\be\label{metaspectrum}
P_{\pm}^{2} = 2(1- N_{\pm}).
\ee
The metrics $\tfrac12(H\pm \eta )$ are  of signature 
$(-1,1^{25}; 0^{26})$ and $(0^{26};-1,1^{25})$ respectively.
We can see that imposing the Hamiltonian and diffeomorphism constraint conditions forces the lattice $\Lambda$ to be {\it integral} and {\it even} with respect to these two metrics.
Indeed  if 
$P_{1},P_{2} \in \Lambda$ then $P_{1}+P_{2} \in \Lambda$ and
\be
P_{1\pm}\dd P_{2\pm}= \frac12 (P_{1\pm}+ P_{2\pm})^{2}
-\frac12 P_{1\pm}^{2} - \frac12 P_{2\pm}^{2} \in \mathbb{Z}.
\ee
This means that $P=(P_{+},P_{-})$ belongs to the dual lattice $\Lambda^{*}$. 
The usual argument\footnote{Whereas the Nakamura construction discussed previously ensures manifest modular invariance of the moduli space parameterization, there is still a condition on the CFT coming from the modular invariance of the CFT torus partition function.} of modular invariance requires this lattice to be also  self-dual
\cite{Giveon:1994fu, Narain:1985jj}. 
There exists two fundamental and remarkable results in lattice theory 
\cite{conwaysphere}: first,  self-dual integral Lorentzian lattices of signature $1,d+1$ exist only when $ d\equiv 0\,\mathrm{mod}(8)$ and second, when they exist  they are  {\it unique}  modulo Lorentz transformation. 
Therefore the space of momenta of the  metastring is given by the {\it unique}  integral even self-dual Lorentzian lattice
\be\label{lambda}
\Lambda_d = \II_{1,d+1}\times \II_{1,d+1}.
\ee
By the no-ghost theorem \cite{Goddard:1972iy},
it is only when  $d=0,8,16, 24$ that the space of states of the metastring is equipped with a positive scalar product. These are therefore the dimensions restricted by the Virasoro symmetry, modular invariance and unitarity. We can also demand criticality (Weyl invariance), that is $d=24$. In this case then, the (flat) metastring is {\it unique}. Note then that from this point of view, the usual Polyakov string in 26 flat dimensions is obtained by a certain limit that we referred to as classical in footnote 17. 

We therefore conclude that at criticality the metastring spectrum is characterized by the Lorentzian lattice $\Lambda_{24}= \II_{1,25}\times \II_{1,25}$. 
What is important here is that this self-dual lattice 
involves monodromies in all directions, spacelike and timelike. 
The possibility of having compactifications in all directions has already been pursued in the literature \cite{Moore:1993zc,Moore:1993qe, Dixon:1988qd,Lizzi:1997xe,Giveon:1992ij} but never promoted to a fundamental perspective. 
In the context of the metastring, it appears that we indeed have a fundamental setting, but this requires a re-interpretation.  Usually one interprets the lattice to mean that there is a (Lorentzian) toroidal compactification of space-time. At least naively, this interpretation would inevitably lead to problems with causality, etc. Instead, we interpret the lattice $\Lambda_{24}$ to be the unique metastring background, involving monodromies in all directions.\footnote{Again we emphasize that this applies to the flat $\sigma$-model. Curving the construction will be considered elsewhere.} 

Usual string backgrounds are recovered by taking particular limits on the allowed sets of monodromies.
For instance to recover the Polyakov flat backgrounds from the metastring, we first parametrize $ (P_+,P_-) \in \Lambda_{24}$ and take a limit in which $(P_++P_-) >> (P_+ - P_-)$.
In this large quantum number limit the spectrum of $ P \equiv (P_++P_-) $ becomes continuous, while the Fourier transform of the fields becomes independent of $(P_+ - P_-)$. The fields becomes independent of the dual coordinates and we recover a space-time description.
{\it This suggests that what is usually considered a decompactification limit is 
really a {\it classical} limit of the fundamental background $\Lambda_{24}$.}
One of the remarkable features of $\Lambda_{24}$ is that it is {\it universal}, 
it possesses no moduli! The diversity of classical backgrounds appears here as the diversity of classical limits that can be taken from a quantum theory, and moduli appear generically in any of these classical limits.
This point of view will be further developed in \cite{flmq}. This formulation is clearly a generalization of the definition of S-matrix,
the quantum amplitude  reverts to an S-matrix in any of these classical limits\footnote{Note that in a continuum limit of the lattice in which the $\Delta$'s become continuous $P$'s, this condition becomes just a space-time functional, which presumably reduces to the usual S-matrix element.}.
We  will come back to these points in the last section.

\section{Classical Observables and the Stringy Poisson Bracket}

We now present the construction of a complete set of physical observables, that is observables that commute with the constraints.
We also analyze their Poisson brackets and we find that 
these observables form an algebra which is a generalization of the Courant algebra, with additional multi-string terms.
This shows that the observables can also be interpreted as the generators of the background symmetries.
 We discuss under what conditions these additional contributions vanish at the classical and the quantum level. In fact, these conditions of  mutual locality are radically different at the classical and the quantum level. At the classical level they define  space-time as a Lagrangian submanifold.
 At the quantum level this notion is quantized into a fundamental lattice.

Gauge invariant observables are defined to be functionals $O(\X)$ that commute with $ L^{\pm}_{N}$. Gauge transformations are defined to be $\delta_N^\pm \equiv \{ L^\pm_N, \cdot\}$.
We denote $\X^{\pm} =\frac12 (J\pm 1)\X$ and  $ \Delta^\pm = \frac12 (J\pm 1)\Delta$.
From now on we work in the gauge where $\S=0$, and in this gauge the results of Section \ref{sold} show that on-shell we have $\X^\pm = \X^\pm (\tau\pm\s)$.
 The chiral diffeomorphism  acts on $\X^\pm$ as 
 \be\label{LN}
 \delta^{\pm}_{N} \X^{\pm}= N\pa_{\s}\X^{\pm},\qquad 
 \delta^{\pm}_{N} \X^{\mp}= 0,
 \ee
 where $N$ is a periodic function. 
 The first consequence of this transformation is that the monodromies are
  gauge-invariant observables: 
 \be
 \delta_N^\pm \Delta =0.
 \ee
Thus any function of $\Delta$ is an observable. 
 
 The second consequence is that there can be no local observables;  all observables must be integrated. This follows from the exponentiation of the conformal transformations $T_{N,\tilde{N}} \equiv \exp \delta^+_N \exp \delta^-_{\tilde{N}}$ which leads to a finite conformal transformation
 \be
T_{N,\tilde{N}}\left[ \X^+ (\s)\right] = \X^+( F_N(\s))
,\qquad
 T_{N,\tilde{N}}\left[ \X^-(\s)\right] = \X^-( F_{\tilde{N}}(\s)) ,
\ee 
 where $F_N$ is a solution of the so-called Julia equation \cite{aschenbrenner2013julia,Schnabl:2002gg}
 \be
 N(\s)\pa_\s F_N (\s) = N(F_N(\s)).
 \ee
 The integrated observables are functionals $O(\X^+,\X^-)$ that 
  are labelled by the number of strings they are supported onto. 
  The transformations (\ref{LN}) generate the group $\mathrm{Diff}(S^{1})\times \mathrm{Diff}(S^{1})$ acting on such functionals.
It is clear that the simplest  observable invariant under   $\mathrm{Diff}(S^{1})\times \mathrm{Diff}(S^{1})$ is  the integral  on $S^1$
of  the  one-form 
 \be\label{xi}
\xi(\X)\equiv \xi^{+}_{A}(\X^{+}) 
\rd \X^{+A}+ \xi^{-}_{A}(\X^{-}) \rd \X^{-A},
 \ee
 where $\xi^{\pm}_{A}(\X^{\pm}) $ are  functionals which depends only on  $\X^+$ or $\X^-$.
This is  a  one-form on phase space $\cal P$ that we call the {\it stringy gauge field}.
Consider  the integral of the pullback of this form on $S^1$:
 \be
 \bra \xi^{\pm}\ket (\X) \equiv \oint  \rd \s\ \xi_{A}^{\pm}(\X^{\pm})(\s) \pa_{\s}\X^{\pm A},
 \ee
 where $\oint\equiv \frac1{2\pi}\int_0^{2\pi}$ as usual.
 We can easily check that  under the gauge transformations (\ref{LN}), we have 
 \be
 \delta_N^\pm \bra \xi^{\pm}\ket = N_0 
 \left[
 \xi_{A}^{\pm}(\X^{\pm}(0) + \Delta^\pm) - \xi_{A}^{\pm}(\X^{\pm}(0))  \right]\pa_\sigma \X^{\pm A}(0) ,
 \ee
 where $N_0\equiv N(0)=N(2\pi)$ and we have rewritten $\xi_{A}^{\pm}(\X^{\pm})(2\pi) $ as $\xi_{A}^{\pm}(\X^{\pm}(0) + \Delta^\pm)$.
Therefore, in order to have  invariance under all gauge transformations, and not just the ones for which $N_0=0= N_{2\pi}$, we have to impose in addition  the periodicity condition
 \be\label{per}
 \xi(\X +\Delta) = \xi(\X),\qquad \Delta \in \Lambda,
 \ee
 where $\Lambda$ denotes the lattice of admissible monodromies.
 We also see that the gauge transformation 
 \be
 \xi \mapsto \xi +\rd \Phi,
 \ee
 where $\Phi$ is a periodic function,
 leaves the string observable $\bra \xi^{\pm}\ket$ unchanged.
 The set of  observables that are supported on one string is therefore in one-to-one correspondence with the set of  periodic abelian gauge fields on $\P$, modulo gauge transformations.
 When $\xi_A$ is a constant, the integral simply becomes 
 $\bra \xi^{\pm}\ket= \xi_A \Delta^{\pm A}/2\pi$, and we recover that $\Delta$ is an observable.
 
 It is convenient to expand the gauge fields in Fourier modes so that the observables 
 $\bra \xi^\pm\ket$ can be written  $ \bra \xi^\pm\ket =\sum_{P\in \Lambda^*} \xi^\pm_A (P) V_P^{\pm A} $ in terms of vertex operators 
 \be
 V_{P_\pm}^{A}(\X) \equiv \oint\rd\s\  e^{  i P_\pm\cdot \X^\pm} \pa_\s \X^{\pm A} .
 \ee
The periodicity condition (\ref{per}) means that $   P_\pm\cdot \Delta^\pm \in 2\pi \mathbb{Z}$, 
which implies  that 
 \be
 P^A(H\pm\eta)_{AB} P_{\Delta}^B \in  \mathbb{Z},
  \ee
 where $P_\Delta =\frac{1}{2\pi}J\Delta$  is associated with the monodromy introduced in (\ref{Ppi}).
 Although we are at the classical level we see that these conditions resemble the ones seen  in Section \ref{metamp}, that followed   from the implementation, at the quantum level, of the Hamiltonian and diffeomorphism constraints. In fact they imply that  $P\in \Lambda^*$, i.e., $P$ belongs to the dual lattice. If one imposes that the lattice is self-dual,  these integrality conditions  imply that  we can choose the momentum $P$ appearing in the expression of the vertex operators to also be in $\Lambda$.
 
The operators that we have just constructed generate the algebra of classical observables. There are two operations we can perform in order to 
obtain multi-loop observables: we can either take products of single loop observables or we can take Poisson commutators of them.
At the quantum level, these two operations are fused since the commutator is obtained by the difference of two products, but at the classical level they are not.

Let's consider first the  product of two gauge field vertex operators.
 Depending on whether we multiply two gauge field operator of the same or opposite chirality we get three different types of higher order vertex operators. For instance, we can consider the non-chiral observable
\be
V^{AB}_{P}(\X) \equiv V_{P_+}^{A}V_{P_-}^{B}(\X).
\ee
This can be written in a suggestive manner as 
\be
V^{AB}_{P}(\X) \equiv \frac14\oint\oint  e^{i P\cdot H \cdot \X( \tau,\s)} e^{i P\cdot \eta \cdot \X( \tau,\s)}
[(\pa_\tau+\pa_\s)\X^{A}] [(\pa_\tau-\pa_\s)\X^{B}] \rd \s_+ \rd \s_-,
\ee
where $\s_\pm=\s\pm \tau$.

\subsection{Poisson bracket of observables }

It turns out that unlike what happens in  the Polyakov string, the set of integral observables just constructed is {\it not} closed under the canonical Poisson bracket. This is a new feature of the metastring: as we will see, the bracket of two single loop observables generates multi-loop observables that cannot be simply written as a product of single loop observables.
The main reason for this novelty is that, unlike the Polyakov string, the coordinates of the metastring at equal time do not commute.
The bracket between two coordinates is given by 
\be
\{\X^A(\s),\X^B(\s')\}=  \eta^{AB} \theta(\s,\s')
\ee
where $\theta(\s,\s')$  is the staircase distribution.
It is antisymmetric and quasi-periodic $ \theta(\s +2\pi ,\s') =\theta(\s,\s') +2\pi$.
Given this Poisson bracket we can construct   higher order observables of the form
$\{ \bra \xi^\pm\ket, \bra \xi'^\pm \ket \}$, while the other brackets $\{ \bra \xi^+\ket, \bra \xi'^- \ket \}$ vanish.
Let us first recall that this bracket is expected to also be a gauge invariant observable due to the validity of the Jacobi identity. It follows from:
\be\label{LA}
\{L_N^\pm,\{\bra \xi^\pm\ket, \bra \xi'^\pm\ket \}\} =  \{\{L_N^\pm,\bra \xi^\pm\ket \}, \bra \xi'^\pm\ket \}
+\{\bra \xi^\pm\ket,  \{L_N, \bra \xi'^\pm\ket \} \} .
\ee
The RHS is expected to vanish since we have that $\{L_N^\pm,\bra \xi^\pm\ket \}=0$ and we therefore expect $\{\bra \xi^\pm\ket, \bra \xi'^\pm\ket \}$ to be an observable. We'll see shortly  that there is a very interesting flaw in this argument.

Before doing so, let us generalize the set of observables we consider to a larger set: the set of  diffeomorphism invariant observables, that commute with the diffeomorphism constraints $D=L^+-L^-$ but not necessarily with the Hamiltonian constraints $H=L^++L^-$.
These observables are encoded into a general  gauge field 
 $\xi(\s) =\xi_{A}\pa_{\s}\X^{A}= \xi_{A}^{+}(\X)\pa_{\s}\X^{+A} + \xi_{A}^{-}(\X)\pa_{\s}\X^{-A} $, which does not necessarily satisfy the chirality constraints $\pa_A^- \xi_B^+= \pa_A^+ \xi_B^-=0$. That is, $\xi_A^+$ is a function of both $\X^+$ and $\X^-$ in general.
 We also introduce the shorthand notation  $\bra \xi\ket =\oint \xi(\s) \rd\s$.
Given two observables $\bra\xi\ket,\bra\xi'\ket$, their bracket is:
\be \label{stringpb}
\{\bra \xi\ket,\bra \xi'\ket \} =  \bra [\xi,\xi']\ket -  \bra \xi_{A}\pa_{\s}\xi'^{A} \ket +
\bra\!\bra \pa^{A}{\xi} \circ \pa_{A}{\xi}' \ket\!\ket. 
\ee
We refer to this as the {\it stringy Poisson bracket}, because it does not preserve the number of loops. It is written as a sum of three terms that we now analyze.

The first term in the stringy Poisson bracket  is essentially the Lie bracket:
\be
\bra [\xi,\xi'] \ket \equiv \bra [\xi,\xi']_{A}\pa_{\s}\X^{A} \ket.
\ee
Given a one-form $\xi_{A}$ we can use the \Pm to convert it to a 
vector field $\xi^{A}=\eta^{AB}\xi_{B}$.
The bracket $[\xi,\xi']_{A}$  is then the Lie bracket on vector fields once we use this duality:
\be
[\xi,\xi']_{C}\eta^{CB} = \xi^{A}\pa_{A} \xi^{'B} - \xi'^{A}\pa_{A}\xi^{B}.
\ee
This term shows that the local observable $\bra \xi \ket$ generates phase space 
diffeomorphisms on the space of zero modes. More precisely $\xi^{\pm}$ generates diffeomorphisms 
along the chiral subspace of $T^*{\cal P}$.

The  second term in the stringy Poisson bracket is reminiscent of a central extension,
\be
\bra \xi_{A}\pa_{\s}\xi'^{A} \ket = \oint \rd\s\ \eta_{AB} \xi^{A}(\s) \pa_{\s} \xi'^{B}(\s).
\ee
The antisymmetry follows after integration by parts.
It appears naturally in the construction of a central extension of the loop diffeomorphism group $\xi(\s)  \to (\xi(\s),c)$ with centrally extended bracket
\be
[(\xi(\s),c), (\xi'(\s),c')] = ([\xi,\xi'](\s), \bra \xi_{A}\pa_{\s}\xi'^{A} \ket ) .
\ee

The last term is another stringy correction to the bracket of diffeomorphisms that introduces 
multi-string observables explicitly given by 
\be
\bra\!\bra \pa^{A}{\xi} \circ \pa_{A}{\xi}' \ket\!\ket
=\oint\oint \rd\s\rd\s'\ \eta^{AB} [\pa_{A} {\xi}(\s) \theta(\s-\s')\pa_{B} {\xi}'(\s')] ,
\ee
where we use the shorthand notation $\pa_{A} \xi(\s) \equiv  \pa_{A}\xi_{B}(\X) \pa_{\s}\X^{B}$.
This term is specific to the metastring,
and this  is one of its remarkable features. A possible interpretation for this is that the first quantization of the metastring
 already contains the composition of loops, and thus in some sense is automatically second quantized! It is a fascinating question to wonder to what extent the appearance of this term is related to the introduction of the string coupling $g_{str}$. In this work we will focus on the consequences of this extra contribution.

Let us first remark that this term can be made to vanish by demanding that 
 the functions $\xi^{\pm}$ depend only on a subset $L$ of phase space $\P$ which is null with respect to  $\eta$. That is $L\subset \P$ is such that 
 if $V,W$ are two vectors in the tangent space  $TL \subset T\P$ then $\eta(V,W)=0$. Because of this property $L$ can be thought of as a Lagrangian submanifold\footnote{ Here we are abusing language since $TL$ is an integrable null subspace of $T\P$  with respect to $\eta$ of maximal dimension, while a Lagrangian submanifold is a null subspace with respect to a symplectic form $\omega$. In the case that $\eta+\omega$ is of maximal dimension as we discussed we can choose $L$ to be null with respect to both. Keeping this caveat in mind  we will refer to L as a Lagrangian in the following.}.
 A typical example is when $\xi$ depends only on $X^{\mu}$ of $\X=(X^{\mu},Y_{\mu})$.
 Denoting $\pa_{\mu}\equiv  \pa_{X^{\mu}}$ and $\tilde{\pa}^{\mu}\equiv \pa_{Y_{\mu}}$ we have that 
 \be\label{proj}
\pa^{A} {\xi}  \pa_{A} {\xi}'  = \pa_{\mu} \xi  \tilde{\pa}^{\mu}\xi'  +
  \tilde{\pa}^{\mu} \xi \pa_{\mu} \xi'  ,
 \ee
 which clearly vanishes when no fields depend on $Y$. This case amounts to restricting the  
 section of $T\P$  to be a section of $ \mathbb{T} L \equiv TL \oplus T^{*}L$ where $L$ represents the space-time manifold. The key point here is that "space-time" is viewed as a Lagrangian. 
 In the following, fields that satisfy the condition (\ref{proj}) are said to be {\it projectable} along $L$ or $L$-projectable.
 $L$-projectable fields are  given by products of functions on $L$ and sections of  $\mathbb{T} L$.
 
In fact, demanding that this sort of non-locality always vanishes is essentially equivalent to demanding the usual space-time description
of canonical string theory.
 We therefore witness that once we restrict the observables to be projectable, hence purely local with respect to a 
fixed space-time, the blurring between first- and second-quantization disappears.

It is important to note that the projectability condition is a mathematical implementation of what we referred to as {\it absolute locality} in the introduction. That is, there is a preferred space-time $L$, here viewed as a Lagrangian manifold,  the common support of all fields.  
In that respect it is interesting to note that the condition of absolute locality follows from  the demand of worldsheet  locality of observables.
The deep relationship between the worldsheet notion of locality and the space-time notion of absolute locality is one of the key and deeply surprising features  of string theory.
It is important to appreciate that this connection between worldsheet locality and target space absolute locality, has been established only at the classical level.
 It is therefore natural to wonder what generalization of locality the quantum metastring provides. This  will be the subject of the following sections.

\subsection{C-Bracket, associativity and generalized diffeomorphism}

In the  previous discussion we have identified the observables as being generated by a projectable one-form $\bra \xi \ket$ where $\xi$  is the chiral one-form defined in (\ref{xi}).
What we emphasize here is the fact that this collection of observables can be understood as generating the background gauge symmetries and that these symmetries contain diffeomorphism symmetry.
Moreover we show that the Poisson bracket of these observables is identified to be the 
{\it C-bracket} introduced in the physics literature by Siegel \cite{Siegel:1993th,Siegel:1993xq}.

Let us first consider
the Poisson bracket of $\bra \xi\ket$ with a projectable tensor $H_{AB}\pa_{\s}\X^{A} \pa_{\s}\X^{B}$. This computation defines a transformation of the projectable field $H_{AB}\to  \mathbb{L}_{\xi} H_{AB}$ which is interpretable as a  generalization of the diffeomorphism transformation:
\be
\{\bra \xi\ket, H_{AB}\pa_{\s}\X^{a} \pa_{\s}\X^{B}(\s) \} = \left( \mathbb{L}_{\xi} H_{AB}\right)
\pa_{\s}\X^{a} \pa_{\s}\X^{B}(\s) ,
\ee
where $ \mathbb{L}_{\xi} $ denotes a generalization of the Lie derivative \cite{Hohm:2010pp} associated with
the projectable form $\xi \in \mathbb{T}L$. It is given by 
\be
\mathbb{L}_{\xi} H_{AB} \equiv \xi^{C}\pa_{C}H_{AB} + F_{A}{}^{C} H_{CB} +   F_{B}{}^{C}H_{AC},
\ee
where  $F_{AB} \equiv \pa_{A}\xi_{B} - \pa_{B}\xi_{A}$ is the curvature of the stringy gauge  field $\xi$.
This transformation has been shown to be a symmetry of the effective action of string theory \cite{Siegel:1993th,Siegel:1993xq,Siegel:1993bj,Hohm:2010pp,Hull:2009zb,Hohm:2010jy,Hull:2009mi,Zwiebach:2011rg,Aldazabal:2013sca,Berman:2013eva} when the background fields are all projectable.  Here this transformation naturally arises as a canonical transformation associated with the 
simplest gauge invariant observables: the integral of the stringy gauge field.

The Jacobi identity for the Poisson structure implies that when applied to projectable fields the commutator of the generalized Lie transform is itself a generalized Lie transformation:
\be
[\mathbb{L}_{\xi},\mathbb{L}_{\xi'}] \Phi = \mathbb{L}_{\dbl \xi,\xi'\dbr}\Phi,
\ee
when $\Phi$ is projectable.
Here we have defined a bracket $ \dbl \xi,\xi'\dbr$ acting on pairs of projectable gauge fields. This bracket is defined to be 
such that 
\be
\{\bra \xi\ket,\bra \xi'\ket \} =  \bra \dbl \xi, \xi'\dbr_{B} \pa_{\s}\X^{B}\ket \equiv  \bra \dbl \xi, \xi'\dbr\ket,
\ee
for projectable $\xi,\xi'$.
Note that it may sound awkward to associate a Lie derivative to form fields. But this is not surprising since 
the \Pm $\eta$ provides an isomorphism  between forms and vector fields, given by $A^{A} =\eta^{AB}A_{B}$, so the generalized Lie derivative and bracket  can also be viewed as being labelled or acting  on vector fields.
By evaluating the commutator explicitly we find that the bracket is given by:
\be
\dbl \xi, \xi'\dbr_{B} \equiv  \xi^{A}\pa_{A} \xi{'}_{B} - \xi'^{A}\pa_{A}\xi_{B} -\tfrac12 \left(\xi^{A}\pa_{B} \xi{'}_{A} - \xi'^{A}\pa_{B}\xi_{A}\right).
\ee
We recognize here the {\it C-bracket} introduced in the physics literature by Siegel \cite{Siegel:1993th,Siegel:1993xq} and further developed by Hull and Zwiebach \cite{Hull:2009zb} to describe the symmetry algebra of the effective geometry of strings (see also
\cite{Losev:2005pu, Halmagyi:2009te, Halmagyi:2008dr, Schulgin:2014gra}). This bracket is a generalization of the {\it Courant bracket} first introduced by Courant \cite{courant1990dirac} and further developed by Weinstein, et al \cite{Liu:1995lsa} to 
unify the pre-symplectic and Poisson geometries (see \cite{Bursztyn:2011kk} for a review).
What is remarkable here is that this elaborate structure appears simply as the Poisson bracket of classical observables.

It is obvious that the C-bracket differs from the Lie bracket,
which appears in the first two terms. What is less obvious is that this bracket  
does not satisfy the Jacobi identity if we do not restrict the fields to be projectable.
The reason for this can be understood as follows:
we have seen that the stringy Poisson bracket of two observables decomposes into the sum of 
single string terms plus a double string term
\be
\{\bra \xi\ket,\bra \xi'\ket \}= \{\bra \xi\ket,\bra \xi'\ket \}_{1}+ \tfrac12 \bra\!\bra \pa^{A}{\xi} \circ \pa_{A}{\xi}' \ket\!\ket.
\ee
The single string contribution is  given by the integral of the C-bracket:
\be
\{\bra \xi\ket,\bra \xi'\ket \}_{1} =  \bra \dbl \xi, \xi'\dbr_{B} \pa_{\s}\X^{B}\ket \equiv  \bra \dbl \xi, \xi'\dbr\ket,
\ee
while the double string contribution vanishes iff the fields are projectable.
From this point of view the C-bracket arises as a truncation of a bracket that satisfies the Jacobi identity
\be
J(\xi,\xi',\xi'')\equiv \{\{ \bra \xi\ket , \bra \xi'\ket \}, \bra \xi''\ket  \} +\mathrm{cycl.} = 0 ,
\ee
where cycl. denotes cyclic permutations of $ (\xi, \xi', \xi'')$.
The LHS of this expression decomposes into a sum of integrals supported on one, two or three strings,
$J=J_{1}+J_{2}+J_{3}$, and each contribution vanishes separately after summing over cyclic permutations.
The vanishing of the  contribution supported on a single string implies that  
\be
\bra \dbl \dbl \xi, \xi'\dbr, \xi'' \dbr\ket + \bra [\xi\dd \pa_{A} \xi'-\xi\dd \pa_{A} \xi'] \pa^{A}\xi'' \ket + \mathrm{cycl.}=0.
\ee
We therefore see explicitly that the integral of the C-bracket does not satisfy the Jacobi identity unless the fields are projectable.
When they are projectable we can check that the C-bracket satisfies the following condition:
\be
\dbl \dbl \xi, \xi'\dbr, \xi'' \dbr_{B} +\mathrm{cycl.} = \tfrac16 \pa_{B}\left( \dbl \xi, \xi'\dbr \dd \xi'' +\mathrm{cycl.}\right) .
\ee
The violation of the Jacobi identity of the C-bracket for projectable fields, is a total derivative which disappears when integrated, since $\bra \pa_{B}\Phi \pa_{\s}\X^{B}\ket =0$.

Here we have established at the classical level that  gauge invariant observables are  canonical generators 
for the  background symmetries when the fields are projectable.
This fundamental relation between single loop observables and background symmetry established here at the classical level is expected to also be true at the quantum level \cite{Evans:1989cs, Evans:1989xq, Evans:1991qf, Moore:1993qe, Schulgin:2014gra}.

\subsection{Classical Anomaly}

Before embarking into an analysis of the quantum case,
 we would like to investigate whether or not the projectability of the background fields, which is assumed in the usual Polyakov framework (and consequently double field theory), is a necessity.
In order to get a deeper understanding of this question, let's consider the multi-string contribution to the Poisson bracket. It is given by 
\be
\bra\!\bra \pa^{A}{\xi} \circ \pa_{A}{\xi}' \ket\!\ket = 
\oint_{S_{1}} \rd \s\oint_{S_{1}} \rd {\s'}\ \xi(\s,\s')\theta(\s,\s') ,
\ee 
where we have introduced 
\bea
\xi(\s,\s') \equiv \tfrac12 \left(\pa^A \xi(\s) \pa_A\xi'(\s') -\pa^A \xi(\s') \pa_A\xi'(\s) \right) ,
\eea
which is skew-symmetric in $(\s,\s')$, with $\pa_A \xi(\s) \equiv \pa_A\xi^{+}_B  \pa_{\s}\X^{+ B } +
\pa_A\xi^{-}_B  \pa_{\s}\X^{- B }$. As explained in (\ref{LA}) we expect 
this bracket to commute with the Virasoro constraints since it appears in the bracket of two observables.
But in fact it does not! This can be seen by evaluating  the variation 
\be\label{dN2}
\delta_N^\pm \bra\!\bra \pa^{A}{\xi} \circ \pa_{A}{\xi}' \ket\!\ket  
= N_0 \oint_{S_{1}}   \left(\xi(0,\s)- \xi(\s,0)\right) \rd \s = 2N_0 \oint_{S_{1}}  \xi (0,\s) \rd \s ,
\ee
which does not vanish in general. This contribution is due to the non-periodicity of the staircase distribution.
This implies that in general the Poisson bracket of two observables is not an observable.
How can this be possible?

This follows from the fact that in order  for $\oint \xi(\s) \rd \s$ to be an observable 
we need to  impose in addition the periodicity condition
\be
\xi_A(\X+\Delta) = \xi_A(\X),
\ee 
since $\Delta$ is a dynamical variable that possesses a non-trivial Poisson bracket with $\X$. The periodicity condition is not preserved by the Poisson bracket.
In other words, even if $\Phi$ is a periodic function, its bracket $\{\Phi,\X\}$ is not. For instance, 
suppose that $\Phi(\X)(2\pi) =\Phi(\X)(0)$; we still have that  the commutator 
\be
\{ \left(\Phi(\X)(2\pi) -\Phi(\X)(0)\right), \X^A(\s) \} =  \pa^A \Phi(\X)(0) \left[\theta(2\pi,\s)-\theta(0,\s) \right] 
= 2\pi \pa^A \Phi(\X)(0) \neq 0,
\ee
doesn't vanish.
Since the bracket doesn't preserve the periodicity conditions, and these are crucial in order to imply that 
 the integrals are Virasoro observables, we have that the bracket of observables is not gauge invariant in general.
 
If we demand for consistency that the bracket of two observables is also gauge invariant we are 
driven towards imposing the projectability constraints again. 
More precisely we need that 
\be
\pa_A\xi(\s) \oint \pa^A\xi'(\s') \rd \s' = \pa_A\xi'(\s) \oint \pa^A\xi(\s') \rd \s',
\ee
which is satisfied when the fields are projectable.
We therefore see that projectability is necessary in order to insure the consistency of the 
{\it classical} metastring. We will show next that this is not the case at the quantum level --- the classical anomaly that we just witnessed is not present at the quantum level. This is one of the miracles of the quantum metastring.

\subsection{Quantum Mutual Locality}

We have just observed that at the classical level the presence of the non-local contribution
to the Poisson bracket is also responsible for the breaking of periodicity and creates an anomaly,
in which brackets of observables are no longer observables. The way to remedy this at the classical level is to 
restrict the fields to be projectable. This ensures that different fields are mutually local and then gauge invariant.
Remarkably this is {\it not} necessary at the quantum level. The quantum theory takes care of itself!
It turns out that the periodicity condition that ensures Virasoro invariance is also the condition necessary to ensure the mutual locality of operators.
Although in this paper, we have not presented all of the details of the quantization of the metastring theory, we will provide here enough of the quantum theory 
to understand the restoration of gauge invariance and mutual locality.
We work in the Heisenberg representation where the commutator of position operators are given by
\be
\Big[\X^A(\s), \X^B(\s') \Big] = \tfrac{1}{ i} \eta^{AB} \theta(\s-\s').
\ee
One of the main ingredients of the quantum theory is the normal-ordering operation, which we can describe simply in the flat Tseytlin model.
It is an operation that removes  singularities in the products of operators.
In order  to define this we introduce the positive and negative frequency distributions
\be 
\theta_+(\s,\s') \equiv {\s} +i \sum_{n=1}^\infty \frac{e^{-in(\s-\s')}}{n},
\qquad 
\theta_-(\s,\s') \equiv \theta_+(\s',\s),
\ee
which satisfies the key identity
\be
e^{i \theta_+(\s,\s') }=  ( e^{i\s}- e^{i\s'}).
\ee
We introduce 
\be
\Theta_\pm^{AB}(\s,\s') \equiv \frac12 {(H+\eta)^{AB}} \theta_\pm(\s,\s') 
+ \frac12 {(H-\eta)^{AB}} \theta_\mp(\s,\s'),
\ee
which satisfy
\be
\Theta_+^{AB}(\s,\s')-\Theta_-^{AB}(\s,\s') = \eta^{AB} \theta(\s-\s').
\ee
General operators in the Heisenberg representation are represented in terms of functionals of the form 
$O(\s)=O(\X,\pa_\s\X,\pa_\s^2\X,\cdots)$.
Here we restrict to operators that depends only on $\X$ and $\pa_\s\X$, that is operators 
$O(\s) =O(\X(\s),\pa_\s\X(\s))$. For such functionals we denote $\pa_A \equiv \tfrac{ \pa}{\pa \X^A}$ and 
$ \delta_A\equiv \tfrac{ \pa}{\pa \mathbb{P}^A} $ with $\mathbb{P}^A =\pa_\s \X^A$.
We introduce the bi-local differential operator 
\be\nonumber
\Delta(\s,\s')\equiv 
\overleftarrow{\pa}_A \Theta_+^{AB}(\s,\s') \overrightarrow{\pa}_B + 
\overleftarrow{\delta}_A \pa_\s\Theta_+^{AB}(\s,\s') \overrightarrow{\pa}_B +
\overleftarrow{\pa}_A \pa_{\s'}\Theta_+^{AB}(\s,\s') \overrightarrow{\delta}_B +
\overleftarrow{\delta}_A \pa_\s\pa_{\s'}\Theta_+^{AB}(\s,\s')\overrightarrow{\delta}_B.
\ee
The normal-ordered product is a {\it commutative} product related to the operator product via the 
expression:\footnote{We could conversely write that 
\be
V(\s) V'(\s')  \equiv:\!V(\s)e^{ \tfrac{1}{i} \Delta(\s,\s') } V'(\s')\!:
\ee It is interesting to note that  this  resembles the relationship between Moyal star product and commutative product that appears in star quantization. This analogy is of course a deep one and is not just accidental. }
\be
:\!V(\s) V'(\s')\!:  \equiv   V(\s)e^{ {i} \Delta(\s,\s') } V'(\s'),\ee 
and the commutativity can be checked directly.

In order to illustrate our main point on mutual locality 
we first focus on scalar vertex operators:
\be
V_{P} (\s) \equiv :\!e^{ i P\cdot \X(\s)}\!:
\ee 
whose product is given by
\be\label{prodz}
V_{P} (\s) V_{Q} (\s')  = ( e^{i\s}- e^{i\s'})^{P_+\cdot Q_+}
( e^{i\s'}- e^{i\s})^{P_-\cdot Q_-} :\!V_{P} (\s)V_{Q} (\s')\!: .
\ee
This expression is valid when $\s\neq\s'$ and it can also be written as 
\be
V_{P} (\s) V_{Q} (\s')  = e^{iP_+\cdot Q_+ \theta_+(\s,\s')}
e^{iP_-\cdot Q_- \theta_-(\s,\s')} :\!V_{P} (\s)V_{Q} (\s')\!: .
\ee
Therefore we have 
\be
  V_{P} (\s) V_{Q} (\s') =  e^{{i}P\cdot Q\ \theta(\s-\s') } \,V_{Q} (\s')V_{P} (\s),\qquad \s\neq \s',
\ee 
where we used that $P\cdot Q = P_+\cdot Q_+- P_-\cdot Q_- $.
If both $P$ and $Q$ belong to $\Lambda$  the condition $P \cdot Q \in \mathbb{Z}$ is satisfied.
Since $\theta(\s) \in \pi \mathbb{Z}$ we see that the two vertex operators 
commute or anti-commute.
They commute if  $P\cdot Q $ is even while they {\it anti-commute} if 
$P\cdot Q$ is odd\footnote{In order to construct fully commuting vertex operators, it is necessary to multiply the naive vertex operators by cocycle factors \cite{Polchinski:1998rq, Hellerman:2006tx}. We will see \cite{flmq} that these cocycles have a natural interpretation in terms of the metastring 2-form $\omega$. We do not develop this further here for simplicity. }.

In general when $P_\pm\cdot Q_\pm$ are not integers,
the expression (\ref{prodz}) contains cuts and is ambiguous. In order to define it, we take 
the extension of this product to  imaginary time. Defining 
$z=e^{-\tau +i\s}$ we get
\be\label{OPE}
V_{P} (i\tau,\s) V_{Q} (i\tau', \s')  = ( z- z')^{P_+\cdot Q_+}
\left( \frac1{\bar{z}'}- \frac1{\bar{z}}\right)^{P_-\cdot Q_-} :\!V_{P} (\s)V_{Q} (\s')\!:
\ee 
which is well defined\footnote{In this case we take 
$ ( z- z') = e^{-\tau+ i\s + \ln(1- {z'}/{z})}$ or equivalently
$(z-z') = e^{i\theta_+(i\tau+\s, i\tau'+\s')}$ and similarly  
$(1/{\bar{z}'}- 1/{\bar{z}})= e^{\tau'+ i\s' + \ln(1- {\bar{z}'}/{\bar{z}})}$
. } when $|z| > |z'|$ or equivalently $\tau < \tau'$.
We see that the OPE  of two scalar vertex operators is {\it local}
provided that $P_\pm\cdot Q_\pm \in \mathbb{Z}$ are integers.
The condition of mutual locality is therefore ensured by the condition that
the momenta belong to the lattice $\Lambda$. 

Let us now consider the case $P_\pm\cdot Q_\pm\in 2\mathbb{Z}$.
From the previous analysis we can conclude that the commutator $[V_{P}(\s), V_{Q} ( \s') ]$ is a distribution which has support on the diagonal $\s=\s'$.
If we desire to compute this distribution one can use the definition of the commutator from the OPE. In the case when the OPE is chiral there is a well-defined prescription given by:
\be
[V(\s), V ( \s') ] \equiv \lim_{\epsilon \to 0} \left(
V_{P}(i\epsilon,\s) V_{Q} (0, \s')  - V_{Q}(i\epsilon,\s') V_{Q} (0, \s)\right).
\ee
If one integrates the first vertex operator this relationship can be written in a more familiar manner \cite{Polchinski:1998rq}  as 
\be
\left[\oint V(w) \rd w, V (z) \right] = \lim_{\epsilon \to 0} \int_{C^\epsilon_z}
 \left(V(w) V (z)\right) \rd w.
\ee
where the integral is over a  circle $C^\epsilon_z$ of size $\epsilon$ entered at $z$.
In the chiral case the integral doesn't depend on $\epsilon$ and the limit is trivial. In the non-chiral case the correspondence is much more subtle.
The naive $\epsilon \to 0$ limit is divergent \cite{Evans:1994sh}; this is due to the appearance of contact terms like $\delta(z)/|z|^2$.
In order to take the $\epsilon \to 0$ limit we therefore need to first renormalize the operators before we can  project consistently the integral onto the $\epsilon^0$ term.
Unfortunately, the theory of non-chiral vertex operator algebras is not as 
developed mathematically as its chiral counterpart, and except for a few scattered heroic attempts \cite{Berenstein:1999jq,Berenstein:1999ip,Ranganathan:1993vj, Cederwall:1995nc, Zeitlin:2007vd, Zeitlin:2006ia, Sen:1990hh} that deal with these issues, the subject is largely untouched and no complete theory is available.
This is in our view one of the stumbling blocks in the way of understanding 
in a deeper manner the space of CFT deformations and of unravelling the full symmetry algebra of the closed string.

 Despite these caveats, we can still easily see from this definition of the commutator, and the form of the OPE,  that 
 two vertex operators commute $[V_P(\s),V_Q(\s')]=0$, for all $(\s,\s')$ provided 
 the momenta  satisfy the spectral  condition $(P,Q)\in S$ with
 \be
 S=\{(P,Q)| P_\pm\cdot Q_\pm \geq 0, \,\,\mathrm{and}\,\,P\cdot Q \in 2\mathbb{Z} \}.
 \ee

{\it We conclude from this analysis that the condition of mutual locality which implies at the classical level that the fields are projectable is replaced at the quantum level by the condition of $\Lambda$-periodicity}, 
 where a function $\Phi(\X)$ is said to be $\Lambda$-periodic with respect to an even self-dual lattice $\Lambda$ if 
$\Phi(\X+ 2\pi P) =\Phi(\X) $ for $P\in \Lambda^*=\Lambda$. 
This suggests that a $\Lambda$-periodic field is a quantum generalization of a field on space-time and that we recover the projectable field as a limit of $\Lambda$-periodic ones.
Let us emphasize that the condition of  $L$-projectability which is implemented in DFT  on the fields \cite{Hull:2009mi} is {\it not} necessary to effectively describe the quantum string; this condition is a classical notion. 
The modification of this condition at the quantum level begs for an interpretation of $\Lambda$-periodicity in terms of a quantum space-time Lagrangian. We give this interpretation in a later section and show that the quantum Lagrangian is a modular space-time.

\subsection{The Quantum Gauge Algebra }
Before doing so we would like to discuss what happens to the gauge invariant operators and their algebra at the quantum level.
We give here only some elements and defer the full discussion to \cite{flmq}.
We have seen that at the classical level, the gauge invariant operators are 
circle integrals of the chiral gauge field $\bra \xi^\pm_A \rd \X^A_\pm\ket$ where $\xi^+$ ( resp. $\xi^-$) is an arbitrary functional of $\X^+$, resp. $\X^-$.
At the quantum level this conclusion is drastically modified: the set of gauge invariant operators is given by circle integrals of an {\it infinite} collection of 
arbitrary spin fields 
\be\label{Qobs}
\bra \xi^\pm_A \rd \X^A_\pm\ket \mapsto 
\left(\bra{\xi^{(0)\pm} } \ket,\quad \bra \xi^{(1)\pm}_A \rd \X^A_\pm \ket,\quad 
\bra\xi^{(2)\pm}_{AB}  \rd \X^A_\pm \rd \X^B_\pm\ket,
\quad 
\bra\xi^{(3)\pm}_{ABC}  \rd \X^A_\pm \rd \X^B_\pm \rd \X^C_\pm\ket,
\quad \cdots \right).
\ee
Here each field $\xi^{(N)\pm}$ is chiral, i-e depends only on $\X^\pm$.
However this infinite collection of fields is  not arbitrary since each one has to satisfy an on-shell condition:
\be
\frac12\Box_\pm \xi^{(N)\pm}= N-1,\qquad \Box_\pm = \frac{1}{2} (H \pm \eta)^{AB}\pa_A\pa_B.
\ee
Here $N$ is the spin of the field $\xi^{(N)}$, i.e., the number of derivatives appearing in its definition.
 These conditions follow from  the computation of  the commutator of $L^\pm(\s)\equiv :\!\pa_\s\X^{\pm A}\pa_\s\X_A^\pm(\s)\!:$ with 
a scalar field $\Phi(\s) =:\!\Phi(\X(\s))\!:$.
The commutator is given by\footnote{
In order to evaluate the ordered products we use the key identities
\be
\left[\theta_+'(\s)\right]^2 =  \frac14 + \frac1{i} \theta_+''(\s), \qquad \theta_+'(\s) -\theta_-'(\s) =2\pi \delta (\s).
\ee} 
\bea
 [L^\pm(\s),  \Phi(\s') ]  &=& \frac{2\pi}{i} \left(2 {\delta(\s,\s')} :\!\pa_\s \X^{\pm A}(\s) \pa_A^\pm \Phi(\s')\!: -  \pa_\s \delta(\s,\s')  \Box_\pm \Phi(\s') \right). \nonumber
\eea
This shows that all the fields in (\ref{Qobs}) are chiral by construction and
of dimension $(1,0)$ or $(0,1)$ and we conclude that their integrals on a circle commute with the Virasoro generators. 
The product of these fields, and hence their commutators, will still commute with the Virasoro generators, and unlike the classical case no anomaly is present, and therefore the commutators of these chiral observables  
form a Lie algebra.
This Lie algebra can be understood as the symmetry algebra of the flat metastring. In particular it can be understood as being part of  the background gauge symmetry \cite{Sen:1990hh}. 
It is the product of two  infinite-dimensional chiral algebras.
Each one of these is a so-called Borcherds algebra \cite{Borcherds1, Borcherds2}. A Borcherds algebra is a generalization of a Kac-Moody algebra whose Cartan matrix has a Lorentzian signature.
Interesting examples are when the root system of such algebras can be identified with an even Lorentzian lattice of dimension   $26$ or an even Lorentzian lattice of dimension $10$.
In our case the symmetry algebra $B$ is the biggest Borcherds algebra  associated  with the self-dual Lorentzian  lattice of dimension 26; it is usually referred to as the monster Lie algebra\footnote{ Also called the fake monster algebra, since the algebra used in the moonshine conjecture is related to the fake monster algebra by orbifolding\cite{Dixon:1988qd}.} \cite{Borcherdsmonster, Frenkel:1988xz}.
The  simple roots of $B$ have been characterized by Conway \cite{Conway} and Borcherds \cite{Borcherdsmonster} in terms of a null vector $\rho$ (the Weyl vector)
\be
\rho\equiv(0,1,2,\cdots,24|70).
\ee
 The simple roots are either of positive norm (also called real simple roots)  or null.  The real simple roots are given by Lorentzian vectors $K \in L $ where 
 \be
 L\equiv \{ K\in \mathrm{I\!I}_{1,25}|\, K^2=2,\, K\cdot\rho =-1 \}
 \ee
 is isomorphic to the Leech lattice\footnote{The Leech lattice $\Lambda_{24}$ is the unique unimodular lattice of dimension 24 which possesses no roots (vectors of norm 2). The isomorphism between the  Leech lattice and $L\subset \mathrm{I\!I}_{1,25}$ can be described
 once we choose a null vector $\tilde{\rho}$ such that $\tilde{\rho}\cdot\!\rho=-1$.
 It is explicitly given by 
 $\lambda \to K_\lambda =\lambda +\tilde{\rho} + (\tfrac{\lambda^2}{2}-1)\rho$
 where $\lambda$ is taken to be orthogonal to $(\rho,\tilde\rho)$.
 This shows that $\mathrm{I\!I}_{1,25} = \mathrm{I\!I}_{1,1} \oplus \Lambda_{24}$. The scalar product of two distinct simple roots is always negative since $K_\lambda\cdot K_\lambda' = 2 -\tfrac12(\lambda-\lambda')^2,$ and $(\lambda-\lambda')^2 \geq 4$ for distinct Leech vectors.
  } \cite{ConSloane}. The null simple roots are of multiplicity $24$, they are labelled by an integer $N$ and given by $ P=N\rho$. 
The Cartan generators of each chiral algebra are simply the translation operators \cite{Frenkel:1988xz}
\be
H_P^\pm = P^\pm\!\cdot\! \pa_\s \X^\pm ,
\ee
where the momenta $P^\pm$ are labeled by an element  of the Leech lattice $L$ and the level $N^\pm$:
\be
P_{K^\pm,N^\pm}^\pm= K^\pm+N^\pm\rho.
\ee
It naturally satisfies the mass shell condition $\tfrac12 P^2_{K,N}= 1-N$.

Since it is chiral and  commutes with $L^\pm$, the double Borcherds algebra is an algebra that fixes the background value of the fields and it contains the duality symmetry transformations\cite{Evans:1995su, Dine:1989vu}. This algebra has no classical analog.
 It is natural to embed  this algebra into a bigger symmetry algebra that acts non-trivially on the background fields and  generalizes  diffeomorphism symmetry.
 At the classical level we have seen that this is achieved by relaxing the chirality conditions on the fields.  This means that we now look for the same collection of fields $\bra \xi\ket  \to \bra \xi^{(N)} \ket$
where the fields are no longer chiral. As in the classical case, they can be used to deform the Virasoro generators
\be
L^{\pm} \to L^{\pm} + \delta L^\pm,\qquad \delta L^\pm = [\bra \xi^{(N)} \ket, L^\pm ].
\ee
This deformation algebra does contain a generalization of diffeomorphism transformation \cite{Evans:1989cs, Evans:1991qf}, but it is in general too unwieldy. Since it is not chiral we do not expect its commutator algebra to be well-defined in general.
We can decide to restrict the study to the non-chiral current algebra generated by operators of  dimensions $(1,0)$  or $(0,1)$.
That is, we impose the mass-shell conditions:
$
\Box_+ \xi^{(N)+}= 2(N_+-1)$ and $ \Box_- \xi^{(N)+}=0$ (and similarly for $\xi^{(N)-}$).
Even if the generators are non-chiral we still have that the commutator
of  two fields  of dimension $(1,0)$ is a field of dimension $(1,0)$, while the commutator of fields of dimension $(1,0)$ with fields of dimension $(0,1)$ involves contact terms, but the algebra can still be expected to close, see e.g. \cite{Ashok:2009xx}.

It is important to note that all the difficulties in defining this string
 symmetry algebra comes from the existence of non-trivial pairing and commutators between the functionals of  $\X$, that is from the terms in the OPE's of the form
 \be
 (z-w)^{\alpha P^+\cdot Q^+} (\bar{z}-\bar{w})^{\alpha P^-\cdot Q^-} ,
 \ee
  where we have introduced a metastring $\alpha$ parameter 
  by rescaling the metrics: $\eta \to \alpha \eta$, $H\to \alpha H$.
  This factor allows us to keep track of the number of phase space derivatives.
It is also important to note, that if one tries  to expand these terms in $\alpha$ we run into trouble, since any such expansion will involve terms of the type 
  $\alpha^n \ln^n(z-w)$ which are non-local on the worldsheet.
  It is only the infinite summation of such terms that allows us to recover 
  worldsheet locality.
  It is therefore {\it inconsistent} to truncate this expansion.
  The string symmetry algebra has to be understood at all levels in $\alpha$ or not at all.
  
  This point seems to have been under-appreciated in the recent literature, where most of the analysis done in the context of DFT \cite{Hohm:2013jaa, Schulgin:2013xya} proposes to  simply "neglect" these non-trivial pairings.
 If we neglect these terms, the algebra generated by $\bra \xi\ket$ corresponds to  a simple deformation of the generalized diffeomorphism algebra we described at the classical level.
 However, there is no rationale that allows us to neglect such terms since they cannot be argued to be negligible in a small $\alpha$ expansion.

In summary, we expect that  at the quantum level,  the symmetry algebra is profoundly modified  by the fact that fields are $\Lambda$-periodic.
Writing the full gauge algebra is a challenging issue that requires the development of a deeper understanding of non-chiral vertex algebras. Such symmetry is very rich since it contains, in particular, the full duality symmetry group of flat backgrounds and also a generalization of diffeomorphism symmetry.

\section{Quantum Lagrangians and Modular Space-time}

We have emphasized that the metastring is not based on a space-time formulation and that space-time is not presumed to be a fixed (i.e. non-dynamical) entity. Here we would like to understand how this intuition manifests itself  and what 
notion of locality  emerges from the metastring.

There are in fact two levels of generalization of space-time involved in the metastring. The first level is purely classical, but it is still a nontrivial step; the second one is purely quantum.
Since the metastring is entirely chiral, its target is naturally interpreted as 
a phase space; its dimension is the double of the usual space-time and it possesses one symplectic form $\omega$ and two metrics $\eta,H$. 
The key point is that from this phase space perspective, space-time appears as a Lagrangian sub-manifold $L \subset \P$. More precisely, $L$ is a submanifold of $\P$ which is null with respect to $\eta$.
Let us recall that from a symplectic perspective the choice of a \Pol metric $\eta$ on a phase space $\P$ amounts to a choice of a bilagrangian structure \cite{Etayo:2004lja, Bilagrangian2}. What is new about this viewpoint is that it allows us to think about the choice of space-time inside phase space as a dynamical question. 

We have seen how this point of view naturally arises when we consider the set of classical diffeomorphism invariant observables.
These observables form a closed algebra if the space of fields is $L$-projectable. This algebra is made of functionals on $\P$  that depend only on a Lagrangian $L$ null with respect to $\eta$.
The condition of $L$-projectability is a target space locality condition 
specifying which submanifold of $\P$ can be understood as a space-time.
What is conceptually remarkable here is that this target space locality condition is selected by demanding worldsheet locality.
This phenomenon is even more drastic at the quantum level and worldsheet locality selects for us what modification of the notion of fields one should consider.
The result is that at the quantum level the notion of $L$-projectability is replaced by the condition of $\Lambda$-periodicity. Before explaining how this generalization is a modification of the notion of fields,
we first want to establish a very important fact: although $L$-projectable fields are obviously different  than $\Lambda$-periodic ones, we have  an {\it isomorphism} between 
$\Lambda$-periodic fields and $L$-projectable fields.

\subsection{Isomorphism  between $\Lambda$-periodic fields and $L$-projectable fields}
To get an intuitive feel for this isomorphism, we consider 
a simple toy model\footnote{This lattice is self-dual but not even. Thus, it does not represent a physically meaningful lattice, and we use it just for illustration. The physical case will be treated elsewhere.}  in which the  lattice is $\Lambda=\mathbb{Z}^d\times\mathbb{Z}^d$. It is then convenient to parameterize
\beq
\X^A =\left( \frac{X^\mu}{\lambda}, \frac{Y_\mu}{\varepsilon} \right)
=\left( \frac{2\pi x^\mu}{R}, R {\tilde{x}_\mu} \right) .
\eeq
These different parameterizations reflect the different points of view one can have on the metastring. The first parameterization in terms of $(X,Y)$ introduces a string length scale and energy scale, whose product is $\hbar$. The periodicity $\X\to\X+2\pi P$ with $P=(n,m)$ then amounts to  
 $X\to X+\lambda n$,  
 $Y\to Y+\epsilon m$. This expresses the fact that the cell is of unit size in the $\X$ variables while it is  fundamentally  Planckian\footnote{We mean that the area of the cell is $\hbar$, even if $\lambda$ and $\epsilon$ are the string length and string energy.} in the $(X,Y)$ picture.  Also, this is the natural metastring parameterization.
 The second parameterization in terms of double coordinates $(x,\tilde{x}) $ which have dimension of length and inverse length introduces an arbitrary length scale, and is the one that appears in the usual string treatment,
 where it is thought of as a string compactification radius\footnote{The periodicity $\X\to\X+2\pi P$ with $P=(n,m)$ then amounts to  
 $x\to x+Rn$,  
 $\tilde{x}\to \tilde{x}+2\pi m/R$.}.
 As we will see this common perspective is deeply misleading.
 At this stage the differences between these parameterizations is merely psychological; they only involve trivial rescalings of the coordinates.
 The difference lies in the limits one takes and what one keeps fixed. For instance, in what we call the ``extensification" limit $R\to \infty$ while keeping $ x$ and $Y$ fixed, the fields are taken to 
 \be
\Phi(x,\tilde{x})=  \Phi\left(x, \frac{Y}{\epsilon R}\right) \to \phi (x).
 \ee
 From the metastring point of view this limit corresponds to a limit where we focus on fields that depend only on $X<<\lambda$.
 We will come back to these different limits presently.

If one chooses the $(x,\tilde{x})$ perspective that is probably more familiar to the reader, the $\Lambda$-periodic fields  $\Phi(x^\mu,\tilde{x}_\nu)$ are doubly periodic functions:  
\be
 \Phi(x^\mu,\tilde{x}_\nu)= \Phi(x^\mu + R n^\mu , \tilde{x}_\nu) =\Phi(x^\mu, \tilde{x}_\nu + 2\pi m_\nu/ R) ,\qquad n^\mu, m_\nu \in \mathbb{Z}.
\ee
In order to describe the isomorphism that we alluded to above, let us define $[x]_R \in [0,R]$ to be 
$x$ modulo $R$:
\be
[x]_R \equiv x \,\mathrm{mod}(R).
\ee
This is the unique element in $[0,R]$ such that 
\be
x= [x]_R + N_x R,\quad N_x \in \mathbb{Z}.
\ee
We also denote $[\bm{x}]_R \equiv ( [x^1]_R,\cdots, [x^d]_R)$ and 
${N}_{\bm x}\equiv ( N_{x^1},\cdots, N_{x^d})$.

Given a $\Lambda$-periodic field $\Phi(\bm{x},\bm{\tilde{x}})$ we define a projectable field $\phi(\bm{x})$ to be given by 
\be
\phi(\bm{x}) \equiv \int_0^{2\pi/R}\rd^d \bm{\tilde{x}} \, e^{-i  R {N}_{\bm x}\cdot \bm{\tilde{x}}} \,\Phi(\bm{x},\bm{\tilde{x}}).
\ee
The projectable field is obtained by a partial Fourier transform on the $\bm{\tilde{x}}$ variables. This map is invertible and its  inverse  is given by 
\be
\Phi(\bm{x},\bm{\tilde{x}}) = R^d \sum_{\bm{n}\in \mathbb{Z}^d} e^{i  R \bm{n} \cdot\bm{\tilde{x}}} \phi([\bm{x}]_R +  R \bm{n}). 
\ee

It is important to emphasize that although we used the familiar notation, the perspective presented here diverges from the usual perspective on compactification. In the usual point of view one assumes that the concept of locality is untouched and one tries to reinterpret the compactified string in terms of compactified fields. This amounts to the truncation of the $\Lambda$-periodic fields into fields that do not depend on $\tilde{x}$: $\Phi(x,\tilde{x}) \to \Phi(x)$, while keeping the periodicity in $x$. Such a truncation projects out degrees of freedom, while the isomorphism just established  shows that the string compactification does not 
project any degrees of freedom. It just rearranges (quite drastically) how one should interpret these degrees of freedom. This is one of the key features of string theory and an important manner in which it fundamentally differs from field theory.

The idea behind the isomorphism $\phi \to \Phi$ is that fluctuations of the projectable field $\phi$ on scales larger than $R$ are encoded in the variable $\bm{y}$, while fluctuations on a scale smaller than $R$ are encoded in the variable $\bm{x}$.  
This shows that, in a precise sense, the notion of $\Lambda$-periodic field is a generalization of the set of projectable fields. 
The $\Lambda$-periodic fields are non-projectable, but  they are isomorphic to projectable fields. This generalization is a consequence of the string quantization and it sharply expresses the relative locality principle behind these constructions. 

Although we have an isomorphism between $L$-projectable fields and $\Lambda$-periodic ones one should not conclude that the two descriptions are interchangeable. 
In order to understand which one defines the proper notion of locality 
one should look at how fields interact. Let us give here only a sketch of the full argument.
It is well known that the interaction vertex of fields is related to the OPE of the corresponding vertex operators \cite{Taylor:2003gn,LeClair:1988sp,LeClair:1988sj}. If one normalizes in the usual manner, the 3-point interaction vertex will have, in momentum space,  the schematic form 
\be
e^{\alpha\sum_i P^2_{i+}}e^{\beta \sum_i P^2_{i-}} \delta^{(2d)}(P_1+P_2+P_3).
\ee 
The exponential terms correspond to a delocalization of the vertex familiar in string interactions. The delta function implies that, once we Fourier transform, the interaction vertex is going to be invariant under translation in
$(x,\tilde{x}) \to (x+a,\tilde{x} +\tilde{a})$.
This shows that the locus of the vertex interaction is given by $([x]_R,[\tilde{x}]_{1/R})$.
The main question lingering now is:  In what sense can we think of this  set 
as quantum space-time?

\subsection{Modular space-time}

In order to answer this question, the first  key 
observation is to take seriously the idea that $\X$ represents phase space points
and understand  the notion of $L$-projectability and $\Lambda$-periodicity from this perspective.
In order to formalize this intuition we define $q\equiv x$ and $p= h \tilde{y}$ and then we promote these variables to operators satisfying the Heisenberg algebra
$[p,q]=i\hbar$.
From this perspective we see that the set of $L$-projectable fields $\phi(q)$ forms a {\it commutative} subalgebra of the Heisenberg algebra.
The condition that $L$ is a Lagrangian translates into the condition that the zero modes of the fields commute.
This is in agreement with the philosophy of non-commutative geometry 
where manifolds are defined in a dual manner in terms of the algebra of functions they generate.  
At the classical level, the algebra is a Poisson algebra and the relevant functions depending only on a Lagrangian submanifold are Poisson-commuting.

If we think along the lines of the Gelfand-Naimark theorem, which states that a commutative star-algebra is the algebra of functions on a compact manifold, we can say that the dual to a Lagrangian submanifold corresponds to a maximal Poisson-commuting subalgebra. This is the algebraic version of the vanishing of the symplectic form on the Lagrangian.
And this is the point of view we are now taking: we define the dual of a quantum Lagrangian to be a maximally commuting subalgebra of the Heisenberg algebra.
According to this definition, a Lagrangian manifold is also a quantum Lagrangian since the set of $\phi(q)$ forms a commutative 
subalgebra.
This is not surprising since the quantization of a commutative Poisson algebra is a commutative algebra. 

The remarkable fact is that at the quantum level  new possibilities become available: we can also consider the algebra generated by \hlt{modular observables}
$([x]_R, [p]_{{h}/{R}})$, where $h=2\pi \hbar$.
The modular variables are defined to be 
\be
[p]_{\tfrac{h}{R}} = p \,\, \mathrm{mod}\left(\frac{h}{R}\right),\qquad 
[q]_R= q \,\, \mathrm{mod}(R).
\ee 
These observables possess two important characteristics. First, the observable $[p]_{{h}/{R}}$ does not survive the classical limit $\hbar \to 0$ --- it has no classical analog.
Second, the set of observables 
\be
\Phi\left([q]_R, [p]_{{h}/{R}}\right), 
\ee
form a commutative subalgebra.  This follows from the fact that such functions can always be expanded in terms of $ e^{ i\tfrac{2\pi x}{ R}}$ and $ e^{i \tfrac{p R}{\hbar}}$. The commutation of two exponential variables is given by
\be
e^{i {\alpha p}} e^{i \beta  q} =  e^{i {\hbar \alpha \beta}} e^{i \beta  q}e^{i {\alpha p}},
\ee
and the phase factor is equal to $1$ for $\alpha = 2\pi/R$ and $\beta = R/\hbar$.
This implies that modular observables $[q]_R$  and 
$[p]_{{h}/{R}}$ {\it commute } with each other.  These modular polarizations are  fundamentally quantum. 
And these are the variables that the metastring is naturally implementing, generalized to Lorentzian geometry.

The central importance of such observables in quantum mechanics has been first recognized by Aharonov. He was the first to ask what type of observables can measure the relative phase of two photons in the two-slit experiment. And 
he realized that no semiclassical observables, constructed as an arbitrary polynomial function of $(p,q)$, can detect this phase! The only observables that can are the \hlt{modular variables} that do not possess classical analogs.
In that respect modular observables capture some of the essence of quantum mechanics. Their physics is described at length in the book of Aharanov and Rohrlich \cite {aharonov2008quantum}, where the emphasis is put on the description of ``quantum effects without classical analogues''.
In particular, one can easily see that, in the Heisenberg picture, the time evolution of such observables is fundamentally non-local, and that this non-locality is subtly hidden from us by the delicate interplay between the uncertainty relation and the compactness of the modular observables.
This dynamical non-locality is the key driving factor behind such phenomena as the Ahranov-Bohm \cite{Aharonov:1959fk} or Aharanov-Casher \cite{Aharonov:1984xb} effects.
 
In the context of quantum mechanics, operators $x$ and $p$ do not commute and thus their eigenvalues cannot be localized within a cell of area $\hbar$. The modular variables $[x]$ and $[p]$, on the other hand, commute with one another, and thus their eigenvalues can be specified precisely. The Heisenberg uncertainty then appears as the statement that if one does specify the eigenvalues of $[x]$ and $[p]$, one cannot know in {\it which} cell the eigenvalues of $x$ and $p$ appear.

Indeed, returning now to the metastring, the modular variables $[\X]$ generate the commutative subalgebra of the non-commutative algebra generated by $\X$, where the non-commutativity is seeded by the two-form $\omega$. In the quantum theory, we say that a cell in ${\cal P}$ coordinatized (continuously) by $[\X]$ is a \hlt{quantum Lagrangian}, or equivalently, a \hlt{modular space-time}.  The (Lorentzian) volume of this cell is fixed to be unity, or if we coordinatize it in terms of $X^\mu,Y_\mu$, the cell has volume $\hbar^{d}$.

\subsection{Causality}

One of the central puzzles one has to face in order to understand the deeper meaning of the modular space-time is the fate of causality.
On one side since the dual space is a  lattice $\Lambda$, and if one takes  the usual interpretation that space-time is a classical Lagrangian, one would conclude that this space-time contains a periodic time direction 
which is clearly unacceptable if we are to interpret it as a causal theory.

The notion of quantum Lagrangian on the other hand gives us another interpretation entirely. In the full quantum theory, the $\Lambda$-periodic fields are functions $\Phi([\X])$. This is an acceptable notion of a field on a space-time because $[\X]$ are commutative. On the other hand, it does not seem to have any obvious relationship with our usual classical notion of space-time. 
{\it This means that the notion of causality has to be extended in a way that generalizes the usual notion.}
To be more precise, in the modular space-time the notion of time $t$ is replaced by the modular pair $t \mapsto ([t]_R,[E]_{h/R})$ where $E$ is the energy.
The compactness of modular observables makes it challenging to imagine a proper generalization of the notion of time-ordering, which is a central notion in field theory linking unitarity and causality \cite{Weinberg:1995mt}.
On the other hand the isomorphism between modular fields and usual fields
allows us to imagine such a generalization. For example, since functions of the modular time pair are  isomorphic to functions of a usual time, we could use this isomorphism to pull back the time-ordering.
Indeed, suppose that  an isomorphism $I: \Phi \to \phi$ 
between modular fields and regular fields is given.
One could define  a modular time-ordering to be
\be
T_I[\Phi \Phi' ] \equiv I^{-1} \left(T [ I(\Phi) I(\Phi') ] \right). 
\ee
It is interesting to note that the concept of modular space-time allows us to resolve one of the key issues that arises in non-commutative field theory. As we have already hinted at, it is plausible at this stage that the effective description of the metastring is naturally formulated in terms of a non-commutative field theory living on a quantum phase space (the $26d$ Heisenberg algebra in our example).
It is however well-known that in usual non-commutative field theory (of theta-like non-commutativity) one cannot have both unitarity and causality \cite{Minwalla:1999px}.
One has to choose between the two. The reason for this is easy to appreciate.
In order to formulate a unitary theory one has to define the interacting amplitudes in terms of the time-ordering of the interacting Hamiltonian, for example
\be\label{Hstar}
H_I(t) =\int \rd^3 x (\phi\star\phi \star \phi)(t,x),
\ee 
where $\star$ is the non-commutative product.
If the non-commutativity affects the time direction, then the time-ordering of the interaction Hamiltonian, $T[H_I(t_1)\cdots H_I(t_n)]$ 
demanded by unitarity, does not imply the time-ordering of the fields within the Hamiltonian which is demanded by causality.
It is therefore impossible to satisfy both unitarity and causality \cite{Bahns:2002vm,Bozkaya:2002at}.
Now the concept of modular space-time suggests a way to avoid this negative conclusion. One can  keep the definition (\ref{Hstar}) of the interaction Hamiltonian in terms of a star-product, but demand that the space of fields entering this non-local product form a commutative subalgebra.  In this way the ordering of the fields and the Hamiltonian is the same.
This shows that the condition of commutativity of the label of fields is a necessary condition for a formulation that reconciles causality and unitarity.

This argument also shows that somehow generalizing the notion of causality goes hand-in-hand with generalizing unitarity.
 It is well-known that for the usual string the expectation values of vertex operators (the natural string observables) are interpretable as the $S$-matrix elements of a unitary theory. And the  individual string modes  correspond to asymptotic particle states, which carry representations of the Poincar\'e group and are solution of $\Box \phi = m^2\phi$ with the mass proportional to the spin. 
 This is one of the magical aspects of string theory at play. Unitarity is not demanded; what is demanded is the string consistency, by which we mean the demand of conformal invariance, modular invariance and mutual locality of vertex operators, and unitarity follows from that. 
 In particular one of the key ingredients for unitarity is the factorization property of the CFT.
 
 Our conjecture is that the metastring consistency naturally allows a generalization of causality and unitarity which is adapted to the modular space-time. 
In the metastring, the vertex operators are generalized and the metastring modes cannot be interpreted as modes of a field in a usual space-time.
We can however interpret them as modes of a field $\Phi(\bm{x},\bm{\tilde{x}})$ on modular space-time. These fields carry a representation of the doubled Poincar\'e group and are solutions of a pair of equations 
\be
\Box\Phi +\tilde{\Box}\Phi =  (m^2 +\tilde{m}^2) \Phi,\qquad 
\pa_\mu \tilde{\pa}^\mu \Phi = m \tilde{m} \Phi , 
\ee
where the masses are proportional to the spins.
In this sense  the metastring vertex operators are in one-to-one correspondence with these modular space-time states.
Now from the string side we can consistently define the  string correlation functions associated to these vertex operators. These amplitudes are then  the natural  generalization of S-matrix elements  for modular space-time.
The usual space-time description emerges when the fields can be assumed to be approximately independent of $\tilde{x}$, that is when $m>\!>\tilde{m}$.
 We expect  the magic of the string to still work for the metastring. 
 That is, we conjecture that such amplitudes  respect a generalization of unitarity and causality  adapted to the modular space-time. 
 This will amount to the statement that modular space-time is a full-fledged generalization of the usual notion of space-time adapted to the string.
  We leave the study of this  challenging and exciting possibility for future work.

\subsection{Extensification limits}

In this subsection we would like to elaborate in what sense the usual notion of space-time can be obtained from the modular space-time as a limit akin to a classical limit.  This limit can be viewed as an {\it extensification} limit in which some dimensions become large. This is the limit $R\to \infty$ of 
$([x]_R,[\tilde{x}]_{2\pi/R})$ in which we recover the usual space-time notion.

Geometrically this limit corresponds to a  `squashing' limit of the modular space-time cell that gives rise to a more familiar picture. For example, if we squash the cell in $d-D$ directions along $[\tilde{x}]$ preserving volume, the cell simultaneously expands along $[x]$. Following this squashing to its limit, we obtain what looks like a non-compact space-time in $d-D$ dimensions, with no remaining extent in $[\tilde{x}]$. The limit then looks like the classical notion of a Lagrangian submanifold, but there is a difference --- we have obtained the limit by preserving symplectic volume. 
In the limit, the extra information about the extent in the direction dual to the space-time directions gets lost. In that sense the `squashing'  limit loses some information about the full theory.
Note that this is exactly the opposite conclusion that is usually drawn.
Usually, it is emphasized that if one compactifies $x\to [x]$ one loses information about the theory since it contains less modes. 
 Here we are seeing that this description, which is inspired by a field theory understanding,  misses one very important point about the string. Any time we try to `compactify' a direction we introduce a delocalization $\emptyset \to [\tilde{x}]$ in the dual direction in a way consistent with the Heisenberg uncertainty principle.
 
 \myfig{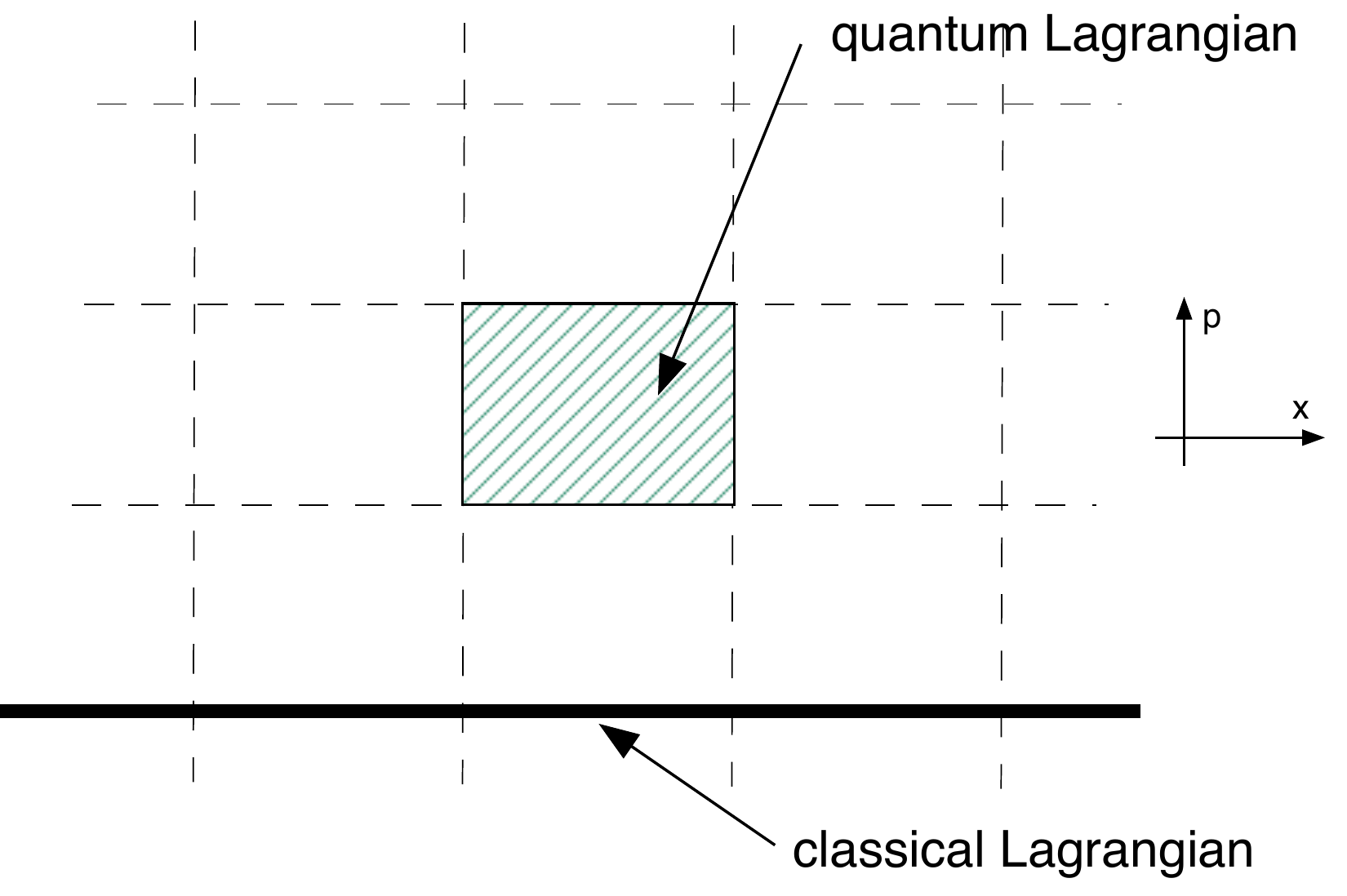}{9}{Quantum and classical Lagrangian submanifolds.}

 Thus, the usual notion of strings propagating in a space-time with $d-D$ non-compact directions and $D$ compact directions is obtained not {\it a priori}, but as a limit of quantum modular space-time.
 If we denote, in momentum space,  by $P$ (resp. $\tilde{P}$) the momenta dual to $x$ (resp. 
 $\tilde{x}$), this limit corresponds to a regime of large quantum numbers in which $P>>\tilde{P}$. This is a classical limit.
 
 The idea that the extensification limit is a classical limit resolves one of the last puzzles we faced.
 As we have seen, the metastring is unique and possesses no moduli: the double Lorentzian lattice determined by $(\eta,H)$ is unique. If one extensifies the metastring along $d-D$ directions,  then one expects to recover the usual description in which there are moduli. Indeed the moduli space of strings compactified on $D$ dimensions  is then
 $\mathrm{O}(D,D;\mathbb{Z})\backslash \mathrm{O}(D,D)/  \mathrm{O}(D)\times \mathrm{O}(D)$.  So where do these moduli appear from and what do they mean then if the fundamental theory has no moduli? The answer is that they appear as a labelling of the ambiguity that exists in taking  the extensification limit. In the same way that there are different classical limits of a quantum theory, there are different extensification limits.
 
 In order to define an extensification limit we first have to decide which directions we extensify and which ones we contract. And 
 in order to do so we choose a pair of maps $(P,\tilde{P}): \P\to \mathbb{R}^{(d-D)}\times \mathbb{R}^{(d-D)}$ which provides an isomorphism between the phase space $\P$ and two copies of $\mathbb{R}^{(d-D)}$. We denote the images by  $x^\mu = R P^\mu{}_A\X^A$,  
 $\tilde{x}_\mu = R^{-1} \tilde{P}_{\mu A}\X^A$. 
  We demand these maps to be such that 
 \be
 P^\mu{}_A P^\nu{}^A =0 = \tilde{P}^\mu{}_A \tilde{P}^\nu{}^A, \qquad 
 P^{\mu A} \tilde{P}_{\nu A} = \delta^\mu_\nu ,
 \ee
 where indices are raised and lowered with $\eta$.
 These conditions imply that the submanifolds $x=\mathrm{const}.$ and $\tilde{x}=\mathrm{const}. $  are two transversal Lagrangian manifolds.
 The maps $(P,\tilde{P})$ encode a  metric and $B$-field via 
 \be
 P^\mu{}_A H^{AB} P^\nu_B = G^{\mu\nu}, \qquad 
 P^\mu{}_A H^{AB} \tilde{P}_{\nu B }= (G^{-1} B) ^{\mu}{}_{\nu}.
 \ee
 These are the moduli that appear only after one has chosen in which way the fundamental cell is  squashed and in which direction
it is extensified. The full extensification is then obtained by taking the limit $R\to \infty$. 
 
  We note that modular space-time has T-duality  built in --- the T-duality operations are just coordinate transformations in ${\cal P}$. This property is shared with double field theory. But our modular space-time point of view differs markedly from double field theory in that it is obtained without truncation from a quantum mechanically consistent string theory. In particular we have seen how the condition of $L$-projectability appears only in the classical description of the string.

\section{Conclusion}

In this paper we have presented certain foundational aspects of metastring theory which involve a dynamical phase space and modular space-time.
We have shown that the background fields  associated with the Tseytlin action all have a  natural geometrical interpretation in term of phase space geometry: they involve a symplectic form $\omega$, a neutral metric $\eta$ that defines a bilagrangian structure and allows to define the classical space-time as a Lagrangian sub-manifold, and a generalized metric $H$ that encodes the geometry along the classical space-time as well as the transverse geometry. Also, in this formulation T-duality exchanges the Lagrangian sub-manifold with its orthogonal complement.

In this new formulation of string theory, the classical notion of space-time (usually seen as a universal fixed structure) is replaced by a Lagrangian subspace embedded in phase space. This allows us
 to think about space-time itself, and not only its geometry, as a dynamical quantity. 
This means that, from a foundational point of view, metastring theory has a built-in notion of relative locality \cite{AmelinoCamelia:2011bm},  
or  {third relativity} \cite{Finkelstein:1984rc}, which simply states
that different observers (or different physical probes) see different space-times.
The metastring goes beyond the naive expression of this idea since the notion of a Lagrangian sub-manifold is identified as a null subspace for a  purely stringy field: the \Pol metric $\eta$. Allowing the notion of space-time to be dynamical goes hand in hand with allowing the $\eta$ metric to become dynamical. In the metastring formulation new CFT deformation modes are allowed, including the winding modes, whose collective excitations correspond to deformations of the $\eta$ background. This links the relaxation of locality with purely stringy excitation modes.

Another important point of this new viewpoint is that the metastring has to be formulated using Lorentzian worldsheets. This follows from the fact that the Tseytlin action describes a chiral formulation and chiral field theories do not admit a canonical Wick-rotated formulation.
This leads to a definition of the metastring amplitude in terms of gluing of strips and potentially opens up the possibility for a new non-perturbative notion of closed string field theory. In this picture string interactions are associated with changes of the strip boundary conditions and the natural diagrams capturing these interactions are stringy generalizations of Feynman diagrams identified as Nakamura graphs. Although the strip description is more reminiscent of an open string picture (as presented in the Appendix), we have shown with great care that the strip description of closed strings is canonically consistent. 
 
 Our main result however is the discovery that quantum strings do not propagate in a usual space-time: the natural arena for the metastring is a modular space-time.  The driving force behind this result is to accept the idea of relative locality, that space-time may not be fundamental but results from sets of relations and interactions of fundamental probes. If one uses relativistic point-like particles as probes, this philosophy lets us rediscover usual space-time as the geometrical realm in which particles propagate and space-time points serve as the loci of particle interactions.
 According to this idea, if one changes the probes from relativistic particles to relativistic strings one should revisit the notion of an effective space-time that emerges from the probe interactions. Conceptually, this is the fundamental insight that has been implemented in this paper. 
 
 The fundamental axioms of string theory are that one should first focus on the  the necessary and sufficient conditions for the 2d world-sheet theory to be defined consistently on all two dimensional surfaces.  
 It is known that these are twofold: crossing symmetry of the four-point functions on the sphere, and  modular invariance of the partition function and the one-point functions on the torus \cite{Moore:1988qv}. Higher genus amplitudes can then be constructed by gluing various punctured spheres and tori together. The above conditions ensure that this procedure gives consistent answers. Therefore   by ``taking string theory seriously'' \footnote{According to Steven Weinberg, ``Our mistake is not that we take our theories too seriously, but that we do not take them seriously enough'' (from his book {\it The First Three Minutes: A Modern View of the Origin of the Universe}).} we  start from the above definition of string theory and then investigate what concept of locality the strings  {\it define}, if any. That is, we define the string geometry  by how probes interact with one another and not the reverse, as is usually assumed.
 This means that we focus on the consistency of string theory: CFT invariance, modular invariance and  world sheet locality, without presupposing an a priori concept of locality.
We found out that these consistency conditions lead, when we perturb around the usual free CFT, to a new notion of target space locality. 
We have used this strategy to let the string define what kind of  space-time the string wants to live in. This space, which emerges from the collection of string mode interactions, we have identified as modular space-time. 
The emergence of a new concept of locality from string consistency has always been  one of the most ``magical'' aspects of string theory. We have witnessed that it is still at play in the metastring formulation and that the usual space-time appears only when we restrict, somewhat artificially, the space of CFT perturbations of the free theory at hand.

Even though many aspects of metastring theory might appear rather novel and unfamiliar, it should be emphasized that many of these new features of the metastring have been discussed and even foreseen in previous literature on string theory in extreme conditions. For instance, both the high energy and high temperature limits of string theory have revealed aspects now found in the metastring.

For example, in the papers of Gross and Mende \cite{Gross:1987ar, Gross:1987kza} on the high energy limit of string theory, one finds evidence for the increase in the effective size of the string as the energy increases. In metastring theory, it is plausible that this physics can be related to dynamical momentum space and Born reciprocity, in which momentum fluctuations are directly related to the spatial fluctuations. This old work also reveals some tantalizing duality between the low
energy and high energy limits of string theory, also naturally  incorporated by Born reciprocity of the metastring.  

Similarly, in the treatment of the high temperature limit of string theory in\cite{Atick:1988si}, one finds evidence for the emergence of a phase space structure in string theory (or more precisely, according to Atick and Witten, for a doubling or complexification of the space-time coordinates in the proper formulation of closed string field theory), which is now made manifest in the metastring. Such complexification of space-time as well as the para-quaternionic structure associated with Born geometry is natural in topological string theory \cite{Witten:1988zd, Neitzke:2004ni}, the topological phase being the conjectured high-energy and high-temperature limit of the canonical string. 
 In previous papers we have emphasized and deepened the geometrical understanding behind this doubling in particular with the notion of Born geometry, a natural unification of complex geometry of quantum theory, symplectic geometry of 
Hamiltonian dynamics and the real geometry of general relativity 
\cite{Gibbons:1991sa, Ashtekar:1997ud, Jejjala:2007rn, Balasubramanian:2002wy, Minic:2003nx,Minic:2003en }.

Concerning our emphasis on Lorentzian worldsheets, let us mention that 
recently Witten has argued  for the importance of having such an understanding and also for a doubling of space-time in his discussion of the $i\epsilon$ in string theory \cite{Witten:2013pra}. 
Of course, the doubling of space-time is a crucial feature of the literature on double field theory
\cite{Siegel:1993th,Siegel:1993xq,Siegel:1993bj,Hohm:2010pp,Hull:2009zb,Hohm:2010jy,Hull:2009mi,Zwiebach:2011rg,Aldazabal:2013sca,Berman:2013eva, Blumenhagen:2014gva}.
However in this literature the fundamental importance of the modularity of the doubling has not been appreciated. Also, we believe that the phase space point of view behind this doubling has been underappreciated as well. We have seen that this doubling is fundamentally quantum and non-perturbative and that it cannot be classically truncated. In our  view the doubling must go hand in hand 
 with the modular space-time interpretation.

There are however important  early  studies of purely stringy phenomena that have shown in a clear manner that the usual concept of geometry
 is profoundly shaken by the quantum string and that certain particular string backgrounds cannot be understood in terms of  usual space-time geometry.
 These are the so-called  ``non-geometric'' backgrounds first introduced by Hellerman and collaborators \cite{Hellerman:2002ax, Hellerman:2006tx} and Hull and collaborators \cite{Hull:2004in, Hull:2006va} under the name of monodrofolds or T-folds.
Similarly, in a series of works focusing on generalizing the notion of T-duality to curved backgrounds the general monodromic aspects of the metastring were intuited previously in works by Klimcik, Severa and collaborators  \cite{Klimcik:1995ux, Klimcik:2000sk}.
We expect the concept of modular space-time to provide a natural geometrical setting for a deeper understanding of these backgrounds.

Let us also note that one of the overarching mysteries of the nature of string theory is that both the high energy and the high temperature limits of string theory point to a drastic reduction of the number of degrees of freedom in some more fundamental formulation of string theory. 
These results rely on the property of T-duality in string theory, which is made covariant in metastring theory. 
This conclusion is also apparent in the early work of
Klebanov and Susskind \cite{Klebanov:1988ba, Karliner:1988hd}. They discovered that the quantum string in light-cone gauge is made of string bits. These string bits appear once we regularize the string and they  have the property 
of being able to construct a continuum world sheet in the limit where the regulator is removed, even though they experience a discrete space-time. In other words, in the limit where the worldsheet distance of two string bits goes to zero their space-time distance goes to a constant: the string length.
This means that there is  a drastic reduction of fundamental short-distance degrees of freedom 
 (see e.g. \cite{Thorn:1991fv}).
This was subsequently explored in the context of the string-black hole transition \cite{Polchinski:1998rq}, in terms of a discretization of space-time. It seems natural to speculate that this could be connected to the concept of modular space-time in metastring theory. Of course, the modern paradigm for such a drastic reduction of degrees of
freedom is given by holography, whether in the context of matrix theory \cite{Taylor:2001vb, Polchinski:1998rr}, or even more precisely, in the context of the AdS/CFT duality. Another persistent notion is a ``stringy uncertainty principle'' 
\cite{Atick:1988si, Amati:1988tn, Amati:1987uf, Amati:1987wq, Witten:2001ib}. 
It seems not too enthusiastic to suggest that the metastring may provide new and interesting interpretations for all of these old problems of string theory.

Let us finally note that one limitation of the point of view developed here is that we have taken here a worldsheet, and hence perturbative, point of view. Most of the developments in string theory over the past twenty years have focused on its remarkable non-perturbative aspects, and many have turned away from the worldsheet formulation in the belief that it cannot be fundamental. Indeed at least in the presence of target space supersymmetry, we know well that fundamental strings rarely are the most relevant degrees of freedom as we move around in parameter space. Our initial comments on open metastrings should make it clear that familiar non-perturbative objects are expected to be present in the theory and should be expected to play a central role. The worldsheet perspective does allow us to extract useful information, such as the symmetry group of the quantum metastring that generalizes space-time diffeomorphism symmetry. Finally, we have considered in this paper a purely bosonic worldsheet theory; it is natural to ask what role supersymmetry might play. We see a range of possibilities here --- the first would involve generalizing the worldsheet theory to a summation over a Lorentzian version of super Riemann surfaces; a second would be to attain target space supersymmetry only concomitantly with a suitable extensification limit. In this second instance, one would then view all of the usual string dualities as special stable properties of a certain subset of extensifications.

We hope to address these and many other issues in future publications. 
We expect that the metastring will offer new paradigms for particle phenomenology, for cosmology and for problems in gravity in which quantum mechanics and locality play a central role, and it may offer new insights into standard problems in string theory, such as vacuum selection. 
On the technical side, our future work will include the full quantization of the worldsheet theory, the further development of Lorentzian worldsheet techniques and the generalization to more general backgrounds. In particular we expect that the interplay between the requirements of quantum consistency (worldsheet Weyl and Lorentz invariance) and the formulation of curved backgrounds including string interactions and non-perturbative structures will tell an interesting story.

\vskip .5cm

{\bf Acknowledgements:}
We thank L. Anderson, P. Argyres, V. Balasubramanian, B. Basso, D. Berman, R. Blumenhagen, M. Buric, F. Cachazo, S. Caron-Huot, L. N. Chang, T. Curtright, M. Cvetic, S. Das, M. Duff,  E. Gimon, J. Gomis, J. Gray, M. G\"{u}naydin, A. Hanany, J. Heckman, P. Horava, Y. Hui-He, C. Hull, V. Jejjala, T. Kephart, M. Kruczenski, D. L\"{u}st, J. Madore, J. Maldacena, L. McAllister, R. Myers, L. Pando Zayas, O. Parrikar, J. Polchinski, R. Plesser, S. Ramgoolam, R. Roiban, W. Siegel,
K. Sfetsos, E. Sharpe, A. Shapere, R. Szabo, T. Takeuchi, W. Taylor,  C. Thorn, A. Tseytlin, C. Tze, H. Verlinde, D. Vaman, P. Vieira,  C. Zachos and  B. Zwiebach for
discussions, comments and communications. {\small RGL} and {\small DM} thank Perimeter Institute for
hospitality. {\small RGL} is supported in part by
the U.S. Department of Energy contract DE-FG02-13ER42001 and
{\small DM}
by the U.S. Department of Energy
under contract DE-FG02-13ER41917. Research at Perimeter Institute for
Theoretical Physics is supported in part by the Government of Canada through NSERC and by the Province of Ontario through MRI.

\appendix\numberwithin{equation}{section}
\section{Appendix: Open strips and open strings}

Although the main purpose of this work is the description of the closed string, in this Appendix we briefly remark that the Nakamura construction can be generalized to the case of internal worldsheet boundaries. We will take this to mean that we can have edges to strips corresponding to boundaries; the appropriate condition at such an edge should be that the symplectic flux vanishes there. 
This condition defines the open string boundary condition. 
Recall that above we quoted the symplectic 2-form that follows from the Tseytlin action
\beqn\label{symp2formopen}
\omega^a=
\frac{1}{4\pi}\epsilon^{ab}\delta\X \dd \pa_b\delta\X
+\frac{1}{8\pi}\epsilon^{ab}\pa_b(\delta\X \omega \delta\X)
+\frac{1}{2\pi}\sigma^a \delta\X\dd\delta\S-\frac{1}{4\pi} \epsilon^{ab}\pa_b\delta\alpha .
\eeqn
Correspondingly, the symplectic flux at an edge ${\bm e}$ is (again, in the gauge $\S=0$) 
\beqn
\Phi_{{\bm e}}&=& \frac{1}{4\pi}\int d\tau
\Big(-\delta\X_{{\bm e}} (\eta+\omega) \pa_\tau\delta\X_{{\bm e}}
+\pa_\tau\delta\alpha_{{\bm e}}
\Big) ,
\eeqn
and a condition on this flux is equivalent to a choice of boundary condition.
 
Consider a strip with an edge that is a worldsheet boundary. 
In this case, we can think of the form $\alpha$ as induced by the addition of boundary terms in the action. 
This will vanish automatically, as long as we take consistent boundary conditions. For example, if we simply take $\alpha_{\bm e}=0$, then we must impose
\beq
\delta\X_{{\bm e}} (\eta+\omega) \pa_\tau\X_{{\bm e}}=0.
\eeq
If we simply impose the boundary equation of motion this reduces to a Dirichlet condition in phase space
\be \label{openDirichlet}
(\eta+\omega) \pa_\tau\X_{{\bm e}}=0.
\ee
The interpretation of this boundary condition depends on the properties of $\eta+\omega$. 
Indeed this condition  means that the worldsheet boundary is associated with a D-brane in phase space which lies  along 
${L} \equiv \ker(\eta +\omega)$. 
Note that since $\S=0$, the boundary condition is equivalent to a Neumann condition
\beq\label{openNeumann}
 (\eta+\omega) J\pa_\s\X_{{\bm e}}=0.
\eeq
The equivalence of these boundary conditions  means that we have {\it both} Dirichlet and Neumann boundary conditions simultaneously. 
The Dirichlet brane is along $L$ while the Neumann boundary conditions are imposed in the directions of $\tilde{L} =J(L)$.
More precisely, we should interpret this to mean that a worldsheet boundary is associated with a D-brane in phase space which lies along a Lagrangian submanifold.\footnote{The Lagrangian of the D-brane does not in general bear any relation to a choice of space-time Lagrangian (here we speak classically for simplicity). Indeed, the choice (\ref{openDirichlet}) could be rotated by an element of $O(d,d)$, yielding a Lagrangian that does not line up with $\ker(\eta+\omega)$. In such a case, the perceived dimensionality of the D-brane in space-time would be given by the intersection of the two Lagrangians.} 

In order to understand these conditions we introduce the following operator and kernels 
\be
K \equiv \eta^{-1}\omega,\quad {L} \equiv \ker(1 + K ),\quad
 \tilde{L} \equiv \ker(1 + JKJ ).
\ee
By construction the chirality operator exchanges $L$ and $\tilde{L}$: $J(L)=\tilde{L}$.
These kernels are relevant since the boundary conditions can be written in term of these as
\be
\pa_{\tau} \X_{{\bm e}} \in {L},\qquad \pa_{\s}\X_{{\bm e}} \in \tilde{L}.
\ee
The key property satisfied by $L$ is that it is a subspace which is null with respect to both $\eta$ and $\omega$ while  $\tilde{L}$ is a subspace which is null with respect to $\eta$.
This follows from the fact that if $X,Y\in L$ then $(\eta+\omega)X,Y=0$ and 
\be
2 X\cdot Y = X^A\eta_{AB}Y^B + Y^A\eta_{AB} X^B=-  X^A \omega_{AB} Y^B - Y^A \omega_{AB} X^B =0.
\ee
This in turn implies that $\omega(X,Y)=0$. Since $J$ is an O$(d,d)$ transformation $\tilde{L}=J(L)$ is also null with respect to $\eta$.

If we also assume that $\eta +\omega$ is maximally degenerate, as is the case for the Polyakov string, this means that $L$ is of maximal dimension.
That is, $L$ is a {\it Lagrangian} submanifold of $\P$. The string is moving freely along this  D-brane. 
Indeed we 
note that $\tilde{L}$ labels the directions transverse to the brane, which carries dynamical degrees of freedom.
The fact that $J(L)=\tilde{L}$ and that $L$ is $\eta$-null implies that 
$\tilde{L}$ is {\it orthogonal} to $L$ with respect to $H$. Indeed if $X\in L$ and $Y=J(\tilde{X}) \in \tilde{L}$ then 
\be
H(X,Y)= \eta(X,\tilde{X})=0.
\ee
In other words, the $\eta$-null subset $\tilde{L}$ is transverse to $\tilde{L}$ and can be thought of as  as momentum space: that is, we have  $L \cap \tilde{L} =\{0\}$ and $ L\oplus \tilde{L}= {\cal P}$.

If $(\eta +\omega)$ is not maximally degenerate the D-brane $L$  is still null with respect to $(\eta,\omega)$ and $\tilde{L}$ is still orthogonal to it.
However $ L\oplus \tilde{L}$ no longer covers $\P$ and the components of $\X$  in ${\cal P}\backslash L\oplus \tilde{L}$ are non-dynamical since they satisfy both Neumann and Dirichlet conditions, but not of maximal dimension.
This implies, in particular, that  if one desires to  maximize the amount of propagating degrees of freedom of the open string, which is half the dimension of phase space, then one needs to impose the  condition that $(\eta +\omega)$ is maximally degenerate.

This interpretation  is consistent with our original  insight regarding the understanding of T-duality
as a Fourier transform \cite{Freidel:2013zga}. Indeed the T-duality transformations $\X\mapsto J(\X)$,
exchanges $L$ and $\tilde{L}$, in agreement with the demand that Fourier transform exchanges space and momentum space.
Our results also show  that the D-brane is always {\it Lagrangian}.

\subsection{Open string symplectic structure}

Let us assume that the boundary condition (\ref{openDirichlet}) is imposed at both edges $\s=0,2\pi$ of the strip. In order to simplify the analysis we also assume that the conditions $L\cap \mathrm{Im}(1\pm J) =\{0\}$ are satisfied. Such conditions are fulfilled by the Polyakov parameterization given in (\ref{etaH}). 
Under such conditions, it can be seen that the general open string solution is then parametrized in terms of an element $X(\s) \in L$ which is $4\pi$-periodic, a monodromy $\Delta \in \tilde{L}$ and is given by
\be
\X(\s,\tau)= x + \frac{\Delta}{2\pi}\s + \frac{J\Delta}{2\pi} \tau + \frac12(1+J)X(\tau+\s) +\frac12(1-J)
X(\tau-\s).
\ee
The open string symplectic structure can then be derived directly from the action.
Since the boundary equations of motion have been imposed, no additional boundary condition is needed in the construction. One simply gets 
\bea
\Omega &=&\frac12\oint \rd\s \delta{\X}^{A}(\eta+ \omega)_{AB} \pa_{\sigma}\delta \X^{B}.
\eea
Using the decomposition of the field  $\X(\sigma,0)=
x + \tfrac{\Delta}{2\pi}\s + \widehat{X}(\s) $, where $x$ is the center of mass of the string,
and introducing the midpoint of the string:
\be
x_{m}\equiv \frac12 (\X(0) + \X(2\pi)),\ee we obtain the decomposition
\bea
\Omega&=&\frac1{2\pi} \eta_{AB} \delta \tilde{x}^{A}\delta \Delta^{B} 
+ 
\frac12 \oint \rd\s  \delta{ X}^{A}(\eta+ \omega)_{AB} \pa_{\sigma}\delta  X^{B} ,
\eea
where $ \tilde{x} \equiv x-\frac12 (1+K) x_{m}$.
This symplectic structure possesses the following gauge invariance
\be
 x_{m} \mapsto x+ a,\qquad x\mapsto x+\frac12(1+K)a.
\ee
We can use this to fix the midpoint value $x_{m}=0$.
If $L\cap \tilde{L} \neq \{0\}$ we have additional invariances where 
\be
\X \mapsto \X + \mathbb{A}, \qquad \mathbb{A} \in L \cap \tilde{L}.
\ee
This is why we restrict the study to the case where $L\cap \tilde{L}=\{0\}$; otherwise this would imply that the open string has less than half the number of degrees of freedom of the closed string.
In this  case, the tensor $(\eta +\omega)$ is invertible on $L$ and we denote the inverse restricted to $L$ by $(\eta +\omega)^{-1}_L$. After gauge fixing the midpoint value to $0$, the symplectic structure is invertible and it leads to the brackets 
\be
\{x^{A},\Delta^{B}\}= 2\pi \eta^{AB},\qquad \{ \X_{n}^{A}, \X_{m}^{B}\}= \frac{\delta_{n+m}}{ i n}\left[(\eta+\omega)^{-1}_L\right]^{AB}.
\ee

%\bibliographystyle{uiuchept}
%\bibliography{metastring}

\providecommand{\href}[2]{#2}\begingroup\raggedright\endgroup

\end{document}